\documentclass[aps,prd,reprint,superscriptaddress,amsmath,amssymb,longbibliography,floatfix,raggedbottom]{revtex4-2}
\usepackage{fontspec,amsthm,mathtools,bm,graphicx,booktabs,array,xcolor,etoolbox}
\usepackage{placeins,hyperref,orcidlink,mathrsfs}
\hypersetup{colorlinks=true,linkcolor=blue,citecolor=blue,urlcolor=blue,
pdfauthor={Kun Zhou and Jin Li},
pdftitle={Full-Stokes response to a stochastic gravitational-wave background in pulsar timing arrays: Fully analytic overlap reduction functions at finite distance and coherence of close pulsar pairs}}
\makeatletter
\AtBeginDocument{\immediate\write\@auxout{\string\citation{apsrev42Control}}}
\makeatother
\newtheorem{theorem}{Theorem}
\newtheorem{proposition}{Proposition}
\newtheorem{corollary}[theorem]{Corollary}
\newtheorem{lemma}{Lemma}
\newcommand{\ii}{\mathrm i}
\newcommand{\dd}{\mathrm d}
\newcommand{\bn}{\hat{\boldsymbol n}}
\newcommand{\bu}{\hat{\boldsymbol u}}
\newcommand{\bv}{\hat{\boldsymbol v}}
\newcommand{\sinc}{\operatorname{sinc}}

\begin{document}
\title{Full-Stokes response to a stochastic gravitational-wave background\protect\\ in pulsar timing arrays: Fully analytic overlap reduction functions\protect\\ at finite distance and coherence of close pulsar pairs}
\author{Kun Zhou~\orcidlink{0009-0001-1099-8623}}
\affiliation{College of Physics, Chongqing University, Chongqing 401331, China}
\affiliation{Department of Physics and Chongqing Key Laboratory for Strongly Coupled Physics, Chongqing University, Chongqing 401331, China}

\author{Jin Li~\orcidlink{0000-0001-8538-3714}}
\email{cqujinli1983@cqu.edu.cn}
\affiliation{College of Physics, Chongqing University, Chongqing 401331, China}
\affiliation{Department of Physics and Chongqing Key Laboratory for Strongly Coupled Physics, Chongqing University, Chongqing 401331, China}

\date{November 4, 2026}
\begin{abstract}
We derive fully analytic overlap reduction functions for the full-Stokes response of a pulsar timing array to an anisotropic tensor gravitational-wave background in general relativity. The response retains both pulsar distances and their complete propagation phases. At each fixed finite sky multipole, the intensity and circular polarization responses reduce to finite combinations involving exponential integrals and logarithms, while the electric and magnetic linear polarization responses require only elementary functions. We evaluate all components through $\ell=8$. Symmetry separates exact nulls from additional Earth-term nulls that propagation can lift. Even an unpolarized background can produce an imaginary cross-spectrum. For a pair with three-dimensional separation $d$, the pulsar terms can remain coherent when $2\pi fd/c\lesssim1$, even if both pulsars are many wavelengths from Earth. A uniform approximation retaining all four Earth and pulsar terms connects this regime to the short-wavelength limit, and a rotationally invariant relation fixes the leading transverse magnetic response. Common-distance errors cancel from the aligned pair phase, while relative-depth errors directly affect coherence. Under Gaussian differential averaging, fixed low-multipole norms and the squared norm summed over all multipoles decay at different rates as the response shifts to higher multipoles. Published pulsar configurations illustrate the roles of distance accuracy, bandwidth, and timing fits. At fixed pointwise intensity, a polarization covariance bound and a relative-depth noise budget quantify when differencing suppresses the background. Independent sky integration and matched-accuracy benchmarks test the analytic expressions over the parameter ranges studied.
\end{abstract}
\maketitle
\section{Introduction}
\label{sec:intro}

Pulsar timing arrays (PTAs) search for gravitational waves through correlations in the arrival-time fluctuations of millisecond pulsars~\cite{Sazhin1978,Detweiler1979}. Evidence for a nanohertz stochastic gravitational-wave background from NANOGrav, the European and Indian PTAs, the Parkes PTA, and the Chinese PTA has broadened interest from the overall spectrum to the angular distribution and polarization of the background~\cite{paper_NANOGrav2023Evidence,paper_EPTA2023Evidence,paper_PPTA2023Evidence,paper_CPTA2023Evidence}. In the short-wavelength limit, the Hellings--Downs curve describes an isotropic, unpolarized tensor background~\cite{paper_HellingsDowns1983}. Departures from this angular pattern can be described by overlap reduction functions (ORFs) that resolve anisotropy and polarization.

Intensity anisotropy describes how power is distributed on the sky. Circular and linear polarization measure, respectively, the power imbalance and the coherence between the two tensor helicities. The Stokes parameters describe these properties in a common language. For tensor waves, $Q\pm iU$ have spin weights $\pm4$ and can be decomposed into electric-type ($E$) and magnetic-type ($B$) multipoles. These observables complement intensity maps in studies of the origin and source populations of the background~\cite{paper_KatoSoda2016,paper_SatoPolitoKamionkowski2022,paper_KumarEtAl2024}. Throughout this paper, polarization refers to the statistical state of the usual $+$ and $\times$ modes.

Two classes of pulsars illustrate why the propagation terms matter. Nearby pulsars, such as PSR~J0437$-${}4715, for which precision timing and an accurate distance are available~\cite{obs_Reardon2024Near}, provide a test of finite path-length effects. Distance priors also enter pulsar-term analyses of continuous waves~\cite{NG15CW}. The size of the propagation correction depends on the product of frequency and distance, so proximity to Earth alone does not guarantee a large effect. Globular clusters, on the other hand, contain pairs that are close on the sky and may also be close in three dimensions. The common response of cluster pulsars has been proposed as a way to supplement PTA searches~\cite{obs_Maiorano2021GC}. A more recent proposal uses differences between the timing residuals of neighboring pulsars to suppress the common background in searches for local dark matter substructure~\cite{obs_Cherukupalli2026DM}. Its projected sensitivity depends directly on the coherence of the background across the pair. Pulsar timing already constrains cluster mass distributions and dark stellar remnants~\cite{obs_Prager2017GC,obs_Smith2024GC}. Studies of environmental correlations also require a reference response to an external background.

These applications raise three questions. How does finite distance change the angular response to each polarization? Which components remain coherent for closely spaced sight lines? How do distance errors and observing windows modify the response? The conditions for the Earth-term approximation are well understood~\cite{paper_Anholm2009,paper_MingarelliMingarelli2018}. Even when both individual pulsar phases are large, the phase difference can remain finite. The additional correlations of spatially close pairs can then alter both isotropic and anisotropic ORFs~\cite{paper_MingarelliSidery2014}. For isotropic tensor backgrounds, power-series and closed-form solutions retaining the pulsar terms are available~\cite{Boitier2021,uniform_asymptotics_Hu2022}. At a given frequency, coherence depends on both transverse separation and relative line-of-sight depth. Geometric nulls differ between polarizations, so a common coherence scale can give rise to different angular behavior.

Previous work has developed optimal cross-correlation statistics, generalized ORFs, and Bayesian inference for anisotropic backgrounds~\cite{paper_Anholm2009,paper_Mingarelli2013,paper_TaylorGair2013}, as well as spherical-harmonic representations of the sky~\cite{paper_Gair2014}. Anisotropy searches illustrate how array geometry and priors affect constraints on higher multipoles~\cite{paper_NANOGrav2023Anisotropy}. The anisotropic response to circular polarization, its estimators, and its observability have also been studied extensively~\cite{paper_KatoSoda2016,Belgacem2020,paper_SatoPolitoKamionkowski2022,next_JimenezCruz2024}. Chu, Liu, and Ng retained the pulsar terms in a line-of-sight treatment of the full-Stokes response~\cite{paper_ChuLiuNg2021}. Liu and Ng formulated this response in harmonic space~\cite{paper_LiuNg2022}, and Bernardo, Liu, and Ng extended the treatment to other propagation speeds and polarization degrees of freedom~\cite{paper_BernardoLiuNg2024}. These finite-distance approaches involve intermediate angular-momentum sums and radial response integrals. Closed forms are also available for the tensor-harmonic response of a single finite baseline~\cite{paper_AndrianovKopeikin2026}, and phase-coherent sky mapping can reconstruct complex polarization amplitudes as functions of direction and frequency~\cite{Curylo2026}. The Stokes multipoles considered here describe the two-point covariance of two baselines. In the Earth-term limit, Kumar and Kamionkowski expressed the intensity and polarization responses in bipolar spherical harmonics~\cite{paper_KumarKamionkowski2024}, and Kumar et al. studied estimators for linear polarization and their sensitivity~\cite{paper_KumarEtAl2024}.

For an unpolarized covariance, Zhou and Li obtained fully analytic anisotropic ORFs that retain both pulsar-distance phases~\cite{paper_ZhouLi2026}. We extend their approach to a general tensor polarization covariance. For each fixed $(\ell,m)$, the full-Stokes sky integral reduces to a closed form. This allows us to compare polarization selection rules and follow the effect of the differential propagation phase on close-pair coherence. Averaging over distance distinguishes errors in the common distance from errors in relative depth. It also distinguishes the attenuation of a fixed low multipole from redistribution across the angular spectrum. Both distinctions matter when specifying distance accuracy or choosing a multipole cutoff.

We consider tensor plane waves in general relativity (GR) propagating at the speed of light, with fixed sight lines and a common receiver at Earth. The two-point Stokes spectrum is local in sky direction and may have arbitrary anisotropy. The response does not require Gaussianity. We use publicly available pulsar configurations to compare responses conditional on specified distance models. A polarization bound, a relative-depth error budget, and a controlled local-signal injection test the conditions for background cancellation. We do not model cluster dynamics, infer dark matter population parameters, or refit timing data.

Section~II defines the response and statistical conventions. Section~III derives its symmetries and reduces the number of independent integrals. Section~IV establishes the analytic structure and illustrates finite-distance effects through angular responses and nearby pulsars. Section~V treats close-pair coherence, distance averaging, observational conditions, and the noise budget for differencing. Section~VI reports accuracy and computational cost. Section~VII summarizes our results. The appendices give proofs and representative closed forms. Our explicit calculations cover $\ell\leq8$, while the closure theorem holds at every fixed finite multipole.

\section{Tensor polarization and the full pulsar response}\label{sec:response}
\subsection{Two-point spectrum and polarization conventions}
Let $\boldsymbol n$ denote the propagation direction of the gravitational wave. We use the Fourier phase
$e^{2\pi\ii f(t-\boldsymbol n\cdot\boldsymbol x/c)}$. The local tangent basis is
$\boldsymbol e_\theta,\boldsymbol e_\phi$, and we choose the polarization tensors
\begin{align}
e^+_{ij}&=e_{\theta i}e_{\theta j}-e_{\phi i}e_{\phi j},\nonumber\\
e^\times_{ij}&=e_{\theta i}e_{\phi j}+e_{\phi i}e_{\theta j},
\qquad e^A:e^{A'}=2\delta^{AA'}.
\end{align}
At positive frequencies we write the two-point spectrum as
\begin{equation}
\begin{split}
\langle h_A(f,\boldsymbol n)h_B^*(f',\boldsymbol n')\rangle
=\frac{\delta(f-f')\delta^2(\boldsymbol n,\boldsymbol n')}{4\pi}
\mathcal P_{AB}(f,\boldsymbol n),
\end{split}
\end{equation}
and define the polarization spectral matrix by
\begin{equation}\label{eq:stokes_matrix}
\mathcal P_{AB}=\frac12
\begin{pmatrix}I+Q&U-\ii V\\U+\ii V&I-Q\end{pmatrix}.
\end{equation}
Positive semidefiniteness requires $I\ge0$ and $I^2\ge Q^2+U^2+V^2$. Stationarity makes the spectrum diagonal in frequency. Locality in sky direction is a separate assumption that excludes correlations between different propagation directions.

Under a passive rotation of the tangent basis by $\psi$, the combinations $Q\pm\ii U$ transform as
$e^{\mp4\ii\psi}(Q\pm\ii U)$. We therefore expand $I,V$ in spin-zero harmonics and the linear polarization in spin-$\pm4$ harmonics. The $E/B$ multipoles start at $\ell=4$, as expected for a tensor Stokes field~\cite{paper_KumarKamionkowski2024,paper_KumarEtAl2024}. Here $B$ denotes the parity-odd multipole of the linear polarization. It differs from the tensor curl mode of the strain field, and its response alone does not determine source parity.

We adopt the Condon--Shortley phase and the normalization
\begin{equation}
\widetilde Y_{\ell m}=\sqrt{4\pi}Y_{\ell m},\qquad
{}_s\widetilde Y_{\ell m}=\sqrt{4\pi}\,{}_sY_{\ell m},
\end{equation}
where $\widetilde Y_{00}=1$ and
\begin{equation}
{}_s\widetilde Y_{\ell m}(\theta,\phi)
=(-1)^s\sqrt{2\ell+1}\,d^\ell_{m,-s}(\theta)e^{\ii m\phi}.
\end{equation}
Spherical harmonics in the response integral are unconjugated. Together with our ordering of the cross-correlation, this fixes the signs under rotation and exchange and the sign of the circular response.

\subsection{Finite-distance response and Stokes projections}
For a pulsar at distance $L_p$ in direction $\boldsymbol p$, write
\begin{equation}
a_p=\frac{2\pi fL_p}{c},\qquad d_p=1+\boldsymbol n\cdot\boldsymbol p,
\qquad \Psi_a(d)=\frac{1-e^{-\ii ad}}d,
\end{equation}
with the continuous extension $\Psi_a(0)=\ii a$. The fractional frequency-shift response is
\begin{equation}\label{eq:pulsar_response}
R_p^A=t_p^A\Psi_{a_p}(d_p),\qquad
t_p^A=\frac12p^ip^je^A_{ij},
\end{equation}
and is regular at $\boldsymbol n=-\boldsymbol p$~\cite{paper_EstabrookWahlquist1975,paper_Anholm2009}.
One time integration converts the fractional frequency shift to a timing residual, introducing a factor $(2\pi f)^{-2}$ in the covariance.

We order the correlation as $R_uR_v^*$ and define
\begin{align}
K_I&=R_u^+R_v^{+*}+R_u^\times R_v^{\times*},\nonumber\\
K_V&=\ii(R_u^\times R_v^{+*}-R_u^+R_v^{\times*}),\nonumber\\
K_Q&=R_u^+R_v^{+*}-R_u^\times R_v^{\times*},\nonumber\\
K_U&=R_u^+R_v^{\times*}+R_u^\times R_v^{+*}.
\end{align}
The full-Stokes ORFs are
\begin{align}
\Gamma^{X;\ell m}_{uv}
&=\frac1{8\pi}\int\dd\Omega\,\widetilde Y_{\ell m}K_X,
\quad X=I,V,\label{eq:IV_response}\\
\Gamma^{\pm;\ell m}_{uv}
&=\frac1{8\pi}\int\dd\Omega\,{}_{\pm4}\widetilde Y_{\ell m}
(K_Q\mp\ii K_U),\label{eq:spin_response}\\
\Gamma^{E;\ell m}_{uv}&=\frac{\Gamma^{+;\ell m}_{uv}+\Gamma^{-;\ell m}_{uv}}2,
\nonumber\\
\Gamma^{B;\ell m}_{uv}&=\frac{\ii}{2}
(\Gamma^{+;\ell m}_{uv}-\Gamma^{-;\ell m}_{uv}).\label{eq:EB_response}
\end{align}
Expanding $I,V$ in $\widetilde Y_{\ell m}$ and adopting
\begin{equation}
Q\pm\ii U=\sum_{\ell m}(E_{\ell m}\pm\ii B_{\ell m})
{}_{\pm4}\widetilde Y_{\ell m},
\end{equation}
gives the redshift cross-spectrum
\begin{equation}
C_{uv}(f)=\sum_{\ell m}\sum_{X=I,V,E,B}
X_{\ell m}(f)\Gamma^{X;\ell m}_{uv}(f).
\end{equation}
The normalization in the $I$ channel agrees with the tensor response of Zhou and Li~\cite{paper_ZhouLi2026}. Contracting the spin harmonics with the polarization response before integration removes dependence on the local polarization basis. The uncontracted $+\times$ components separately depend on that choice.

\subsection{Pulsar-pair frame and the four propagation terms}
We evaluate the explicit formulas in the symmetric frame
\begin{equation}\label{eq:pair_frame}
\begin{aligned}
\boldsymbol u&=(C,-S,0),\quad\boldsymbol v=(C,S,0),\\
C&=\cos\frac\gamma2,\quad S=\sin\frac\gamma2.
\end{aligned}
\end{equation}
The distances $L_u,L_v$ remain attached to their respective directions. Writing
$\tilde x=\boldsymbol n\cdot\boldsymbol u$ and $\tilde y=\boldsymbol n\cdot\boldsymbol v$, we obtain the propagation product
\begin{equation}\label{eq:T_transfer}
T=(1-e^{-\ii a_ud_u})(1-e^{\ii a_vd_v}).
\end{equation}
Let $W^X_{\ell m}$ denote the complete polarization contraction with the common propagation factor removed. For example,
\begin{align}
W^I_{\ell m}&=\widetilde Y_{\ell m}(t_u^+t_v^++t_u^\times t_v^\times),\nonumber\\
W^V_{\ell m}&=\ii\widetilde Y_{\ell m}(t_u^\times t_v^+-t_u^+t_v^\times),\nonumber\\
W^\pm_{\ell m}&={}_{\pm4}\widetilde Y_{\ell m}
(t_u^+\mp\ii t_u^\times)(t_v^+\mp\ii t_v^\times),
\end{align}
and we combine $W^E,W^B$ according to Eq.~\eqref{eq:EB_response}. Thus
\begin{equation}\label{eq:common_integral}
\Gamma^{X;\ell m}_{uv}
=\frac1{8\pi}\int\frac{W^X_{\ell m}T}{d_ud_v}\,\dd\Omega.
\end{equation}
separates the propagation factor from the polarization geometry.

Expanding Eq.~\eqref{eq:T_transfer} gives the four terms $EE,EP,PE,PP$. The first letter refers to endpoint $u$ and the second to endpoint $v$. Thus $EP$ carries the phase at $v$, and $PE$ carries the phase at $u$. We include the minus signs in $EP,PE$, so the four contributions add directly. Equal-distance collinearity is an ideal geometric limit for two distinct pulsars. Equal signal responses permit independent intrinsic noises, and the cross-only examples below use no auto-power information.

\section{Symmetries of the overlap reduction functions}\label{sec:symmetry}
\label{sym:scope}\label{sym:reflection}\label{sym:exchange}\label{sym:zeros}
An anisotropic ORF depends on both the opening angle of a pulsar pair and its orientation relative to the sky multipole. We distinguish components related by a change of frame, nulls enforced by polarization geometry, and additional nulls of the Earth-term approximation. The first two reduce the number of independent integrals. Propagation phases can lift the third.

With the unconjugated harmonic projection of Sec.~II, write
\begin{equation}
 p_I=p_E=1,\qquad p_V=p_B=-1.
 \label{sym:parity-label}
\end{equation}
Except where we specify the $EE$ contribution, the following results retain the full propagation factor. Proofs are given in Appendix~\ref{sym:appendix}.

A common rotation preserves the relative geometry of the two light paths. It changes only their orientation with respect to the sky basis. The rotation law for each multiplet therefore determines the response for an arbitrarily oriented pair.
\begin{theorem}[Rotational covariance]\label{sym:prop-spatial}
Under an active rotation $R$, the full response obeys
\begin{equation}
 \begin{split}
 &\Gamma^{X;\ell m}(R\bu,R\bv;a_u,a_v)\\
 &\quad=\sum_{k=-\ell}^{\ell}[D^\ell_{mk}(R)]^*
 \Gamma^{X;\ell k}(\bu,\bv;a_u,a_v),
 \end{split}\label{sym:rotation}
\end{equation}
where $D^\ell_{mk}(\alpha,\beta,\chi)=e^{-\ii m\alpha}d^\ell_{mk}(\beta)e^{-\ii k\chi}$.
\end{theorem}
It suffices to integrate in the standard pair frame and apply a finite-dimensional rotation matrix. At fixed $\ell$, rotation mixes the $m$ components. A zero of one component can therefore depend on the frame, whereas the norm of the multiplet is independent of the polar axis. We use this norm below to characterize coherence in the magnetic sector.

Reflection in the pair plane preserves both endpoint phases in the standard frame. The reflection parity of the polarization contraction then restricts the sky multipoles that can couple to the pair.
\begin{theorem}[Reflection in the pair plane]\label{sym:thm-reflection}
In the standard pair frame $\bu=(C,-S,0)$, $\bv=(C,S,0)$, the full response satisfies
\begin{equation}
 \Gamma^{X;\ell m}=p_X(-1)^{\ell+m}\Gamma^{X;\ell m}.
 \label{sym:selection}
\end{equation}
\end{theorem}
\begin{corollary}[Nulls and independent components]\label{sym:cor-zeros}
The $I/E$ sectors allow only even $\ell+m$, and the $V/B$ sectors allow only odd $\ell+m$. In any frame,
\begin{align}
 \Gamma^{V;00}&=0,\label{sym:V00}\\
 \Gamma^{V;\ell m}=\Gamma^{B;\ell m}&=0,
 \qquad\gamma=0,\pi.\label{sym:collinear}
\end{align}
\end{corollary}
At each multipole the selection rule leaves $\ell+1$ even-sector and $\ell$ odd-sector components. Summing the components for $\ell\le8$ in the order $(I,V,E,B)$ gives
\begin{equation}
 \begin{gathered}
 (81,81,65,65)\longrightarrow(45,36,35,30),\\
 \text{total}:\quad292\longrightarrow146.
 \end{gathered}\label{sym:counts}
\end{equation}
Here $E/B$ start at $\ell=4$. Thus 146 potentially nonzero integrals remain for a generic pair. The number of parameters an array can measure is a separate question. The vanishing $V$ monopole and the collinear $V/B$ null survive rotations and finite distance. They can be excluded before integration, and the near-collinear response at fixed finite distance must approach the corresponding null.

The $I$ branch reproduces the unpolarized tensor result of Zhou and Li~\cite{paper_ZhouLi2026}. The odd reflection parity of $V/B$ pertains to the Stokes decomposition, independently of the vector and breathing mode labels used in that work. We follow the established spin-four and $E/B$ conventions~\cite{paper_KumarKamionkowski2024,paper_KumarEtAl2024}.

\begin{figure*}[t]
\centering\includegraphics[width=\textwidth]{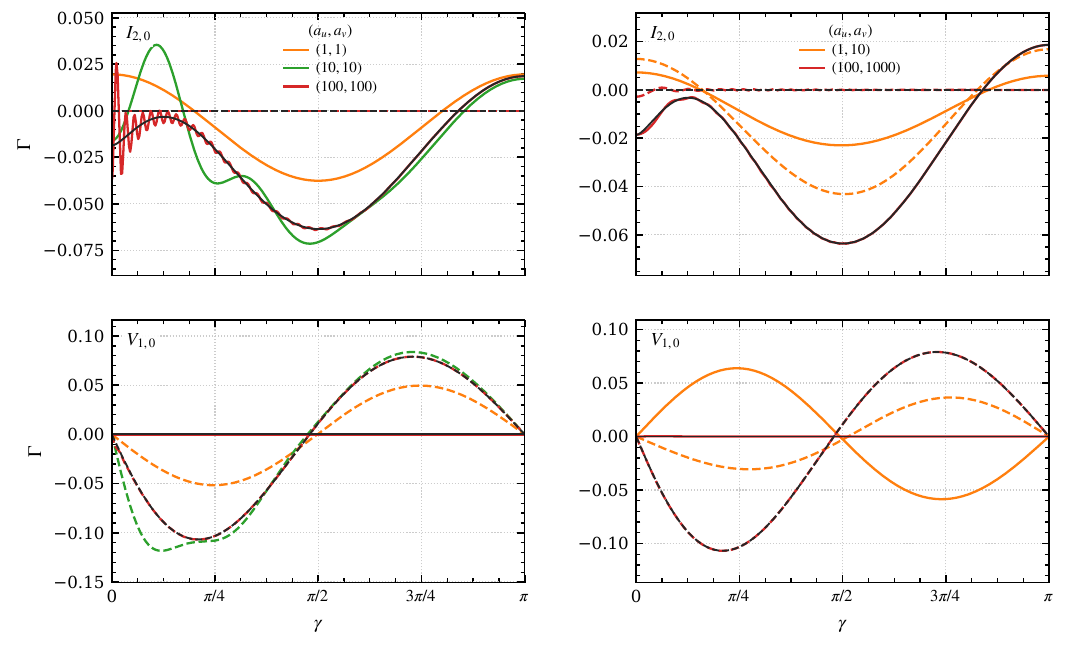}
\caption{Representative angular responses to intensity and circular polarization. Solid and dashed curves show the real and imaginary parts, respectively. Black curves give the Earth term. The distance parameters indicated in each column apply throughout that column.}\label{fig:angular_I}\label{fig:angular_V}
\end{figure*}

\begin{figure*}[t]
\centering\includegraphics[width=\textwidth]{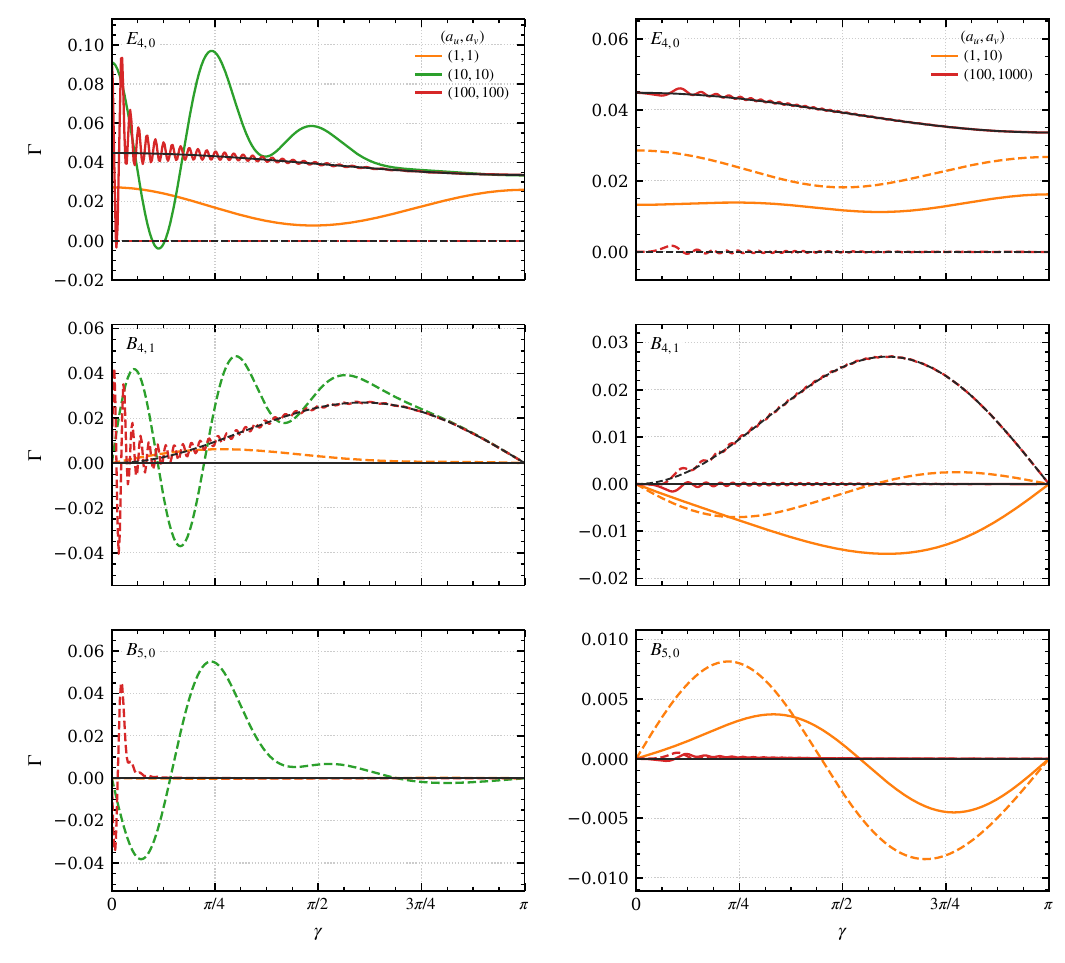}
\caption{Representative angular responses to linear polarization. Solid and dashed curves show the real and imaginary parts, respectively. Black curves give the Earth term. Distance parameters are shared within each column.}\label{fig:angular_E}\label{fig:angular_B}
\end{figure*}

At finite distance the response is generally complex. Interchanging both endpoints of an ordered pair invokes Hermiticity of its cross-spectrum. Swapping only the distances at fixed geometry requires an additional spatial symmetry.
\begin{theorem}[Hermiticity for ordered pairs]\label{sym:prop-Hermitian}
In a fixed global frame, interchanging the sight lines together with their distances gives
\begin{equation}
 \begin{split}
 &\Gamma^{X;\ell m}(\bu,\bv;a_u,a_v)\\
 &\quad=(-1)^m[\Gamma^{X;\ell,-m}(\bv,\bu;a_v,a_u)]^*.
 \end{split}\label{sym:ordered}
\end{equation}
\end{theorem}
\begin{corollary}[Distance exchange and equal-distance response]\label{sym:cor-distance}
Swapping only the distances in the standard pair frame gives
\begin{equation}
 \Gamma^{X;\ell m}(a_u,a_v;\gamma)
 =p_X[\Gamma^{X;\ell m}(a_v,a_u;\gamma)]^*.
 \label{sym:distance}
\end{equation}
For equal distances it follows that
\begin{equation}
 \Gamma^{I/E;\ell m}\in\mathbb R,\qquad
 \Gamma^{V/B;\ell m}\in\ii\mathbb R,
 \qquad a_u=a_v.
 \label{sym:equal-distance}
\end{equation}
\end{corollary}
This corollary combines ordered exchange, $R_x(\pi)$, and reflection in the pair plane. One distance ordering thus determines the other. Equation~\eqref{sym:ordered} has no extra sign for $V/B$. The factor $p_X$ in Eq.~\eqref{sym:distance} comes from the spatial transformation. These geometric and conjugation relations force either the real or imaginary part to vanish at equal distances. Unequal distances can restore it, as illustrated in Figs.~\ref{fig:angular_I} and \ref{fig:angular_E}. The restriction holds in the specified frame and by itself cannot identify polarization (Sec.~\ref{ua:polarization_identification}).

The Earth-term antenna kernel has an additional reflection symmetry. We now ask which of the resulting nulls survive when the propagation phases are restored.
\begin{theorem}[Additional Earth-term reflection]\label{sym:prop-Earth}
In the standard pair frame, retaining only $EE$ gives
\begin{equation}
 \begin{gathered}
 \Gamma_{EE}^{X;\ell,-m}=\eta_X(-1)^m\Gamma_{EE}^{X;\ell m},\\
 \eta_I=\eta_V=\eta_E=1,\qquad\eta_B=-1.
 \end{gathered}\label{sym:earth-negative-m}
\end{equation}
\end{theorem}
\begin{corollary}[Axisymmetric magnetic null of the Earth term]\label{sym:cor-Earth-B0}
In the standard pair frame,
\begin{equation}
 \Gamma_{EE}^{B;\ell0}=0.\label{sym:earth-B0}
\end{equation}
\end{corollary}
This null follows by setting $m=0$ in Eq.~\eqref{sym:earth-negative-m}. Under the same reflection, the full propagation factor obeys
\begin{equation}
 T(\theta,-\phi;a_u,a_v)=[T(\theta,\phi;a_v,a_u)]^*,
 \label{sym:transfer-reflection}
\end{equation}
which generally differs from the original $T$. The additional $EE$ null can therefore be lifted. Equation~\eqref{sym:selection} still excludes even $\ell$. The lowest allowed component is $B_{50}$, shown at finite distance in Fig.~\ref{fig:angular_B}. Its appearance changes the allowed angular structure itself. The exact collinear null persists, so this component requires noncollinear sight lines.

\section{Fully analytic anisotropic overlap reduction functions at finite distance}\label{sec:methods}\label{sec:analytic}
\subsection{Analytic method and closure}\label{sec:integration}\label{sec:direct}
Symmetry identifies the independent integrals. To obtain their dependence on frequency, distance, and opening angle, we apply the real-projection method of Zhou and Li~\cite{paper_ZhouLi2026} to the complete Stokes contraction, retaining both distance phases. For $0<\gamma<\pi$, the real projections onto the two sight lines transform the sky integral into
\begin{equation}\label{eq:direct-ellipse}
\begin{split}
\Gamma^{X;\ell m}_{uv}={}&\frac1{4\pi}\int_{-1}^{1}\dd\tilde x
\int_{\tilde y_-}^{\tilde y_+}\dd\tilde y\,
\frac{\overline W^X_{\ell m}T}{(1+\tilde x)(1+\tilde y)\sqrt{\mathcal D}},\\
\mathcal D={}&\sin^2\gamma-\tilde x^2-\tilde y^2
 +2\cos\gamma\,\tilde x\tilde y,\\
\tilde y_\pm={}&\cos\gamma\,\tilde x
\pm\sin\gamma\sqrt{1-\tilde x^2}.
\end{split}
\end{equation}
Here $\overline W$ is the average over the two hemispheres of the full numerator in Eq.~\eqref{eq:common_integral}. At fixed finite multipole it is a polynomial. Only this numerator distinguishes the Stokes channels. The path denominators and propagation factors have the same form in each channel. For each propagation term in Sec.~II, we align a real polar axis with its phase vector and perform the azimuthal and polar integrals in succession. At collinearity we take the continuous limit of the original sky integral. Appendix~\ref{app:elementary-closure} gives the required intervals and finite exponential moments.

A finite expression for each sky multipole removes the need to truncate an intermediate angular-momentum sum when scanning frequency or distance. At fixed multipole, the $I,V$ response is a finite combination of exponential integrals and logarithms. For linear polarization the spin-four contraction cancels the coefficients of these special functions, giving a stronger closure result.
\begin{theorem}[Elementary closure of the linear polarization response]\label{thm:elementary-closure}
In the GR tensor model of Sec.~II, for any fixed integers $\ell$ and $m$ satisfying $\ell\ge4$ and $|m|\le\ell$, the full $E,B$ coefficients have finite elementary expressions at positive finite distances and any opening angle, with no residual angular integral. The collinear endpoints are obtained by continuity.
\end{theorem}
Appendix~\ref{app:elementary-closure} gives the proof. Even with their spin structure, the linear polarization coefficients require only finitely many functions for each specified sky mode. The theorem applies at every fixed finite multipole. We evaluate the coefficients explicitly through $\ell=8$. Whether a sky model requires higher orders depends on its spectral tail. Elementary closure does not ensure numerical stability under direct substitution. Degenerate parameters require consistent branches, continuous limits, or local expansions.

The components below have distinct reflection parities and endpoint behavior. They show when propagation modifies an existing angular pattern and when it produces structure absent from the Earth term.

\subsection{Intensity and circular polarization}
Intensity and circular polarization measure the total power and the imbalance of the two tensor helicities, respectively. Their angular correlations therefore have different reflection parities. To isolate propagation effects, we first give their Earth-term limits. With $r=\sin^2(\gamma/2)$, the Earth-term intensity monopole for distinct pulsars is
\begin{equation}\label{eq:earth_I00}
\Gamma^{I,EE}_{00}=\frac{2-r}{12}+\frac r2\ln r,
\end{equation}
one third of the conventional Hellings--Downs curve~\cite{paper_HellingsDowns1983}. The low-order intensity Earth terms agree component by component with Zhou and Li~\cite{paper_ZhouLi2026}. The circular monopole vanishes exactly. With the polar axis normal to the pair plane, the lowest nonzero circular Earth term is
\begin{equation}\label{eq:earth_V10}
\Gamma^{V,EE}_{10}=-\frac{\ii\sqrt3 S}{6C}
\left[2(1-r)+3r\ln r\right],
\end{equation}
with continuous endpoint values. After matching frames and normalization, this agrees with the known circular response~\cite{paper_KatoSoda2016}.

Figure~\ref{fig:angular_I} compares $I_{20}$ and $V_{10}$ for the same path phases. At equal distances $I$ is real and $V$ is purely imaginary, as required by Corollary~\ref{sym:cor-distance}. At unequal distances the other quadrature appears. Thus the same polarized sky can give a different cross-spectral phase when the endpoint phases are mismatched. The positions of the zeros and the oscillatory pattern also vary with distance, precluding a universal amplitude rescaling of the Earth term. The parameters illustrate the propagation transition rather than a measured PTA distance distribution.

As the phases grow, coherent departures from the Earth term narrow toward collinearity. The two sectors have different endpoints. Nearby paths can retain coherent pulsar terms in intensity, whereas the antisymmetric circular contraction vanishes for coincident sight lines. The circular response must return to zero. Small angular separation can thus make propagation important without enhancing every polarization. The real part of $V$ shown here is a response to a circularly polarized sky. Conversely, full propagation can give intensity an imaginary response. The quadratures alone do not distinguish the polarization templates.

\label{ua:polarization_identification}
A collinear pair isolates the imaginary intensity response from circular polarization. Setting $\Delta=a_v-a_u$, we can write the full intensity monopole as
\begin{align}
 J(t)&=\frac{\ii}{4t}+\frac1{4t^2}
   +\frac{\ii(e^{2\ii t}-1)}{8t^3},\quad J(0)=\frac16,\label{ua:J}\\
 \Gamma^I_{00}&=J(0)-J(a_v)-J(-a_u)+J(\Delta).
 \label{ua:collinear_J}
\end{align}
The four terms correspond to $EE,EP,PE,PP$, respectively. At equal distances,
\begin{equation}
 \Gamma^I_{00}(a_u,a_u,0)=\frac13-\frac1{2a_u^2}
                         +\frac{\sin(2a_u)}{4a_u^3},
 \label{ua:equal_endpoint}
\end{equation}
the $1/a_u$ contributions from $EP,PE$ cancel, leaving an $1/a_u^2$ correction. The rate at which the full response approaches its limit depends on the relative phases of the terms.
These expressions agree with earlier collinear results. Equation (17) of Ref.~\cite{uniform_asymptotics_Hu2022} equals our $3\Gamma^I_{00}$, and Eq.~(43) of Ref.~\cite{paper_ZhouLi2026} gives $J(t)$ at $a_0=-t$.

At unequal distances the conjugate terms no longer cancel. The imaginary propagation response and its possible impact on circular polarization estimation have been discussed previously~\cite{paper_SatoPolitoKamionkowski2022}, and close-pair analyses have included the complex covariance at unequal distances~\cite{obs_Cherukupalli2026DM}. Equation~\eqref{ua:collinear_J} identifies the finite-distance origin of the effect and the conditions under which it survives. For $I_{00}>0$ and $Q=U=V=0$, one obtains an example independent of the choice of complex harmonic basis,
\begin{align}
 h(t)&=\operatorname{Im}J(t)=\frac{t^2-\sin^2t}{4t^3},\quad h(0)=0,\label{ua:imaginary_h}\\
 \operatorname{Im}\Gamma^I_{00}&=h(a_u)-h(a_v)+h(\Delta).
 \label{ua:imaginary_finite}
\end{align}
The contributions on the right come from $PE,EP,PP$, respectively. All three enter at finite phase.
As $a_u,a_v\to\infty$ at fixed $\Delta$, the single-pulsar terms vanish, but
\begin{align}
 \operatorname{Im}\Gamma^I_{00}&\longrightarrow h(\Delta)\ne0
 \quad(\Delta\ne0),\label{ua:imaginary_intensity}\\
 h(\Delta)&=\frac{\Delta}{12}+O(\Delta^3),\nonumber
\end{align}
where the last expression is the expansion for $\Delta\to0$. At a fixed unequal distance ratio, $|\Delta|\to\infty$ and the imaginary part decays. The surviving imaginary part measures a finite radial phase difference and can occur in an entirely unpolarized background. The collinear $V$ response vanishes identically. In the real time domain the imaginary cross-spectrum corresponds to the covariance antisymmetric under reversal of time lag [Eq.~\eqref{obs:time_cov}].
A collinear pair alone cannot test leakage between $I$ and $V$ templates. In an array, the complex intensity response must be fitted alongside the polarization templates. We illustrate the projection for the NANOGrav 15-year (NG15) geometry below.

\subsection{Linear polarization}\label{sec:collinear}
Linear polarization separates into a parity-even electric sector and a parity-odd magnetic sector. We consider $E_{40}$, $B_{41}$, and $B_{50}$. The first two have nonzero Earth terms, while the additional Earth-term reflection forbids the third. They test coherent enhancement, distortion of an existing magnetic pattern, and the lifting of an Earth-term null, respectively. The corresponding Earth terms are
\begin{align}
\Gamma^{E,EE}_{40}&=\frac{3\sqrt{70}}{2240}(4-r),\label{eq:earth_E40}\\
\Gamma^{B,EE}_{41}&=\frac{3\ii\sqrt{14}}{160}Cr,
\qquad\Gamma^{B,EE}_{50}=0.\label{eq:earth_B41}
\end{align}
We use the spin-four electric and magnetic decomposition of the Stokes field~\cite{paper_KumarKamionkowski2024,paper_KumarEtAl2024}. Figure~\ref{fig:angular_E} compares all three responses at the same distances. Corollary~\ref{sym:cor-distance} fixes the allowed quadrature at equal distances. At unequal distances both quadratures can contribute in either sector.

The equal-distance endpoint enhancement of $E_{40}$ narrows as the propagation phase grows. Setting $a_u=a_v=a$, the two collinear limits are
\begin{equation}\label{eq:E40-endpoint-limits}
\Gamma^{E;40}(a,a,0)=
\begin{cases}
\displaystyle\frac{\sqrt{70}}{280}a^2+O(a^4),&a\to0,\\[3pt]
\displaystyle\frac{3\sqrt{70}}{280}+O(a^{-4}),&a\to\infty.
\end{cases}
\end{equation}
At small phase the endpoint differences suppress the response, which starts at order $a^2$. At large phase the two pulsar terms still have zero relative phase at exact collinearity and equal distance. Their contribution equals the Earth term, and the total approaches twice the collinear Earth term. The narrowing feature therefore survives at large distance. Taking the large-phase limit at fixed nonzero angle misses the endpoint. Appendix~\ref{app:real-angle-example} gives the full endpoint expression, and Appendix~\ref{app:compact-collinear} gives a radial representation at any finite multipole.

The change in $B_{41}$ can be compared directly with its nonzero Earth term. By contrast, the full $B_{50}$ response appears because propagation breaks the additional reflection symmetry. Both magnetic components still obey the full reflection rule and vanish at exact collinearity. The $B_{50}$ pattern thus requires both propagation and suitable geometry. Whether the additional $B_{50}$ template can be estimated independently depends on the array, the distances, and the noise.

For a real sky, combining $m,-m$ makes the Earth-term contributions of intensity and linear polarization real and that of circular polarization imaginary. A single complex harmonic coefficient need not obey this separation, and full propagation generally removes it even for a real sky.

Section~\ref{sec:coherence} treats the width and polarization dependence of the small-angle feature. We first relate the finite-distance phases to observed pulsars and timing frequencies.

\subsection{Observational scales of finite-distance effects}
\label{obs:nearby}
For a nearby pulsar, the relevant quantities are its propagation phase and the uncertainty in that phase. From $a=2\pi fL/c$,
\begin{equation}
 a=64.7\left(\frac{f}{1\,{\rm nHz}}\right)
       \left(\frac{L}{100\,{\rm pc}}\right).
 \label{obs:nearby_phase}
\end{equation}
Using 26 years of Parkes observations, Reardon et al. measured a distance of $156.96\pm0.11\,{\rm pc}$ for PSR~J0437$-${}4715~\cite{obs_Reardon2024Near}. At $1\,{\rm nHz}$ this corresponds to a mean phase of about $101.5$ and a standard deviation of about $0.0711$. A large phase can thus be predicted fairly precisely. A tight distance prior reduces losses from phase averaging, but oscillatory cancellation in the sky integral remains. The value of accurate distances for continuous-wave searches does not directly establish the observability of finite-distance corrections to a stochastic background.

In Figs.~\ref{fig:angular_I} and \ref{fig:angular_E}, the phases $a=1,10,100$ at $1\,{\rm nHz}$ correspond to distances of approximately $1.55,15.5,155\,{\rm pc}$, respectively. The small-phase curves illustrate the transition in the full path response. Present nearby timing pulsars probe the larger phases in this example. The exact expressions apply at any positive finite distance, including that of a closer stable pulsar if one is found. Its utility would depend on frequency coverage, distance posterior, and noise. The large-parameter conditions below apply only to the asymptotic approximations.

\label{obs:ng15}
For the NG15 example we use 67 pulsars, 2211 distinct pairs, and frequencies $f_k=k/(16.03\,{\rm yr})$ for $k=1,\ldots,14$. Positions come from version 2.1.0 of the 15-year narrowband timing release~\cite{NG15Timing,NG15TimingData}. We use 38 parallax and 29 dispersion-measure distance summaries from the continuous-wave analysis~\cite{NG15CW,NG15CWData} and effective spectra from the detector noise budget and associated data products~\cite{uniform_asymptotics_uniform_NG15Noise2023,NG15NoiseData}. The distance summaries do not specify a joint posterior.
For $S_0(f)\propto f^{-13/3}$, define
\begin{align}
 W_{uvk}&=\frac{T_{uv}S_0(f_k)^2}{P_u(f_k)P_v(f_k)},\nonumber\\
 \epsilon^2&=\frac{\sum_{u<v,k}W_{uvk}|\Delta\Gamma_{uvk}|^2}
 {\sum_{u<v,k}W_{uvk}|\Gamma^{I,EE}_{00,uvk}|^2},\label{obs:epsilon}\\
 b_{A^2}&=\frac{\sum_{u<v,k}W_{uvk}\Gamma^{I,EE}_{00,uvk}\operatorname{Re}\Delta\Gamma_{uvk}}
 {\sum_{u<v,k}W_{uvk}(\Gamma^{I,EE}_{00,uvk})^2},\label{obs:amplitude_projection}
\end{align}
where $\Delta\Gamma=\Gamma^{I,\mathrm{full}}_{00}-\Gamma^{I,EE}_{00}$, $T_{uv}$ is the overlap of the observing spans, and $P_i$ is the effective residual spectrum. The weights approximate the covariance as diagonal in pair and frequency. The quantity $b_{A^2}$ projects onto the squared-amplitude template with other parameters held fixed. The effective spectra already account for the timing fit, so we apply no further idealized timing projection. At the quoted point estimates of distance,
\begin{equation}
 \epsilon=1.40339\times10^{-3},\qquad
 b_{A^2}=1.15355\times10^{-4}.\label{obs:ng15_numbers}
\end{equation}
The smallest individual pulsar phase is $113.254$ and the smallest pair-separation phase is $128.985$, consistent with the small low-order intensity correction. With the same weights, project $\operatorname{Im}\Delta\Gamma$ onto the three $V$ Earth-term templates for the global basis $\sqrt3n_j$. The maximum absolute value of the resulting dipole $V/I=\sqrt3\boldsymbol b_V\cdot\bn$ is approximately $0.1452\%$. Since the input contains only unpolarized intensity, this measures the conditional projection onto circular polarization if the imaginary propagation response is omitted. A parameter-bias calculation would require the time windows and cross-covariances of Appendix~\ref{obs:error_interface}. This example neither refits the data nor sets a common error for other polarization channels.

\FloatBarrier
\section{Coherence of close pulsar pairs}\label{sec:coherence}\label{sec:physics}
Globular clusters contain pulsars at a large common distance but small projected separation. This geometry has been proposed both for probing coherence between pulsar terms~\cite{obs_Maiorano2021GC} and for using residual differences to suppress a common background while retaining a local dark matter signal~\cite{obs_Cherukupalli2026DM}. Both applications depend on which propagation terms survive, how relative depth and its uncertainty affect coherence, and how the relevant scales vary with polarization. We address these questions in a common pair geometry and use published timing solutions for 47~Tucanae (47~Tuc) to illustrate the requirements. Correlations produced by the cluster environment constitute a separate physical contribution.

\subsection{Three-dimensional separation and the coherence condition}
\label{ua:scales}
\label{obs:geometry}
With $\boldsymbol r_i=L_i\boldsymbol p_i$ and $\boldsymbol R=\boldsymbol r_v-\boldsymbol r_u$, define the phase length by
\begin{equation}
 L_f=\frac{c}{2\pi f},\qquad a_i=\frac{L_i}{L_f},\qquad
 \kappa=\frac{|\boldsymbol R|}{L_f},
 \label{obs:Lf}
\end{equation}
where $L_f=1.5463\,{\rm pc}\,(f/{\rm nHz})^{-1}$. With $\Delta=a_v-a_u$ as above, the common-distance and separation phases obey
\begin{align}
 A&=\min(a_u,a_v),&a_c&=\frac{a_u+a_v}{2},\nonumber\\
 \kappa^2&=|a_v\bv-a_u\bu|^2
 =\Delta^2+4a_ua_v\sin^2\frac\gamma2.
 \label{ua:q}
\end{align}
The parameter $A$ controls oscillations of each single-pulsar term, whereas $\kappa$ controls the angular phase of $PP$. The former may be large while the latter remains finite~\cite{paper_MingarelliSidery2014}. For $L_{u,v}=L_c\mp z/2$,
\begin{align}
 \kappa^2&=\frac{4L_c^2\sin^2(\gamma/2)+z^2\cos^2(\gamma/2)}{L_f^2},
 \label{obs:depth}\\
 \varphi_{PP}(\bn)&=\frac{z+\bn\cdot\boldsymbol R}{L_f}.
 \label{obs:PP_phase}
\end{align}
Thus both transverse separation and relative depth must be small to reach the usual $\kappa\lesssim1$ coherence regime. This phase condition alone does not define a detection threshold, which also depends on polarization, noise, and the frequency window. The $PP$ term behaves differently at fixed nonzero angle and when the opening angle shrinks as $a_c^{-1}$. A uniform approximation must cover both limits.

\subsection{Coherent response in each polarization sector}
\label{ua:theorem_section}
\label{ua:endpoints}
\label{core:sec-bcoherence}
The near-pair response contains both $PP$ coherence and cancellation of $EP/PE$ against the Earth term. To approximate all four terms, we define $\Gamma_U$ to retain the exact $EE$ response. For the other three propagation terms we match the symmetric coaxial kernels in $I,E,B$ and the first-order geometric kernel in $V$. The full definition is given in Eq.~\eqref{ua:composite}.
\begin{theorem}[Uniform four-term approximation for a fixed sky mode]
\label{ua:theorem}
Fix finite $(\ell,m)$ and $X=I,V,E,B$. There is a constant $C_{\ell mX}$, depending only on the mode and convention, such that for all $a_u,a_v>0$ with $A\geq64$,
\begin{equation}
 \sup_{\bu,\bv\in S^2}|\Gamma^X_{\ell m}-\Gamma^X_{U,\ell m}|
 \leq C_{\ell mX}\frac{\log(2+A)}{A^{2/3}}.
 \label{ua:global_bound}
\end{equation}
The bound is uniform in the positive distance ratio. For fixed $Q$, throughout $\kappa\leq Q$ and at sufficiently large $A$, the replacement error in each of $EP,PE,PP$ obeys the following sector-dependent bound,
\begin{equation}
 R_{I,E,B}=O_{\ell m,Q}(A^{-2}\log A),\qquad
 R_V=O_{\ell m,Q}(A^{-2}).
 \label{ua:coherent_bound}
\end{equation}
At exact alignment ($\gamma=0$), $\Gamma_U$ is exact for all finite positive $a_u,a_v$.
\end{theorem}
See Appendix~\ref{ua:appendix} for the proof. The bound controls absolute error at a fixed multipole. It gives neither relative accuracy at a zero nor uniformity as $\ell\to\infty$. The lower bound $A\geq64$ is sufficient for the proof. At fixed $\gamma\geq\gamma_0>0$, the result recovers the short-wavelength limit~\cite{paper_MingarelliMingarelli2018},
\begin{equation}
 \Gamma^X_{\ell m}=\Gamma^{X,EE}_{\ell m}
 +O_{\ell m,\gamma_0}\!\left(\frac{\log(2+A)}A\right).
 \label{ua:outer}
\end{equation}
If $\Delta$ stays finite and $\gamma\propto a_c^{-1}$, the separation phase $\kappa$ does not grow with the common distance, and $PP$ coherence persists. Circular polarization requires a further distinction between its exact collinear null and its near-collinear slope. A coaxial zero kernel misses the leading $O(A^{-1})$ response. Matching the first-order geometric kernel reduces the remainder to $O(A^{-2})$. Figure~\ref{ua:error_figure} compares the four-term approximation with different truncations.

Asymptotic expansions must use the actual phase combinations, including $\kappa$ and $\kappa\pm\Delta$. No unsuppressed logarithm of distance remains in the total tensor response at fixed nonzero angle. The bounded-kernel hypothesis used here need not hold for non-GR responses such as a longitudinal antenna.

For the magnetic response we seek the scaling away from the collinear endpoint without introducing the polar-axis dependence of a single $m$ component. First take the resolved limit
\begin{equation}
 a_c\longrightarrow\infty,\qquad\gamma=\frac\rho{a_c},\qquad
 \Delta=\delta\quad\hbox{fixed},
 \label{core:eq-resolved-limit}
\end{equation}
then expand at small $\rho$. A deterministic radial mismatch gives $f(x)=e^{\ii\delta(1+x)}$. An axisymmetric distance average can be absorbed into the same weight $f$, where $x$ is the direction cosine relative to the common sight line. Write $A_\ell[f]=E_{\ell0}^{\rm PP}[f,0]$ and introduce the rotationally invariant norm
\begin{equation}
 N_\ell^X[f,\rho]=\left(\sum_{m=-\ell}^{\ell}|X_{\ell m}^{\rm PP}[f,\rho]|^2\right)^{1/2},
 \quad X=E,B.
 \label{core:eq-multiplet-norm}
\end{equation}
\begin{theorem}[Transverse magnetic coherence]
\label{core:thm-b-coherence}
If $(1-x)^2f(x)$ is integrable, then for each fixed finite $\ell\geq4$,
\begin{equation}
 (N_\ell^B[f,\rho])^2=\frac{8|A_\ell[f]|^2}{\ell(\ell+1)}\rho^2+O(\rho^4).
 \label{core:eq-B-invariant}
\end{equation}
If $A_\ell[f]=0$, then $N_\ell^B=O(\rho^2)$.
\end{theorem}
\begin{corollary}[Coherence slope and equal-distance response ratio]
\label{core:cor-b-slope}
For $A_\ell[f]\ne0$,
\begin{equation}
 N_\ell^B[f,\rho]=|\rho|\sqrt{\frac8{\ell(\ell+1)}}|A_\ell[f]|+O(|\rho|^3).
 \label{core:eq-B-slope}
\end{equation}
At equal distances with no relative-distance uncertainty ($f=1$),
\begin{equation}
 \frac{N_\ell^{B,\rm coh}(\rho,0)}{N_\ell^{E,\rm full}(\rho,0)}
 =|\rho|\sqrt{\frac2{\ell(\ell+1)}}+O(|\rho|^3).
 \label{core:eq-B-E-full-ratio}
\end{equation}
\end{corollary}
Appendix~\ref{core:app-coherence} gives the proof. In this limit the collinear $EE$ magnetic response vanishes and $EP/PE$ drop out. The leading magnetic $PP$ term is therefore also the leading full response, whereas the electric response retains equal $EE$ and $PP$ contributions. A transverse mismatch generates a linear magnetic response. Radial mismatch sets its size through $A_\ell[f]$. For deterministic distances and fixed $\ell$, $A_\ell(\delta)=O(|\delta|^{-3})$, so both components of phase matching matter. Equation~\eqref{core:eq-B-invariant} gives the invariant near-collinear scale of the feature in Fig.~\ref{fig:angular_B}. It relates the electric and magnetic projections of the response kernel, and does not describe a conversion of sky polarization power. At finite distance the four-term response still controls the remainder.

\begin{figure*}[t]
 \centering
 \includegraphics[width=\textwidth]{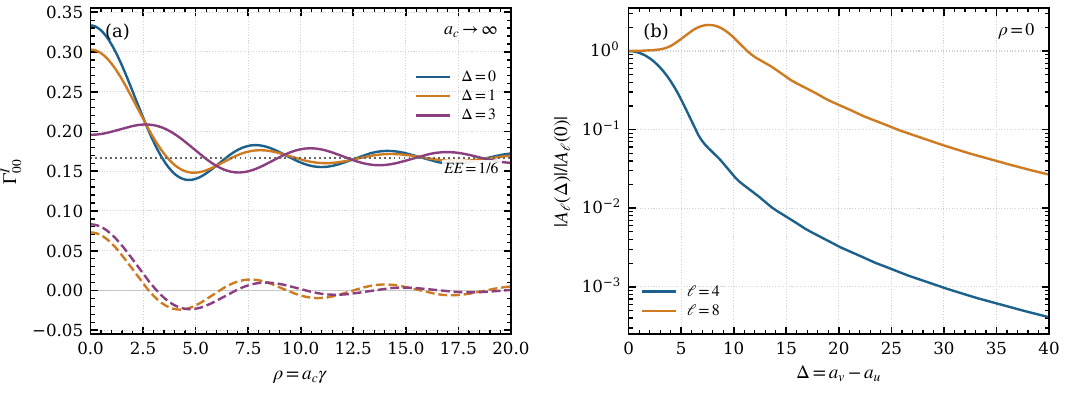}
\caption{Response in the resolved near-collinear limit. Left: the $EE+PP$ intensity limit. Solid and dashed curves show the real and imaginary parts. The gray dotted curve is $EE$, and identically zero imaginary parts are omitted. Right: the radial electric coefficient at exact collinearity. This coefficient also sets the leading transverse magnetic slope, while the magnetic response itself vanishes at collinearity.}
 \label{fig:coherence_behavior}
\end{figure*}

Figure~\ref{fig:coherence_behavior} shows these effects directly. On the left, the equal-distance aligned intensity limit is $1/3$, twice the Earth term $1/6$. With increasing transverse mismatch, $PP$ decays with oscillations and the response returns toward the Earth term. Radial mismatch changes the enhancement and allows unpolarized intensity to retain an imaginary part. On the right, a large radial mismatch suppresses the fixed-order coefficient $A_\ell$ and the leading magnetic slope even at zero opening angle. Redistribution across multipoles can make the transition nonmonotonic. The figure uses the limit in Eq.~\eqref{core:eq-resolved-limit}, where $EP/PE$ vanish at fixed mode. At finite parameters one must retain all four terms or bound their remainder using Theorem~\ref{ua:theorem}.

\subsection{Distance uncertainties and polarization decoherence}
\label{sec:distance}
\label{core:sec-distance-average}
\label{core:sec-common-filter}
\label{core:sec-decoherence}
\label{core:sec-spectral-migration}
We now average the predicted response over a specified joint distance model. Setting $A_u=1+\bn\cdot\bu$ and $A_v=1+\bn\cdot\bv$ gives
\begin{equation}
 Q=\frac{A_u+A_v}{2},\qquad D=A_v-A_u,
 \label{core:eq-common-difference-geometry}
\end{equation}
and the $PP$ phase is $\Delta Q+a_cD$. For jointly Gaussian common and differential distance phases with variances $\sigma_c^2=\operatorname{Var}a_c$ and $s^2=\operatorname{Var}\Delta$ and covariance $c_{c\Delta}$,
\begin{equation}
 \begin{split}
 \langle T_{\rm PP}\rangle&=e^{\ii(\bar\Delta Q+\bar a_cD)}e^{-V_{\rm PP}/2},\\
 V_{\rm PP}&=s^2Q^2+\sigma_c^2D^2+2c_{c\Delta}QD.
 \end{split}
 \label{core:eq-PP-variance}
\end{equation}
At exact alignment $D=0$, so the common-distance error cancels exactly. Near alignment its effect is suppressed by the opening angle. In distance units we obtain the following sufficient condition.
\begin{corollary}[Sufficient condition for preserving phase coherence]
\label{cor:distance_protection}
Suppose $(L_c,z)$ are jointly Gaussian with standard deviations $(\sigma_{L_c},\sigma_z)=(L_f\sigma_c,L_fs)$. For any allowed covariance and any $0<\eta<1$,
\begin{align}
 &2\sin(\gamma/2)\sigma_{L_c}+[1+\cos(\gamma/2)]\sigma_z
 \leq L_f\sqrt{-2\log\eta}\nonumber\\
 &\hspace{3em}\Longrightarrow
 |\langle e^{\ii\varphi_{PP}(\bn)}\rangle|\geq\eta\quad\hbox{for all }\bn.
 \label{obs:protection}
\end{align}
\end{corollary}
The proof is given in Appendix~\ref{coreapp:appendix}. For negligible common-distance error and $\gamma\ll1$, a sufficient bound for retaining $\eta=0.9$ is $\sigma_z\lesssim0.22952L_f$, or about $0.03549\,{\rm pc}$ at $10\,{\rm nHz}$. This condition protects the averaged phase factor. A multipole that crosses zero need not retain the same fraction. The distance model must account for correlations between common distance and relative depth. For a general joint distribution, the appropriate average is the joint characteristic function.

Take the resolved limit at fixed $\rho=\bar a_c\gamma$. If the common transverse uncertainty and its cross term are negligible, the averaged weight reduces to
\begin{equation}
 f_s(x)=e^{\ii\bar\Delta q-s^2q^2/2},\qquad q=1+x.
 \label{core:eq-radial-Gaussian}
\end{equation}
At finite geometry a sufficient condition is $4\sin^2(\gamma/2)\sigma_c^2+8|\sin(\gamma/2)c_{c\Delta}|\ll1$. We take the resolved limit at fixed error parameters before considering $s\to\infty$. For $s\gg1$, the dominant response lies in an antipodal cap of angular radius $O(s^{-1/2})$. The transverse phase variation across the cap is controlled by $\rho/\sqrt{s}$.

\begin{figure*}[t]
 \centering
 \includegraphics[width=\textwidth]{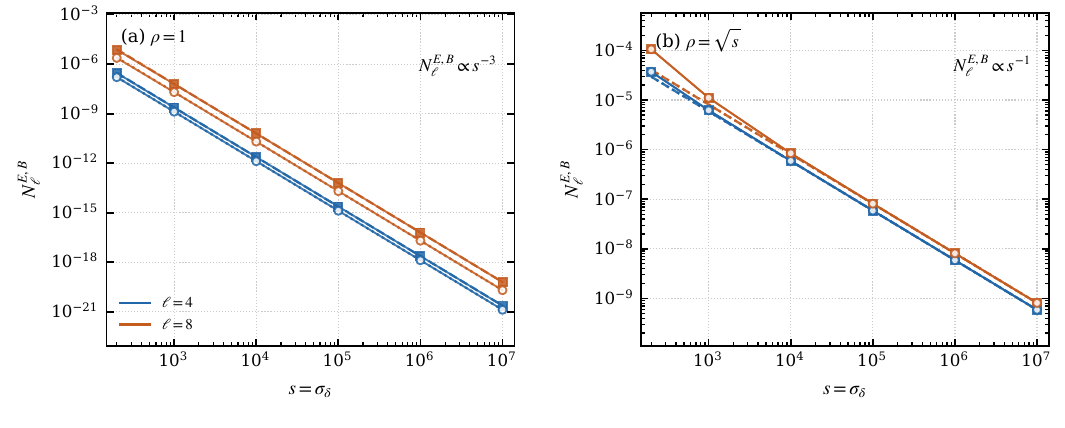}
\caption{Transition between transverse scales under Gaussian distance averaging. The mean radial mismatch is zero and $s$ is the standard deviation of the differential phase. Squares and open circles show numerical integrals for the electric and magnetic norms over the full $m$ multiplet. Dashed and dotted curves give their leading terms, and solid curves connect the numerical points. Only the $PP$ correction is shown, in the resolved limit with negligible common transverse uncertainty.}
 \label{core:fig-decoherence}
\end{figure*}

\begin{theorem}[Gaussian decoherence on two transverse scales]
\label{core:thm-distance-decay}
\label{core:cor-transverse-scale}
In the model of Eq.~\eqref{core:eq-radial-Gaussian}, for fixed finite $\ell\geq4$, $\rho$, and $\bar\Delta$,
\begin{equation}
 \begin{aligned}
 N_\ell^{E,\rm PP}&=C_\ell^E(\rho)s^{-3}+O(s^{-4}),\\
 N_\ell^{B,\rm PP}&=C_\ell^B(\rho)s^{-3}+O(s^{-4}),
 \end{aligned}
 \label{core:eq-fixed-l-decay}
\end{equation}
where $C_\ell^E>0$ and, for $\rho>0$, $C_\ell^B>0$. The magnetic response vanishes identically at $\rho=0$. Closed forms are given in Eqs.~\eqref{core:eq-Gaussian-E-norm} and \eqref{core:eq-Gaussian-B-norm}. The intensity monopole and a representative circular-response slope in the pair frame obey
\begin{equation}
 \begin{aligned}
 \Gamma^{I,\rm PP}_{00}&=\frac{\sqrt{\pi/2}}{4s}+O(s^{-2}),\\
 \frac{\Gamma^{V,\rm PP}_{10}}\gamma&=-\frac{\ii\sqrt{3\pi/2}}{4s}+O(s^{-2}).
 \end{aligned}
 \label{core:eq-IV-decoherence}
\end{equation}
The expansion at fixed $\rho$ requires $\rho^2/s\ll1$. Taking instead $\rho=\lambda\sqrt{s}$ at fixed $\lambda>0$ gives
\begin{equation}
 N_\ell^{E,\rm PP}\sim N_\ell^{B,\rm PP}
 \sim\frac{\sqrt{2(2\ell+1)}}{8s}|F_4(\lambda)|,
 \label{core:eq-second-transverse-scale}
\end{equation}
where $F_4$ is defined in Eq.~\eqref{core:eq-F4}. At a zero of this function the next term is needed.
\end{theorem}
Figure~\ref{core:fig-decoherence} compares the two limits. At fixed transverse mismatch the endpoint zeros of the linear polarization projection give an $s^{-3}$ decay in both electric and magnetic norms, faster than the decay of the intensity response and circular slope. When $\rho\propto\sqrt{s}$, the transverse phase stays finite across the cap and both linear polarization norms recover an $s^{-1}$ leading term. The circular response itself also carries a factor $\gamma$. These powers describe the $PP$ correction. The distance average leaves $EE$ unchanged. See Appendix~\ref{core:app-distance} for the proof.

\begin{figure*}[t]
 \centering
 \includegraphics[width=\textwidth]{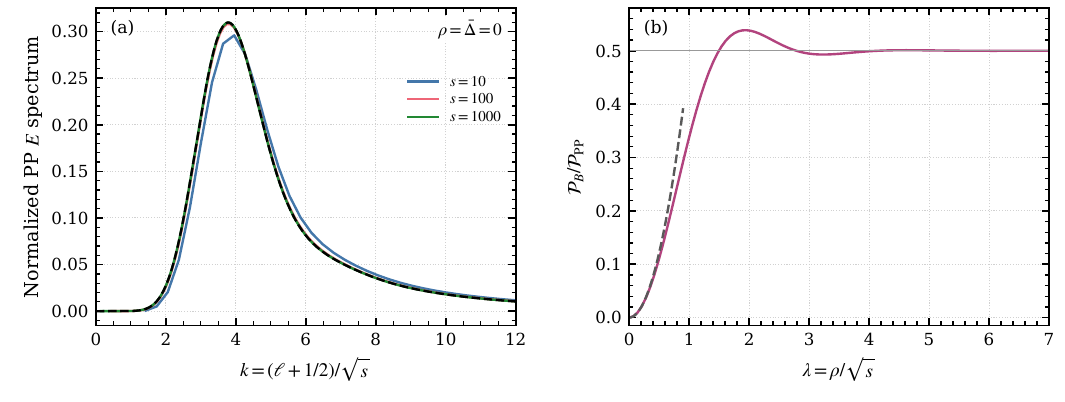}
\caption{The $PP$ angular spectrum and its electric and magnetic fractions after Gaussian differential averaging. Left: rescaled electric spectrum at $\rho=\bar\Delta=0$. The black dashed curve gives the continuum limit in Eq.~\eqref{core:eq-continuous-E-spectrum}. Right: total magnetic fraction in the subdomain of Eq.~\eqref{core:eq-joint-subdomain}. The dashed curve is the quadratic approximation at small $\lambda$. Both panels are normalized by the squared kernel norm, without sky-spectrum or noise weights.}
 \label{core:fig-high-l}
 \label{core:fig-joint-angular}
\end{figure*}

Attenuation at low multipoles is accompanied by redistribution of the angular spectrum. Define $\mathcal P_X=\sum_{\ell m}|\Gamma^{X,\rm PP}_{\ell m}|^2$ and $\mathcal P_{\rm PP}=\mathcal P_E+\mathcal P_B$. Then
\begin{theorem}[Total PP norm and dominant angular scale]
\label{core:cor-spectral-migration}
In the resolved near-collinear limit a purely radial or transverse propagation phase redistributes the angular spectrum while preserving $\mathcal P_{\rm PP}=1/20$. After Gaussian differential averaging,
\begin{equation}
 \mathcal P_{\rm PP}(s)=\frac{\sqrt\pi}{16s}+O(s^{-2}).
 \label{core:eq-PP-total-asymptotic}
\end{equation}
When $\rho=\bar\Delta=0$, only $E_{\ell0}=A_\ell(s)$ is nonzero and the dominant multipoles have $\ell\sim\sqrt{s}$. For any fixed cutoff $L\geq4$ the retained fraction of the squared norm decays as $s^{-5}$.
\end{theorem}
The left panel of Fig.~\ref{core:fig-high-l} shows the continuum limit in $k=(\ell+1/2)/\sqrt{s}$. A fixed-order norm falls as $s^{-3}$, whereas the squared norm summed over all multipoles falls only as $s^{-1}$. The surviving structure thus moves to higher multipoles. For a sky restricted to $\ell\leq L$, the exact ORFs at those orders still give the complete projection. The fraction of the kernel lost to truncation does not directly measure lost signal or information. Appendices~\ref{core:app-extended} and~\ref{core:app-spectrum} give the full-Stokes sum rules, continuum spectrum, and remainders.

The transverse phase also sets the partition between electric and magnetic norms. For independent Gaussian errors in the subdomain with fixed $\lambda=\rho/\sqrt{s}$, $\gamma\sqrt{s}\to0$, and $\gamma\sigma_c/\sqrt{s}\to0$, the magnetic fraction tends to $\mathcal B(\lambda)$. At small $\lambda$ it is $(4-\pi)\lambda^2/\sqrt\pi+O(\lambda^4)$, and in the subsequent large-$\lambda$ limit it tends to $1/2$. The right panel of Fig.~\ref{core:fig-joint-angular} shows the transition away from electric dominance. At fixed bounded $\rho$ the result is $\mathcal P_B/\mathcal P_{\rm PP}\sim(4-\pi)\rho^2/(\sqrt\pi s)$. These fractions refer to the response kernel. Inferring polarization information requires array and noise weights. Appendix~\ref{core:sec-joint-angular} gives the full theorem, its domain, and the remainder at fixed $\rho$.

The leading coherent term alone does not recover the high-multipole tail. At finite parameters $EP/PE$ must cancel against $EE+PP$ to restore regularity of the full kernel. Nor can the fixed-mode large-distance limit be interchanged directly with the sum over the spectrum (Appendix~\ref{core:sec-regularity-boundaries}). The $s^{-3}$ law also depends on the distance prior. Appendix~\ref{core:app-priors} gives the characteristic-function conditions to check for an observational posterior.

\subsection{Observing windows and close pulsar pairs}
\label{sec:observation}
\label{obs:windows}
\label{obs:47tuc}
Finite bandwidth and timing fits further filter the coherent response. Write $\tau_i=L_i(1+\bn\cdot\boldsymbol p_i)/c$. The terms $EE,PE,EP,PP$ then have delays $0,-\tau_u,\tau_v,\tau_v-\tau_u$, respectively. A flat frequency band gives the factor $\sinc(\pi\Delta f\,\tau)$, with
\begin{equation}
 |\tau_v-\tau_u|\leq\frac{2|\boldsymbol R|}{c},\qquad
 |\pi\Delta f(\tau_v-\tau_u)|\leq\frac{\Delta f}{f_0}\kappa(f_0).
 \label{obs:bandwidth}
\end{equation}
A narrow fractional bandwidth can thus preserve $PP$ coherence for $\kappa\lesssim1$. Realistic source spectra and windows enter through the frequency integral. Timing-model attenuation depends on $fT$, while the propagation response depends on $fL/c$. For continuous uniform sampling and an equal-weight fit removing $1,t,\ldots,t^q$, the retained power fraction is~\cite{struct_TPvanHaasteren2013,struct_TPHazboun2019}
\begin{equation}
 \mathcal T_q(fT)=1-\sum_{j=0}^{q}(2j+1)j_j^2(\pi fT),
 \label{obs:timing_transfer}
\end{equation}
where $j_j$ is a spherical Bessel function. At $fT=1$, $\mathcal T_2=0.234067$ and $\mathcal T_3=0.0424209$. Irregular sampling and actual timing fits require the individual pulsar design matrices. Appendix~\ref{obs:error_interface} gives the corresponding statistical expressions.

\begin{figure*}[!t]
 \centering
 \includegraphics[width=\textwidth]{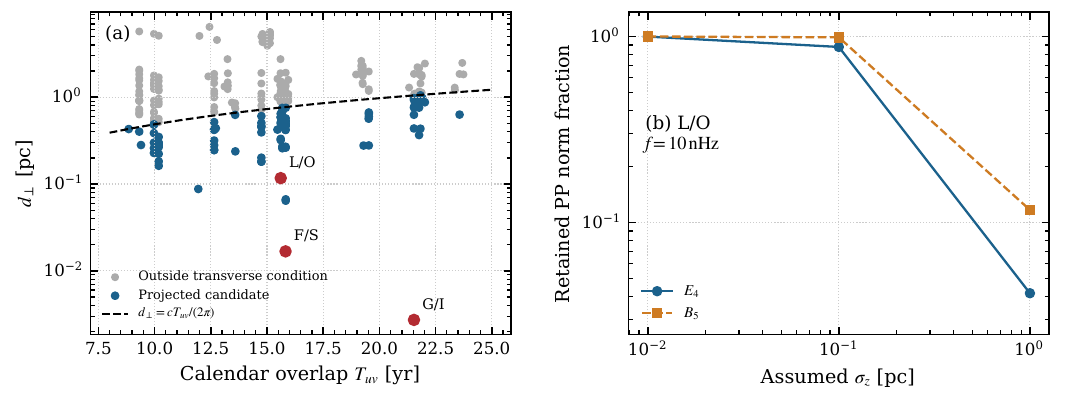}
\caption{Transverse geometry and conditional $PP$ response for 47~Tuc. Left: inverse overlap of the published observing spans, $1/T_{uv}$, used as a diagnostic frequency. The relative depth is unconstrained, and overlapping spans do not imply simultaneous samples. Right: fraction of the $PP$ kernel norm over all $m$ retained relative to zero depth error, at zero mean depth and negligible common-distance error. The curves omit $EE/EP/PE$ and finite-distance remainders and do not measure the retained signal-to-noise ratio.}
 \label{obs:47_compact_figure}
\end{figure*}

We use 23 phase-connected timing solutions for 47~Tuc published in 2016--2018. In the \href{https://www3.mpifr-bonn.mpg.de/staff/pfreire/47Tuc/index.html}{published parameter table}, the solutions for W and X are from Ridolfi et al.~\cite{Ridolfi2016}, that for aa is from Freire and Ridolfi~\cite{FreireRidolfi2018}, and the other 20 are from Freire et al.~\cite{paper_Freire2017Cluster}. We propagate positions to MJD~51600 using each pulsar's proper motion, adopting $D=4690\,{\rm pc}$ and $d_\perp=2D\sin(\gamma/2)$.
Of 253 pairs, 98, 31, 6, and 2 satisfy $d_\perp/L_f\leq1$ at $2,5,10,30\,{\rm nHz}$, respectively. At the pair-specific diagnostic frequency $1/T_{uv}$, 88 pairs pass (left panel of Fig.~\ref{obs:47_compact_figure}). This selection constrains only transverse separation. The mean depth must also satisfy $|z|\leq\sqrt{L_f^2-d_\perp^2}/\cos(\gamma/2)$, and its uncertainty enters separately through Eq.~\eqref{obs:protection}. At $10\,{\rm nHz}$ the upper limits on mean depth for the three pairs in Table~\ref{obs:pairs_table} are $0.15460,0.15372,0.10112\,{\rm pc}$, respectively.

\begin{table}[t]
\caption{Transverse scales for representative 47~Tuc pairs. The last column gives the frequency at which $d_\perp=L_f$.}
 \label{obs:pairs_table}
 \centering
 \begin{tabular}{lrrr}
 \toprule
Pair & $\gamma$ (arcsec) & $d_\perp$ (pc) & $f$ (nHz)\\
 \midrule
 G/I & 0.119736 & 0.00272252 & 567.96\\
 F/S & 0.737556 & 0.01677038 & 92.20\\
 L/O & 5.144982 & 0.11698538 & 13.22\\
 \bottomrule
 \end{tabular}
\end{table}

The right panel of Fig.~\ref{obs:47_compact_figure} uses the measured transverse geometry, zero mean depth, and Gaussian relative-depth errors. At $10\,{\rm nHz}$ and $\sigma_z=0.1\,{\rm pc}$ the fractions of the coherent $PP$ norms retained over all $m$ are similar for the three pairs: about $87.5\%$ for $E_4$ and $99\%$ for $B_5$. The $B_5$ norms at zero depth error are $0.00090104,0.00554548,0.03706661$, respectively. L/O has a larger transverse phase and conditional magnetic response. The smallest angular separation therefore need not yield the largest polarized response.

All these timing solutions fit F2, and the solution for H also fits F3. They remove phase polynomials through at least cubic and quartic order, respectively. Low-frequency geometric gains must therefore be weighed against timing-fit attenuation. Close pairs also share signal modes. At identical distance and direction, the signal covariance before projection is a single common rank-one mode. Its rank and degeneracies must be reassessed after individual timing projections. A count of geometrically selected pairs is therefore not a count of independent measurements.

Timing measurements of acceleration and jerk constrain the cluster potential and local environment~\cite{obs_Prager2017GC,obs_Smith2024GC}. Newly discovered cluster pulsars extend the range of accessible geometries~\cite{obs_Chen2026GC}, and continuum observations suggest a population of millisecond pulsars that remains incompletely detected~\cite{obs_Urquhart2026GC}. A population estimate alone does not establish a precision timing sample. These fitted parameters carry different information from the stochastic cross-covariance of postfit residuals. Modeling environmental correlations requires separate dynamics and noise models. The full ORF derived here supplies a reference response to an external background against which such contributions can be compared.

\subsection{Background suppression and distance accuracy for close-pair differencing}
\label{sec:dm_difference}
Dark matter searches based on close-pair differences exploit the different spatial scales of a gravitational-wave background and a local flyby~\cite{obs_Cherukupalli2026DM}. The proposed method already includes unequal distances, orientation, complex covariance, and multipulsar channels. We use the physical constraints on the Stokes covariance and a joint distance average to test how polarization and unknown relative depth affect cancellation.
For the normalized difference $r_-=(r_u-r_v)/\sqrt2$,
\begin{equation}
 S_-^{\rm GW}=\frac{S_{uu}^{\rm GW}+S_{vv}^{\rm GW}}2
                  -\operatorname{Re}S_{uv}^{\rm GW}.
 \label{obs:DM_difference}
\end{equation}
The gain in detectability depends on the change in both residual background and local signal template. Reducing the background below the other noise imposes a stronger requirement.

Fix the intensity $I(\bn,f)$ at every direction and frequency, and compare a physical polarization spectrum $P$ with the unpolarized spectrum $P_0=I\mathbf1_2/2$. Positive semidefiniteness in Eq.~\eqref{eq:stokes_matrix} gives
\begin{equation}
 0\preceq P\preceq2P_0,\qquad
 0\preceq\mathsf K[P]\preceq2\mathsf K[P_0],
 \label{dm:covariance_bound}
\end{equation}
where $\mathsf K$ is the background covariance constructed with the full response. The first inequality follows from the eigenvalues $(I\pm\sqrt{Q^2+U^2+V^2})/2$ of $P$. Response contraction and integration preserve the matrix order, giving the second inequality. The same linear observing map, timing projection, or distance average applied to both sides preserves the bound, without a low-multipole cutoff or an Earth-term approximation.
Adding the same positive-definite noise $\mathsf N$ and using the same deterministic projected template $\boldsymbol s$ gives
\begin{align}
 {\rm SNR}_{P}^{2}
 &\geq\boldsymbol s^\dagger[\mathsf N+2\mathsf K[P_0]]^{-1}\boldsymbol s
 \nonumber\\
 &\geq\tfrac12{\rm SNR}_{0}^{2}.
 \label{dm:snr_bound}
\end{align}
This matched-filter bound assumes \emph{known covariance} and equal pointwise $I$. Equation~\eqref{dm:covariance_bound} applies to the physical covariance, including its diagonal, and cannot be applied to a cross-only matrix with the diagonal removed. Fixing only $I_{00}$ while changing the sky distribution, introducing nonlocal directional coherence, or jointly inferring unknown polarization and geometry requires a separate signal-to-noise calculation.

Relative depth also controls whether the residual background falls below the white noise. In the common-sight-line, large-common-distance limit of Eq.~\eqref{core:eq-resolved-limit}, take an isotropic unpolarized reference and set
$\widehat\Gamma=3\Gamma^{I,\mathrm{coh}}_{00}$, which normalizes each pulsar auto-covariance to unity. With the limiting values $\rho=d_\perp/L_f$ and $\Delta=z/L_f$,
\begin{equation}
 \begin{split}
 \widehat\Gamma(\rho,\Delta)=\frac12+\frac3{16}\int_{-1}^{1}\!\dd\mu\,
 &(1-\mu)^2J_0\!\left(\rho\sqrt{1-\mu^2}\right)\\[-2pt]
 &\times e^{\ii\Delta(1+\mu)}.
 \end{split}
 \label{dm:normalized_kernel}
\end{equation}
This expression retains the complete separation phase within the limiting model. At finite distance the $EP/PE$ terms and geometric remainder require separate control.
The phase-adjusted difference $(r_u-e^{\ii\phi}r_v)/\sqrt2$ has power $1-\operatorname{Re}(e^{-\ii\phi}\widehat\Gamma)$. For known depth the choice $\phi=\arg\widehat\Gamma$ gives $\lambda_{\rm known}=1-|\widehat\Gamma|$.

Suppose $\Delta\sim\mathcal N(\bar\Delta,\sigma_\Delta^2)$, where $\sigma_\Delta=\sigma_z/L_f=s$, and neglect the common transverse uncertainty. A channel fixed at each frequency before the true depth is known cannot adjust its phase for each realization. Its minimum expected power satisfies
\begin{align}
 \lambda_{\rm fixed}
 &=1-|\langle\widehat\Gamma\rangle|\nonumber\\
 &=\frac{3\rho^2}{40}+\frac{11\bar\Delta^2}{160}
       +\frac{\sigma_\Delta^2}{10}+O(\epsilon^4),
 \label{dm:fixed_expansion}\\
 \lambda_{\rm fixed}-\langle\lambda_{\rm known}\rangle
 &=\frac{\sigma_\Delta^2}{32}+O(\epsilon^4).
 \label{dm:phase_penalty}
\end{align}
Here $\epsilon^2=\rho^2+\bar\Delta^2+\sigma_\Delta^2\ll1$. The exact gap is $\langle|\widehat\Gamma|\rangle-|\langle\widehat\Gamma\rangle|\geq0$. Of the total contribution $\sigma_\Delta^2/10$, $11\sigma_\Delta^2/160$ arises from the geometry distribution itself. The additional $\sigma_\Delta^2/32$ is the cost of using one phase for all depths (Appendix~\ref{app:dm_checks}). Distance is a static latent variable. These expressions optimize the \emph{expected power} of a fixed channel. A distance-mixture likelihood generally differs from a Gaussian likelihood evaluated with the mean covariance.

\begin{figure*}[t]
 \centering
 \includegraphics[width=\textwidth]{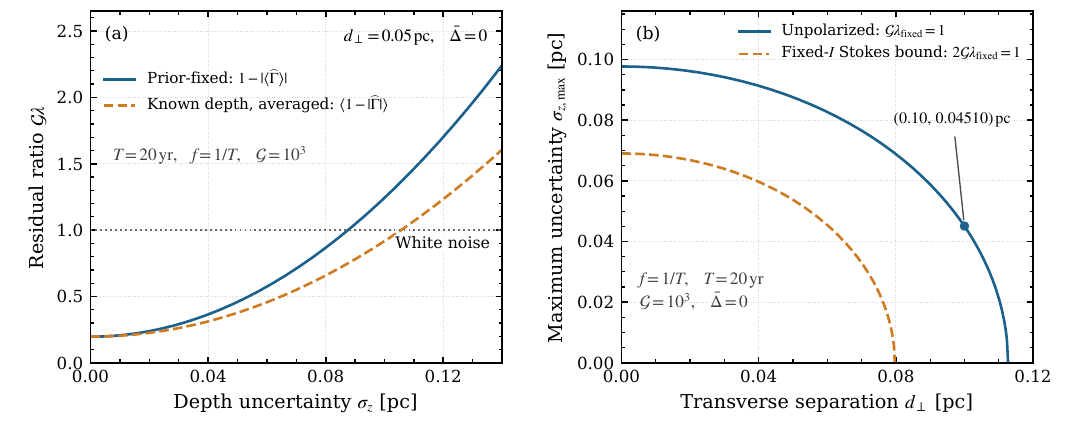}
\caption{Relative-depth noise budget in the common-sight-line $EE+PP$ model, with $\bar\Delta=0$. (a) Expected background power in a fixed channel and average power after adjusting the phase at each known true depth. The latter is an ideal reference. The horizontal line marks the white-noise level. (b) Maximum relative-depth standard deviation for the fixed channel. The full-Stokes curve is a sufficient bound at fixed pointwise $I$. Beyond its endpoint, the bound no longer ensures white-noise dominance. The curves use the unexpanded kernel and do not marginalize the dark matter template over distance.}
 \label{dm:budget_figure}
\end{figure*}

Define $\mathcal G(f)=S_{uu}^{\mathrm{GW},0}/S_{uu}^{\rm white}$, where the superscript $0$ denotes the unpolarized reference, and assume equal, independent white noise in the two pulsars. At small phase the single-frequency condition $\mathcal G\lambda_{\rm fixed}\leq1$ gives
\begin{equation}
 \sigma_z\lesssim L_f\left[
 \frac{10}{\mathcal G}-\frac34\rho^2-
 \frac{11}{16}\bar\Delta^2\right]^{1/2},
 \label{dm:depth_budget}
\end{equation}
provided the bracket is nonnegative. For a sufficient condition valid for every physical Stokes sky of the same pointwise intensity, Eq.~\eqref{dm:covariance_bound} replaces $10/\mathcal G$ by $5/\mathcal G$. A particular sky can meet the white-noise target outside this sufficient region.
Figure~\ref{dm:budget_figure} evaluates the Gaussian average of Eq.~\eqref{dm:normalized_kernel} directly. It compares a fixed channel with the ideal known-depth result and gives the noise budget from the unexpanded kernel. At $T=20$ yr, $f=1/T$, $\mathcal G=10^3$, $d_\perp=0.1$ pc, and $\bar z=0$, the unpolarized reference requires $\sigma_z\lesssim0.0451$ pc. This single-frequency budget depends on background strength more directly than a requirement on coherence alone. A detection threshold would require the broadband timing likelihood.

Appendix~\ref{app:dm_checks} checks the full four-term response for three equal-distance transverse configurations and applies the same timing projection to a local Doppler injection. These tests evaluate the mechanism within the specified model. A dark matter sensitivity forecast must also propagate the distance dependence of the signal, individual timing designs, intrinsic red noise, and baryonic flybys. Background coherence or Eq.~\eqref{dm:snr_bound} alone does not give that sensitivity.

\section{Numerical accuracy and computational cost}
\label{sec:validation}

Analytic closure replaces the sky integral by finitely many function evaluations. Practical use also depends on cancellation near degenerate parameters and on whether geometric preparation can be amortized over repeated evaluations. We test accuracy by independent integration and compare evaluation times at matched accuracy.

We construct the reference integrand directly by contracting the pulsar directions with transverse polarization tensors. We integrate over the sky and project onto harmonics without using the integrated coefficients. We use Gauss--Legendre quadrature in the polar cosine and an equally spaced periodic grid in azimuth. Direct integration includes positive and negative $m$ and components forbidden by Corollary~\ref{sym:cor-zeros}, covering all 292 components with $\ell\leq8$.

Across five noncollinear geometries, two exact endpoints, and one zero-transfer case, the largest absolute difference in 2336 comparisons is $3.34\times10^{-14}$. A separate endpoint test at five distance settings per endpoint yields a maximum absolute difference of $1.01\times10^{-14}$ in 2920 comparisons. Refining the quadrature changes the two sets by at most $3.50\times10^{-14}$ and $8.48\times10^{-15}$, respectively. The noncollinear tests reach phase $(a_u,a_v)=(20,30)$ with a minimum angular distance of $0.002$ from an endpoint. The dedicated endpoint tests reach phase $(60,80)$. These results establish accuracy at the sampled parameters. We assess weak responses by absolute error and draw no inference about floating-point stability at arbitrarily large phase.

At fixed angle and distance ratio, the geometric coefficients in the $E/B$ expressions can be reused in a frequency scan. For two fixed geometries, 520 double-precision evaluations of 40 $E/B$ components at $\ell=4,5$ agree with high-precision references to a maximum absolute difference of $3.58\times10^{-17}$. Table~\ref{val:tab-runtime} times three of these components using single-threaded numerical kernels on an Intel Xeon Platinum 8573C. Both methods cache frequency-independent quantities and must satisfy
\begin{equation}
 \max_{X\in\{E_{40},B_{41},B_{50}\}}
 |\Gamma_X^{\rm eval}-\Gamma_X^{\rm ref}|<10^{-12}.
 \label{val:eq-runtime-tolerance}
\end{equation}

\begin{table}[!t]
\caption{Combined repeat-evaluation time for three selected $E/B$ components. A denotes the finite analytic expression and Q direct sky quadrature. Both meet Eq.~\eqref{val:eq-runtime-tolerance}. Times are medians over seven batches and exclude geometry preparation, loading, and initial compilation.}
\label{val:tab-runtime}
\setlength{\tabcolsep}{3pt}
\begin{ruledtabular}
\begin{tabular}{rrrcc}
$a_u$ & $a_v$ & $\gamma$ & $t_{\rm A}$ [$\mu{\rm s}$] & $t_{\rm Q}$ [ms]\\
\hline
100&80&1.1&7.43&2.92\\
300&240&1.1&7.20&11.43\\
1000&800&1.1&6.87&262.91\\
1000&1000&$10^{-3}$&11.40&212.85\\
\end{tabular}
\end{ruledtabular}
\end{table}

Across the four tasks the ratios of repeat-evaluation times range from $3.93\times10^2$ to $3.83\times10^4$. The savings come from reusing the geometric coefficients and eliminating oscillatory sky quadrature. Preparing each table of geometric coefficients takes about $2.3$--$2.5\,{\rm s}$. Changing angle or distance ratio requires rebuilding the table, a cost that must enter the total time. The comparison covers only the three listed components and four settings.

\section{Conclusions}\label{sec:conclusion}
Extending the real-projection integration method of Zhou and Li to a general tensor polarization covariance, we obtained closed forms for the anisotropic full-Stokes PTA response at finite pulsar distances in GR. For each fixed finite sky mode, $I/V$ require at most a finite combination of exponential integrals, logarithms, and elementary functions, whereas $E/B$ are elementary. We evaluated all responses through $\ell=8$. The closure proof holds at every fixed finite multipole. Symmetry reduces the number of independent integrals. Independent sky integrations and timing comparisons at matched accuracy test the expressions and quantify the benefit of reuse over the parameter ranges studied.

Finite distance changes both the angular response and the phase of the cross-spectrum. The full response preserves the pair-plane selection rule, the $V$ monopole null, and the collinear $V/B$ nulls. Propagation can, however, lift additional Earth-term nulls and generate the $B_{50}$ angular template. An unpolarized intensity can itself have a complex cross-spectrum, so circular polarization must be identified using the full templates. The nearby-pulsar and NG15 examples place these results on observed phase scales. Their impact on inference also depends on the distance posterior, observing window, and noise weights.

Close-pair coherence is controlled by three-dimensional separation. A large common distance can coexist with a finite separation phase. Transverse proximity and radial phase matching must both be tested. The uniform $EE/EP/PE/PP$ approximation connects the coherent region to the short-wavelength regime at general opening angles and bounds the replacement error in each propagation term. Polarizations obey different collinear limits. An invariant relation connects the leading transverse magnetic response to a radial electric coefficient. At finite distance, cancellations involving $EP/PE$ remain important, particularly in the high-multipole tail.

Distance uncertainty attenuates coherence and redistributes the angular spectrum. Common-distance uncertainty cancels in the collinear $PP$ phase, whereas relative-depth uncertainty directly limits coherence. In the resolved near-collinear limit with Gaussian differential averaging and negligible common transverse uncertainty, the fixed-order linear polarization $PP$ norm and the squared norm summed over all multipoles decay at different rates. The remaining angular structure shifts to $\ell\sim\sqrt{s}$. At fixed transverse mismatch the total magnetic fraction vanishes, but it can remain finite if the transverse phase grows as $\sqrt{s}$. These statements concern the response kernel. The effect of angular truncation on signal or information also depends on the sky model.

The 47~Tuc timing solutions illustrate how measured transverse separations select candidate pairs. Their mean relative depths and depth errors determine phase matching, and bandwidth and timing fits further limit the surviving signal. For close-pair differences, preserving phase coherence and reducing residual background below the noise impose distinct requirements. At fixed pointwise intensity the full covariance of any physical Stokes state is bounded above by twice the unpolarized covariance. In the small-phase Gaussian model, depth uncertainty raises the expected residual power of a fixed channel. The second-order cost of being unable to adjust the phase at each true depth is $\sigma_\Delta^2/32$. Direct four-term calculations and a local injection propagated through the timing fit test this interpretation within the model. A dark matter sensitivity calculation would additionally require depth-dependent signal templates, pulsar-specific noise, and baryonic perturbations. Inferring the cluster environment requires a dynamical model in addition to the external-background response derived here.

\begin{acknowledgments}
This work was supported by the National Natural Science Foundation of China (Grant Nos. 12575072, 12547101 and 125B2102), the Fundamental Research Funds for the Central Universities Project (Grant No. 2024IAIS-ZD009), and the National Key Research and Development Program of China (Grant No. 2021YFC2203004).
\end{acknowledgments}
\section*{Data and analytic results}
Section~\ref{sec:analytic} and the appendices give representative analytic expressions, collinear limits, and their derivations. Section~\ref{sec:validation} describes the independent sky quadrature, its parameter range, and the errors for all 292 components through $\ell=8$. Section~\ref{sec:analytic} cites the positions, distance summaries, and noise spectra used for NG15. The sources of the 47~Tuc timing solutions are cited in Sec.~\ref{sec:coherence}.
\FloatBarrier
\appendix
\section{Proofs of the symmetry relations}\label{sym:appendix}
\begin{proof}[Proof of Theorem~\ref{sym:prop-spatial}]
Make the change of variables $\bn\mapsto R\bn$ in the response integral. Since
\[
 (R\bn)\cdot(R\boldsymbol p)=\bn\cdot\boldsymbol p,
 \qquad \dd\Omega_{R\bn}=\dd\Omega_{\bn},
\]
the propagation factor is invariant. The basis phases of the antenna contraction and spin harmonic cancel. The harmonic rotation law then gives Eq.~\eqref{sym:rotation}.
\end{proof}

\begin{proof}[Proof of Theorem~\ref{sym:thm-reflection}]
In the standard pair frame, $\mathcal M_z:(X,Y,Z)\mapsto(X,Y,-Z)$ preserves the propagation factor. The $F_p^+$ response is even and the $F_p^\times$ response odd. From
\begin{equation}
 {}_s\widetilde Y_{\ell m}(\pi-\theta,\phi)
 =(-1)^{\ell+m}{}_{-s}\widetilde Y_{\ell m}(\theta,\phi),
 \quad s=0,\pm4,
 \label{sym:scalar-reflection}
\end{equation}
the $I/V$ contractions acquire the even and odd signs, respectively. The spin branches interchange, and their $E/B$ combinations acquire $p_E=1,p_B=-1$, respectively. Applying $\mathcal M_z$ in the integral gives Eq.~\eqref{sym:selection}.
\end{proof}

\begin{proof}[Proof of Corollary~\ref{sym:cor-zeros}]
The selection rule and the $V$ monopole null follow from Eq.~\eqref{sym:selection}. The monopole is rotation invariant. For parallel or antiparallel sight lines, $\bv=\epsilon\bu$ and $\epsilon=\pm1$, with
\begin{equation}
 F_v^A=\frac{1+\bn\cdot\bu}{1+\epsilon\bn\cdot\bu}F_u^A,
 \qquad A=+,\times.\label{sym:parallel-antenna}
\end{equation}
Hence the antisymmetric $V$ antenna contraction vanishes pointwise. With the common axis as polar axis, axial symmetry allows only $m=0$, and ${}_4\widetilde Y_{\ell0}={}_{-4}\widetilde Y_{\ell0}$ implies $B=0$. At the exceptional point we take the removable limit of the full response. Rotational covariance extends the result to any frame.
\end{proof}

\begin{proof}[Proof of Theorem~\ref{sym:prop-Hermitian}]
For $X=I,V,Q,U$, we have $K_X(u,v)=K_X(v,u)^*$, with the distances included in the pair labels. Combining this relation with
\begin{equation}
 {}_s\widetilde Y_{\ell m}^*=(-1)^{s+m}{}_{-s}\widetilde Y_{\ell,-m},
 \qquad s=0,\pm4,\label{sym:spin-conjugation}
\end{equation}
immediately gives Eq.~\eqref{sym:ordered} for $I/V$. For linear polarization one first obtains
\begin{equation}
 \Gamma_{uv}^{\pm;\ell m}=(-1)^m[\Gamma_{vu}^{\mp;\ell,-m}]^*.
 \label{sym:helicity-Hermitian}
\end{equation}
Substitution into the $E/B$ definitions gives the same conclusion. Complex conjugation acts on the explicit $\ii$ in both $V$ and $B$, leaving no additional sign.
\end{proof}

\begin{proof}[Proof of Corollary~\ref{sym:cor-distance}]
In the standard pair frame, $R_x(\pi)$ interchanges the sight lines, and
\begin{equation}
 \widetilde Y_{\ell m}(R_x(\pi)\bn)=(-1)^\ell\widetilde Y_{\ell,-m}(\bn).
 \label{sym:half-turn}
\end{equation}
Combining this with Eq.~\eqref{sym:ordered} gives
\[
 \Gamma^{X;\ell m}(a_u,a_v)
 =(-1)^{\ell+m}[\Gamma^{X;\ell m}(a_v,a_u)]^*.
\]
For a nonzero component, use Eq.~\eqref{sym:selection} to replace $(-1)^{\ell+m}$ by $p_X$. This gives Eq.~\eqref{sym:distance}, which also holds for the null components. Setting $a_u=a_v$ gives Eq.~\eqref{sym:equal-distance}.
\end{proof}

\begin{proof}[Proof of Theorem~\ref{sym:prop-Earth} and Corollary~\ref{sym:cor-Earth-B0}]
For $\mathcal M_y:\phi\mapsto-\phi$,
\begin{equation}
 \begin{aligned}
 F_u^+(\theta,-\phi)&=F_v^+(\theta,\phi),\\
 F_u^\times(\theta,-\phi)&=-F_v^\times(\theta,\phi).
 \end{aligned}
 \label{sym:second-reflection}
\end{equation}
Thus the $K_I,K_Q,K_V$ part of $EE$ is even and the $K_U$ part odd. Equation~\eqref{sym:spin-conjugation} and the definition of $E/B$ give Eq.~\eqref{sym:earth-negative-m}. Setting $m=0$ yields $\Gamma_{EE}^{B;\ell0}=-\Gamma_{EE}^{B;\ell0}=0$.
\end{proof}

Full propagation transforms according to Eq.~\eqref{sym:transfer-reflection}. With $K_{U,EE}$ denoting the real antenna cross-kernel, pairing reflected directions gives
\begin{equation}
 \begin{split}
 \Gamma^{B;\ell0}={}&\frac1{16\pi}\int{}_{4}\widetilde Y_{\ell0}K_{U,EE}\\
 &\times[T(\phi)-T(-\phi)]\,\dd\Omega.
 \end{split}
 \label{sym:B0-antisymmetric}
\end{equation}
For $EE$ the bracket vanishes. At equal distances with full propagation it is $2\ii\operatorname{Im}T$, so odd-order $B_{\ell0}$ may be nonzero and purely imaginary. Even orders remain forbidden by Eq.~\eqref{sym:selection}.

\section{Analytic closure and the collinear response}\label{app:elementary-closure}
\subsection{Elementary closure}
Write the four propagation terms of Sec.~II as $w_\alpha e^{\ii\boldsymbol K_\alpha\cdot\bn}$, with weights and phase vectors
\begin{equation}\label{eq:four-physical-phases}
\begin{aligned}
(w_{EE},\boldsymbol K_{EE})&=(1,\boldsymbol0),\\
(w_{PE},\boldsymbol K_{PE})&=(-e^{-\ii a_u},-a_u\boldsymbol u),\\
(w_{EP},\boldsymbol K_{EP})&=(-e^{\ii a_v},a_v\boldsymbol v),\\
(w_{PP},\boldsymbol K_{PP})&=(e^{\ii(a_v-a_u)},-a_u\boldsymbol u+a_v\boldsymbol v).
\end{aligned}
\end{equation}
It suffices to show that, with each nonzero $\boldsymbol K_\alpha$ chosen as a real polar axis, the azimuthal average is polynomial on at most three intervals and has degree at most $\ell+2$. For zero phase we may choose any real axis in the pair plane.
\begin{proof}[Proof of Theorem~\ref{thm:elementary-closure}]
Represent $H_{\ell m}(\boldsymbol r)=r^\ell\widetilde Y_{\ell m}(\widehat{\boldsymbol r})$ by the symmetric trace-free solid-harmonic tensor $\mathcal Y^{\ell m}_{i_1\cdots i_\ell}$ and define
\begin{equation}
 \boldsymbol v_p^\sigma=\boldsymbol p-(\bn\cdot\boldsymbol p)\bn
 +\sigma\ii\bn\times\boldsymbol p,\qquad\sigma=\pm1.
 \label{eq:elementary-helicity-vector}
\end{equation}
The full numerator of Eq.~\eqref{eq:spin_response} is
\begin{equation}
 \begin{aligned}
 W^\sigma_{\ell m}&=C_\ell\mathcal Y^{\ell m}_{ijkrL}
 (v_u^\sigma)^i(v_u^\sigma)^j\\
 &\quad\times(v_v^\sigma)^k(v_v^\sigma)^r n^L,\\
 C_\ell&=\frac{\ell!}{4(\ell-4)!}
 \sqrt{\frac{(\ell-4)!}{(\ell+4)!}}.
 \end{aligned}\label{eq:elementary-helicity-numerator}
\end{equation}
Here $L$ contains $\ell-4$ indices. With $\boldsymbol q_\sigma=\boldsymbol e_\theta+\sigma\ii\boldsymbol e_\phi$,
$\boldsymbol v_p^\sigma=(\boldsymbol p\cdot\boldsymbol q_{-\sigma})\boldsymbol q_\sigma$. The four tangent contractions give the spin-$4\sigma$ projection, and $\boldsymbol q_\sigma^2=\boldsymbol q_\sigma\cdot\bn=0$ removes the traces. Thus $W^{E/B}$ is a polynomial of degree at most $d=\ell+4$. Average over reflection in the pair plane and eliminate even powers of the normal component using $\bn^2=1$. The resulting $\overline W(\tilde x,\tilde y)$ has the same degree bound.

Write $c_\gamma=\bu\cdot\bv$. The three algebraic zeros we need are
\begin{equation}
 \begin{gathered}
 \overline W(-1,-c_\gamma)=\overline W(-c_\gamma,-1)=0,\\
 \overline W(-1,-1)=0.
 \end{gathered}
 \label{eq:elementary-three-zeros}
\end{equation}
The first two follow from $\boldsymbol v_p^\sigma(-\boldsymbol p)=0$. For the third, continue the polynomial to the complex unit sphere, with the dot product extended bilinearly. At $\bn\cdot\bu=\bn\cdot\bv=-1$, the vectors $\boldsymbol a=\bu+\bn$ and $\boldsymbol b=\bv+\bn$ satisfy $\boldsymbol a^2=\boldsymbol b^2=0$ and $\boldsymbol a\cdot\boldsymbol b=c_\gamma-1\ne0$ and hence lie on the two distinct null lines of the complex tangent plane. The eigenvalues of $J\boldsymbol t=\bn\times\boldsymbol t$ on these lines are $\pm\ii$, so $1+\sigma\ii J$ must annihilate one of them. Equation~\eqref{eq:elementary-helicity-numerator} then vanishes. This continuation establishes an algebraic identity while leaving the direction integral real.

Choose any physical phase vector as a real polar axis and write $\boldsymbol p=g_p\boldsymbol e_1+h_p\boldsymbol e_3$ and $g_p^2+h_p^2=1$ in the pair plane. Set $x=\bn\cdot\boldsymbol e_3$, $s^2=1-x^2$, $t=s\cos\phi$, and $b_p=1+h_px$, so that the directional denominator is $d_p=b_p+g_pt$. The basic azimuthal integral is
\begin{equation}
 \frac1{2\pi}\int_0^{2\pi}\frac{\dd\phi}{b_p+g_ps\cos\phi}
 =\frac1{\sqrt{b_p^2-g_p^2s^2}}=\frac1{|x+h_p|}.
 \label{eq:elementary-real-angular-identity}
\end{equation}
Splitting at $-h_u,-h_v$, the azimuthal average of a degree-$d$ numerator $P(x,t)$ divided by $d_ud_v$ is rational. Its only possible finite denominator factors are
\begin{equation}
 x+h_u,\quad x+h_v,\quad L(x)=g_vb_u-g_ub_v.
 \label{eq:elementary-possible-factors}
\end{equation}
At generic positions all are simple. The average of an ordinary angular polynomial involves only powers of $s^2$ and grows no faster than $x^{d-2}$ at infinity.

At a candidate pole $x_0=-h_u$ or $L(x_0)=0$, the singular coefficient is a linear functional of $P(x_0,t)$ and annihilates any polynomial multiple of $d_u(x_0,t)$. After cancellation of $d_u$, the only possible pole of the single-denominator integral, $-h_v$, differs from $x_0$. The coefficient therefore depends only on $P(x_0,t_0)$ evaluated at $d_u(x_0,t_0)=0$. Interchanging $u,v$ gives the analogous result. These three evaluations are precisely the zeros in Eq.~\eqref{eq:elementary-three-zeros}, so all finite poles cancel. The rational function is then a polynomial of degree at most $d-2\le\ell+2$. The third zero is essential even when it lies outside the real integration domain. Boundedness on the real sphere alone cannot establish this cancellation.

Degenerate axes follow by taking a limit from generic positions. The full spin contraction is bounded, so dominated convergence gives pointwise convergence on every surviving open interval. A polynomial of degree at most $\ell+2$ is determined by $\ell+3$ interior points, preserving the degree bound. For exactly parallel or antiparallel directions, direct azimuthal averaging gives
\begin{equation}
 \begin{aligned}
 \langle W^E_{\ell m}\rangle_\phi
 &=\frac{\widetilde Y_{\ell m}(\bu)}4
 \sqrt{\frac{(\ell-4)!}{(\ell+4)!}}
 (1-x^2)^4P_\ell^{(4)}(x),\\
 \langle W^B_{\ell m}\rangle_\phi&=0.
 \end{aligned}\label{eq:elementary-collinear-numerator}
\end{equation}
The explicit zeros cancel $(1+x)(1+\epsilon x)$ ($\epsilon=\pm1$), again leaving a polynomial of degree at most $\ell+2$.

If the azimuthal average on an interval is $G_\alpha(x)=\sum_jg_{\alpha j}x^j$, then
\begin{equation}
 \Gamma^{E/B;\ell m}_{uv}
 =\frac14\sum_\alpha w_\alpha\sum_{\text{intervals},j}
 g_{\alpha j}M_j(k_\alpha;b_-,b_+).
 \label{eq:finite-linear-response}
\end{equation}
with finite exponential moments
\begin{equation}
 \begin{aligned}
 M_j(k;b_-,b_+)&=[F_j(k,x)]_{b_-}^{b_+},\quad k\ne0,\\
 F_j(k,x)&=e^{\ii kx}\sum_{r=0}^{j}
 \frac{(-1)^rj!x^{j-r}}{(j-r)!(\ii k)^{r+1}},\\
 M_j(0;b_-,b_+)&=\frac{b_+^{j+1}-b_-^{j+1}}{j+1}.
 \end{aligned}\label{eq:finite-exponential-moments}
\end{equation}
Each interval contributes finitely many monomials, and there are only four propagation phases. Thus Eq.~\eqref{eq:finite-linear-response} is a finite elementary expression with no remaining angular integral. The last expression gives its continuous extension at $k=0$.
\end{proof}

\subsection{A representative integral}\label{app:real-angle-example}
In the collinear limit $\bu=\bv=\boldsymbol e_x$ of the standard pair frame, the lowest electric component is
\begin{equation}
 \begin{split}
 \Gamma^{E;40}_{uv}={}&\frac{9\sqrt{70}}{2048}\int_{-1}^{1}
 (1-x)^4(1+x)^2\\
 &\times(1-e^{-\ii a_u(1+x)})(1-e^{\ii a_v(1+x)})\,\dd x.
 \end{split}\label{eq:E40-direct-example}
\end{equation}
The coefficients follow from $\widetilde Y_{40}(\boldsymbol e_x)=9/8$, $P_4^{(4)}=105$, and Eq.~\eqref{eq:elementary-collinear-numerator}. Applying the finite exponential moments and setting $a_u=a_v=a$ gives
\begin{equation}\label{eq:compact-E40}
\begin{split}
\Gamma^{E;40}(a,a,0)={}&\frac{3\sqrt{70}}{280}
+\frac{27\sqrt{70}}{64a^7}\big[4a^3-20a\\
&+(15-2a^2)\sin2a-10a\cos2a\big].
\end{split}
\end{equation}
The negative powers cancel at the origin. The two endpoint expansions are given in Eq.~\eqref{eq:E40-endpoint-limits}.

\subsection{Collinear radial functions}\label{app:compact-collinear}\label{app:collinear}
Let $\bv=\epsilon\bu$ ($\epsilon=\pm1$). Axial symmetry reduces the $I/E$ response at each order to radial functions. Define
\begin{equation}
 \begin{aligned}
 H_\ell^{I,\epsilon}(x)&=\frac{(1-x^2)^2P_\ell(x)}{(1+x)(1+\epsilon x)},\\
 H_\ell^{E,\epsilon}(x)&=\sqrt{\frac{(\ell-4)!}{(\ell+4)!}}
 \frac{(1-x^2)^4P_\ell^{(4)}(x)}{(1+x)(1+\epsilon x)}.
 \end{aligned}\label{eq:compact-polynomials}
\end{equation}
The numerators cancel the denominators exactly. Expanding the resulting finite polynomials as
$H_\ell^{X,\epsilon}(x)=\sum_{r=0}^{\ell+2}h_{\ell r}^{X,\epsilon}P_r(x)$ and using the spherical Bessel integral identity gives
\begin{equation}
 F_\ell^{X,\epsilon}(k)
 =\frac18\sum_{r=0}^{\ell+2}h_{\ell r}^{X,\epsilon}\ii^rj_r(k).
 \label{eq:compact-radial-finite}
\end{equation}
The full endpoint response is
\begin{equation}\label{eq:compact-main-combination}
\begin{split}
\Gamma^{X;\ell m}_{uv}=\widetilde Y_{\ell m}(\boldsymbol u)
\big[&F_\ell^{X,\epsilon}(0)-e^{-\ii a_u}F_\ell^{X,\epsilon}(-a_u)\\
&-e^{\ii a_v}F_\ell^{X,\epsilon}(\epsilon a_v)\\
&+e^{\ii(a_v-a_u)}F_\ell^{X,\epsilon}(\epsilon a_v-a_u)\big].
\end{split}
\end{equation}
At the lowest intensity order the two radial functions are
\begin{align}\label{eq:compact-I0-kernels}
F_0^{I,+}(k)&=\frac{2j_0(k)-3\ii j_1(k)-j_2(k)}{12},\\
F_0^{I,-}(k)&=\frac{j_0(k)+j_2(k)}{12}.
\end{align}
Equation~\eqref{eq:compact-radial-finite} requires no singular branch at $k=0$. Equation~\eqref{eq:E40-direct-example} gives an independent integral representation for the lowest electric component. All $V/B$ endpoint components vanish by Corollary~\ref{sym:cor-zeros}. The change of basis at each order involves only finite polynomials.

\section{Proofs for the coherent response and distance averaging}
\label{core:app-coherence}
\label{core:app-distance}
\label{coreapp:appendix}

\subsection{Magnetic coherence and the radial coefficient}

Take the common sight line as the polar axis and set $x=\cos\theta$ and $h=\sqrt{1-x^2}$. In the resolved limit the collinear antenna patterns are $F^+=(1-x)/2$ and $F^\times=0$, giving
\begin{equation}
 K_Q^{\rm PP}=\frac{(1-x)^2}{4}f(x)e^{\ii\rho h\cos\varphi},
 \qquad K_U^{\rm PP}=0,
 \label{core:eq-resolved-spin-kernel}
\end{equation}
and
\begin{equation}
 A_\ell[f]=\frac{\sqrt{2\ell+1}}{16}\int_{-1}^{1}
 (1-x)^2d^\ell_{0,-4}(\theta)f(x)\,\dd x.
 \label{core:eq-A-radial}
\end{equation}

\begin{proof}[Proof of Theorem~\ref{core:thm-b-coherence} and Corollary~\ref{core:cor-b-slope}]
The azimuthal integral gives
\begin{equation}
 \begin{split}
 B_{\ell m}={}&\frac{\ii\sqrt{2\ell+1}}{32}
 \int_{-1}^{1}(1-x)^2f(x)\\
 &\times[d^\ell_{m,-4}(\theta)-d^\ell_{m,4}(\theta)]
 \,\ii^mJ_m(\rho h)\,\dd x.
 \end{split}
 \label{core:eq-B-bessel}
\end{equation}
The angular-momentum recurrence and Wigner reflection identity give
\begin{equation}
 h(d^\ell_{1,-4}-d^\ell_{1,4})
 =h(d^\ell_{-1,-4}-d^\ell_{-1,4})
 =-\frac{8d^\ell_{0,-4}}{\sqrt{\ell(\ell+1)}}.
 \label{core:eq-Wigner-step}
\end{equation}
Using $J_1(z)=z/2+O(z^3)$ yields
$B_{\ell,\pm1}=2\rho A_\ell[f]/\sqrt{\ell(\ell+1)}+O(\rho^3)$.
The remaining components satisfy $B_{\ell0}=0$ and $B_{\ell m}=O(\rho^{|m|})$.
By $B_{\ell m}(-\rho)=(-1)^mB_{\ell m}(\rho)$ the squared norm contains only even powers. Summing at fixed $\ell$ gives Eq.~\eqref{core:eq-B-invariant}. If $A_\ell\ne0$, taking the square root gives Eq.~\eqref{core:eq-B-slope}. Otherwise $m=\pm2$ may contribute at order $O(\rho^2)$. At equal distances $N_\ell^{E,\rm full}=2|A_\ell(0)|+O(\rho^2)$, giving Eq.~\eqref{core:eq-B-E-full-ratio}.
\end{proof}

For deterministic distances the radial coefficient is a finite sum of spherical Bessel functions. Set
$n_\ell=\sqrt{(\ell-4)!/(\ell+4)!}$ and expand
$Q_\ell=(1-x)^2P_\ell^4(x)=\sum_nq_{\ell n}x^n=\sum_jb_{\ell j}P_j(x)$. Then
\begin{align}
 b_{\ell j}&=(2j+1)\!
 \sum_{\substack{j\leq n\leq\ell+2\\n-j\ {\rm even}}}
 \frac{q_{\ell n}n!}{(n-j)!!(n+j+1)!!},\nonumber\\
 A_\ell(\delta)&=\frac{\sqrt{2\ell+1}n_\ell}{8}e^{\ii\delta}
 \sum_{j=0}^{\ell+2}b_{\ell j}\ii^jj_j(\delta).
 \label{core:eq-radial-finite-Bessel}
\end{align}
At equal distances, four integrations by parts of $Q_\ell=(1-x)^4(1+x)^2P_\ell^{(4)}$ give
$\int_{-1}^1Q_\ell\dd x=16(-1)^\ell[\ell(\ell+1)-12]$, so
\begin{equation}
 A_\ell(0)=(-1)^\ell
 \sqrt{\frac{(2\ell+1)(\ell-4)!}{(\ell+4)!}}
 [\ell(\ell+1)-12].
 \label{core:eq-A-equal}
\end{equation}
In the variable $q=1+x$, the radial integrand has a second-order zero at the lower endpoint and a fourth-order zero at the upper endpoint. Hence at fixed $\ell$,
$A_\ell(\delta)=O(|\delta|^{-3})$.

\subsection{Joint averaging and cancellation of common errors}

For an arbitrary joint phase distribution, the characteristic function and the full averaged response are
\begin{align}
 \Phi(t_u,t_v)&=\langle e^{\ii(t_ua_u+t_va_v)}\rangle,
 \label{core:eq-joint-characteristic}\\
 \langle T\rangle&=1-\Phi(-A_u,0)-\Phi(0,A_v)+\Phi(-A_u,A_v).
 \label{core:eq-general-average}
\end{align}
For a Gaussian distribution with means $\bar a_u,\bar a_v$, variances $v_u,v_v$, and covariance $c_{uv}$,
\begin{equation}
 \begin{split}
 \langle T\rangle={}&1-e^{-\ii\bar a_uA_u-v_uA_u^2/2}
 -e^{\ii\bar a_vA_v-v_vA_v^2/2}\\
 &+e^{\ii(\bar a_vA_v-\bar a_uA_u)}
 e^{-(v_uA_u^2+v_vA_v^2-2c_{uv}A_uA_v)/2}.
 \end{split}
 \label{core:eq-Gaussian-average}
\end{equation}
Changing variables to $a_c,\Delta$ gives
$s^2=v_u+v_v-2c_{uv}$, $\sigma_c^2=(v_u+v_v+2c_{uv})/4$, 
$c_{c\Delta}=(v_v-v_u)/2$, which yields Eq.~\eqref{core:eq-PP-variance}.

\begin{proof}[Proof of the coherence bound]
Using $0\leq Q\leq2$, $|D|\leq2\sin(\gamma/2)$, and
$|c_{c\Delta}|\leq s\sigma_c$, 
$\sqrt{V_{\rm PP}}\leq sQ+\sigma_c|D|$ gives Eq.~\eqref{core:eq-allsky-protection}.
In the physical distance variables of Corollary~\ref{cor:distance_protection}, the same variance bound reads
\[
 \operatorname{Var}\phi_{\rm PP}
 \leq\frac{\{2\sin(\gamma/2)\sigma_{L_c}+[1+\cos(\gamma/2)]\sigma_z\}^2}{L_f^2}.
\]
Combining this with $|\langle e^{\ii\phi}\rangle|=e^{-\operatorname{Var}\phi/2}$ gives the stated sufficient condition.
\end{proof}

At exact alignment and equal mean distances, independent Gaussian common and differential errors can be written as
\begin{equation}
 a_u=\bar a_c+\epsilon_c-\epsilon_d/2,\qquad
 a_v=\bar a_c+\epsilon_c+\epsilon_d/2,
 \label{core:eq-common-prior}
\end{equation}
where $\operatorname{Var}\epsilon_c=\sigma_c^2$ and $\operatorname{Var}\epsilon_d=s^2$. Denote the marginal width in this appendix by $b=(\sigma_c^2+s^2/4)^{1/2}$. Then
\begin{equation}
 \langle T(q)\rangle=1+e^{-s^2q^2/2}
 -2e^{-b^2q^2/2}\cos(\bar a_cq).
 \label{core:eq-common-filter-transfer}
\end{equation}
Define the auxiliary integral
\begin{equation}
 W(b,k)=\frac1{128}\int_0^2(2-q)^4e^{-b^2q^2}\cos(kq)\,\dd q,
 \label{core:eq-W}
\end{equation}
to obtain
\begin{equation}
 \mathcal P_{\rm PP}=W(s,0),\qquad
 \mathcal P_{\rm EP+PE}=2[W(b,0)+W(b,2\bar a_c)].
 \label{core:eq-single-leg-norm}
\end{equation}

\begin{proof}[Proof of selective distance averaging]
Direct differentiation gives
\begin{equation}
 \frac{\partial\mathcal P_{\rm EP+PE}}{\partial\sigma_c^2}
 =-\frac1{32}\int_0^2(2-q)^4q^2e^{-b^2q^2}
 \cos^2(\bar a_cq)\,\dd q\leq0.
 \label{core:eq-filter-monotone}
\end{equation}
Replacing $(2-q)^4$ by 16 and extending the upper integration limit gives an absolute error $O(b^{-2})$ uniformly in $\bar a_c/b$. Set $r=\bar a_c/s$ and $\tau=b/s$. Then
\begin{equation}
 \frac{\mathcal P_{\rm EP+PE}}{\mathcal P_{\rm PP}}
 =\frac2\tau(1+e^{-r^2/\tau^2})[1+O(s^{-1}+b^{-1})].
 \label{core:eq-relative-single-leg}
\end{equation}
By $1\leq1+e^{-r^2/\tau^2}\leq2$, the ratio tends to zero exactly when $b/s\to\infty$, which gives Eq.~\eqref{core:eq-common-filter-criterion}.
The modulus of the averaged near-collinear PP phase is
$\exp[-\sigma_c^2D^2/2-s^2Q^2/2]$.
Taking $s\to\infty$ and $s\gamma^2\to0$ at fixed $\eta=\gamma\sigma_c/\sqrt{s}$, azimuthal integration over the cap gives
\begin{align}
 \mathcal P_{\rm PP}&\sim\frac{\sqrt\pi}{16s}R(\eta),\nonumber\\
 R(\eta)&=\frac2{\sqrt\pi}\int_0^\infty
 e^{-t^2-\eta^2t}I_0(\eta^2t)\,\dd t
 =1-\frac{\eta^2}{\sqrt\pi}+O(\eta^4).
 \label{core:eq-cap-protection}
\end{align}
The integrand is the azimuthal average of $e^{-t^2-2\eta^2t\cos^2\varphi}$. Thus $R$ is nonincreasing, and $\eta\ll1$ protects the dominant PP norm.
\end{proof}

\subsection{Endpoint expansion for polarization decoherence}

\begin{proof}[Proof of Theorem~\ref{core:thm-distance-decay}]
Denote the two PP helicity coefficients locally by $P^\pm$, so that
\begin{equation}
 \begin{split}
 P^\pm_{\ell m}={}&\frac{\sqrt{2\ell+1}}{16}
 \int_{-1}^{1}(1-x)^2e^{\ii\bar\Delta q-s^2q^2/2}\\
 &\times d^\ell_{m,\mp4}(\theta)\ii^mJ_m(\rho\sqrt{1-x^2})\,\dd x.
 \end{split}
 \label{core:eq-distance-helicity-integral}
\end{equation}
At fixed $\ell,\rho,\bar\Delta$ the lowest endpoint powers of the Wigner and Bessel factors are
$q^{|m\mp4|/2}$ and $q^{|m|/2}$, respectively. Their product starts at order $q^2$ or higher. Set $H=\sqrt{\pi/2}$. Then
\begin{equation}
 D_{\ell m}=\frac1{(4-m)!m!}
 \sqrt{\frac{(\ell+4)!(\ell-m)!}{(\ell-4)!(\ell+m)!}},\quad0\leq m\leq4,
 \label{core:eq-D-coefficient}
\end{equation}
Substitution of $J_m(z)=(z/2)^m/m!+O(z^{m+2})$ and $t=sq$ gives, for $0\leq m\leq4$,
\begin{equation}
 P^+_{\ell m}=\frac{(-1)^{\ell+m}\ii^m\sqrt{2\ell+1}H}{16s^3}
 D_{\ell m}\rho^m+O(s^{-4}).
 \label{core:eq-helicity-leading}
\end{equation}
The remaining positive-helicity components are at least one order smaller. Reflection gives the negative-helicity components.
Forming $E=(P^++P^-)/2$ and $B=\ii(P^+-P^-)/2$ and summing gives
\begin{align}
 N_\ell^{E,\rm PP}
 &=\frac{\sqrt{2\ell+1}H}{16s^3}
 \left[D_{\ell0}^2+\frac12\sum_{m=1}^{4}D_{\ell m}^2\rho^{2m}\right]^{1/2}\nonumber\\
 &\quad+O(s^{-4}),\label{core:eq-Gaussian-E-norm}\\
 N_\ell^{B,\rm PP}
 &=\frac{\sqrt{2\ell+1}H}{16s^3}
 \left[\frac12\sum_{m=1}^{4}D_{\ell m}^2\rho^{2m}\right]^{1/2}\nonumber\\
 &\quad+O(s^{-4}).\label{core:eq-Gaussian-B-norm}
\end{align}
The endpoint Taylor remainder is uniformly bounded, and the Gaussian controls the integral outside the endpoint region. The resulting remainder is $O(s^{-4})$.
The endpoint constants for $I_{00}$ and $V_{10}/\gamma$ are $1/4$ and $-\ii\sqrt3/4$, respectively. The same substitution yields Eq.~\eqref{core:eq-IV-decoherence}.

When $\rho=\lambda\sqrt{s}$, one must retain $J_m(\lambda\sqrt{2t})$. Define
\begin{equation}
 F_4(z)=\int_0^\infty e^{-t^2/2}J_4(z\sqrt{2t})\,\dd t.
 \label{core:eq-F4}
\end{equation}
Only the pole values of the $m=\pm4$ components contribute at order $s^{-1}$. Combining the two helicities then gives Eq.~\eqref{core:eq-second-transverse-scale}. The bound $|d^\ell_{mn}|,|J_m|\leq1$ and the Gaussian envelope justify taking the limit inside the integral. At a zero of $F_4$ the leading term vanishes.
\end{proof}

\section{Angular-spectrum limits and regularity of the full kernel}
\label{core:app-spectrum}

\subsection{Full-Stokes sum rules and redistribution across multipoles}

\begin{proof}[Proof of Theorem~\ref{core:thm-stokes-budget}]
Let $(a_1,a_2)$ and $(b_1,b_2)$ be the endpoint antenna patterns in a common real polarization basis. Then
\begin{align}
 K_I&=T(a_1b_1+a_2b_2),&K_Q&=T(a_1b_1-a_2b_2),\nonumber\\
 K_V&=\ii T(a_2b_1-a_1b_2),&K_U&=T(a_1b_2+a_2b_1).
 \label{core:eq-pointwise-Stokes}
\end{align}
It follows that $|K_I|^2+|K_V|^2=|K_Q|^2+|K_U|^2$. Our Parseval normalization reads
$\sum_{\ell m}|\Gamma_{\ell m}|^2=(16\pi)^{-1}\int|K|^2\dd\Omega$. The total electric plus magnetic norm is half the sum over the two helicities. This gives Eq.~\eqref{core:eq-Stokes-sum-rule} with common value
\begin{equation}
 \mathcal P[T]=\frac1{256\pi}\int
 (1-\bn\cdot\bu)^2(1-\bn\cdot\bv)^2|T|^2\,\dd\Omega.
 \label{core:eq-general-budget}
\end{equation}
\end{proof}

\begin{proof}[Proof of Theorem~\ref{core:cor-spectral-migration}]
Applying Parseval's identity to Eq.~\eqref{core:eq-resolved-spin-kernel} gives
\begin{equation}
 \mathcal P_{\rm PP}[f,\rho]
 =\frac1{128}\int_{-1}^{1}(1-x)^4|f(x)|^2\,\dd x.
 \label{core:eq-PP-Parseval}
\end{equation}
A pure phase therefore preserves the total. For $|f|=1$ it equals $1/20$, while Gaussian averaging gives
$\mathcal P_{\rm PP}=128^{-1}\int_0^2(2-q)^4e^{-s^2q^2}\dd q
=\sqrt\pi/(16s)+O(s^{-2})$.
To obtain the spectrum, set
\begin{equation}
 k=\frac{\ell+1/2}{\sqrt{s}}.
 \label{core:eq-angular-k}
\end{equation}
The Jacobi representation
\begin{equation}
 \begin{split}
 d^\ell_{0,-4}(\arccos(q-1))={}&(-1)^\ell
 \frac{\sqrt{(\ell+4)!(\ell-4)!}}{16\ell!}\\
 &\times[q(2-q)]^2P_{\ell-4}^{(4,4)}(1-q),
 \end{split}
 \label{core:eq-Jacobi-representation}
\end{equation}
and the Mehler--Heine limit~\cite{joint_angular_limit_JAL_DLMF18} give, on compact intervals of nonzero $k$,
$(-1)^\ell d^\ell_{0,-4}\to J_4(k\sqrt{2t})$. The bound $|d|\leq1$ and the Gaussian envelope control the integral over $t=sq$ and give
\begin{equation}
 \frac{4s(-1)^\ell A_\ell(s)}{\sqrt{2\ell+1}}
 \longrightarrow F_4(k_0),\qquad k\longrightarrow k_0>0.
 \label{core:eq-high-l-transition}
\end{equation}
This absolute convergence holds even at zeros of $F_4$, and yields
\begin{equation}
 \frac{|A_\ell(s)|^2}{\mathcal P_{\rm PP}(s)}
 =p_E(k)\Delta k+o(\Delta k),\qquad\Delta k=s^{-1/2}.
 \label{core:eq-spectral-measure}
\end{equation}
The function $F_4$ is the order-four Hankel transform of $e^{-r^4/8}$. Hence
$\int_0^\infty kF_4(k)^2\dd k=\sqrt\pi/2$. Convergence on compact intervals, positivity, and the exact total exclude the escape of a finite fraction of the norm from the $\ell\sim\sqrt{s}$ scale. This argument gives no uniform relative error at arbitrary finite $s$.
For fixed $L$, Eq.~\eqref{core:eq-helicity-leading} and
$\sum_{\ell=4}^L(2\ell+1)(\ell+4)!/(\ell-4)!=(L+1)(L+5)!/[5(L-4)!]$ give
\begin{equation}
 \frac{\sum_{\ell=4}^{L}|A_\ell(s)|^2}{\mathcal P_{\rm PP}(s)}
 \sim\frac{\sqrt\pi(L+1)(L+5)!}{92160s^5(L-4)!}.
 \label{core:eq-fixed-cutoff-fraction}
\end{equation}
\end{proof}

\subsection{Total magnetic norm and the joint angular scale}

\begin{proof}[Proof of Theorem~\ref{core:cor-fixed-rho-budget}]
The radial spin-four operator and its Green function are
\begin{align}
 \mathcal L_4&=-\frac{\dd}{\dd x}\left[(1-x^2)\frac{\dd}{\dd x}\right]
              +\frac{16}{1-x^2},\label{core:eq-L4}\\
 G_4(x,y)&=\frac18
 \left[\frac{(1+x_<)(1-x_>)}{(1-x_<)(1+x_>)}\right]^2.
 \label{core:eq-Green4}
\end{align}
The homogeneous solutions $[(1+x)/(1-x)]^{\pm2}$ have weighted Wronskian 8. The derivative jump and regularity at both endpoints determine $G_4$. Its spectral decomposition gives
\begin{align}
 C(s)&=8\sum_{\ell\geq4}\frac{|A_\ell(s)|^2}{\ell(\ell+1)}\nonumber\\
 &=\frac1{64}\int_0^2\frac{(2-r)^4}{r^2}e^{-s^2r^2/2}
 \left[\int_0^r q^2e^{-s^2q^2/2}\dd q\right]\dd r.
 \label{core:eq-Green-B-coefficient}
\end{align}
Rescaling by $t=sr,u=sq=vt$ gives the leading term
$[4s^2]^{-1}\int_0^1v^2/(1+v^2)\dd v=(4-\pi)/(16s^2)$.
Since $n_x^2\leq2q$, the squared $L^2$ norm of the order-$j$ term in the transverse exponential is $O(\rho^{2j}s^{-j-1})$.
The first- and second-order magnetic terms have only odd and even $m$, respectively, so their cross term vanishes. Cauchy's inequality bounds the cross term between first and third order by $O(\rho^4s^{-3})$. The remaining terms are no larger. This gives Eq.~\eqref{core:eq-fixed-rho-B-total} on bounded sets of $\rho$.
\end{proof}

\begin{proof}[Proof of Theorem~\ref{core:thm-joint-angular}]
In Eq.~\eqref{core:eq-joint-subdomain} the limiting spin field on the antipodal cap is
$f_\pm(r,\varphi)=e^{-r^4/8}e^{\ii\lambda r\cos\varphi}e^{\pm4\ii\varphi}$.
Use the planar Fourier transform normalized by $(2\pi)^{-1}$ and set
$\boldsymbol q=\boldsymbol k-\lambda\boldsymbol e_x$, $q=|\boldsymbol q|$, $\alpha=\arg\boldsymbol q$, and $\beta=\arg\boldsymbol k$. The azimuthal integral
$\int_0^{2\pi}e^{\pm4\ii\varphi-\ii qr\cos(\varphi-\alpha)}\dd\varphi
=2\pi J_4(qr)e^{\pm4\ii\alpha}$ and the translation identity yield
\begin{equation}
 \widehat E=F_4(q)\cos4(\alpha-\beta),\qquad
 \widehat B=F_4(q)\sin4(\alpha-\beta).
 \label{core:eq-flat-EB}
\end{equation}
Writing $\vartheta_k=\alpha-\beta$ gives the normalized spectra
\begin{align}
 p_E(k;\lambda)&=\frac{k}{\pi^{3/2}}\int_0^{2\pi}
 F_4(|ke^{\ii\beta}-\lambda|)^2\cos^2(4\vartheta_k)\,\dd\beta,
 \label{core:eq-joint-pE}\\
 p_B(k;\lambda)&=\frac{k}{\pi^{3/2}}\int_0^{2\pi}
 F_4(|ke^{\ii\beta}-\lambda|)^2\sin^2(4\vartheta_k)\,\dd\beta.
 \label{core:eq-joint-pB}
\end{align}
For each fixed azimuthal order the Jacobi--Bessel limit gives convergence on the sphere. On compact caps the tail of the azimuthal Bessel series vanishes uniformly. Outside them a quartic exponential envelope bounds the remainder. The planar Parseval norm equals the exact spherical total, giving convergence of the full spectral mass.

Let $t=\lambda/q$. The azimuthal average of $\cos8\vartheta$ for the angle between the two wave vectors is
\begin{align}
 M_4(t)&=\frac1{2\pi\ii}\oint_{|z|=1}
 \frac{z^3(1+tz)^4}{(z+t)^4}\dd z,\label{core:eq-M4-residue}\\
 &=\begin{cases}
 1-16t^2+60t^4-80t^6+35t^8,&0\leq t<1,\\
 0,&t\geq1.
 \end{cases}\label{core:eq-M4}
\end{align}
In the first case take the residue at $z=-t$. In the second there is no interior pole, and $t=1$ has a continuous limiting value. Hence
\begin{equation}
 \mathcal B(\lambda)=\frac1{\sqrt\pi}\int_0^\infty
 qF_4(q)^2[1-M_4(\lambda/q)]\,\dd q.
 \label{core:eq-B-fraction}
\end{equation}
The bound $F_4(q)=O(q^4)$ controls the region $q\lesssim\lambda$ in the small-$\lambda$ expansion. The leading term agrees with the Green-function calculation. As $\lambda\to\infty$, dominated convergence gives $\mathcal B\to1/2$. Since $M_4$ need not be nonnegative, the magnetic fraction can exceed $1/2$.
\end{proof}

Finite-geometry corrections follow from the same cap limit. Write
$d=\gamma\sqrt{s}/2$ and $\eta=\gamma\sigma_c/\sqrt{s}$, both fixed and finite. On the antipodal cap set $z=x_\perp+\ii y_\perp$ and $r=|z|$. Then
$A_u=|z-d|^2/(2s)+O(s^{-2})$, $A_v=|z+d|^2/(2s)+O(s^{-2})$.
The averaged envelope is $\exp[-(r^2+d^2)^2/8-\eta^2x_\perp^2/2]$. Bounded antenna amplitudes, convergence on compact caps, and the quartic tail bound imply convergence of the rescaled field in $L^2$. For $\eta=0$, Eq.~\eqref{core:eq-general-budget} gives
\begin{align}
 \mathcal P_{\rm PP}&\sim\frac1{16\pi s}
 \int_{\mathbb R^2}e^{-(r^2+d^2)^2/4}\dd^2z,
 \label{core:eq-cap-total-proof}\\
 &=\frac{\sqrt\pi}{16s}\operatorname{erfc}(d^2/2).
 \label{core:eq-finite-geometry-total}
\end{align}

\subsection{Four-term cancellation and high-multipole regularity}
\label{core:app-regularization}

\begin{proof}[Proof of Eqs.~\eqref{core:eq-spectral-cancellation} and \eqref{core:eq-noncommuting-norms}]
Equation~\eqref{core:eq-common-filter-transfer} obeys
$\langle T(q)\rangle=(\bar a_c^2+\sigma_c^2-s^2/4)q^2+O(q^4)$. Hence
$\langle T\rangle=q^2H(q)$, with $H$ entire. The radial spin-four kernel is
$F(x)=(1-x^2)^2H(1+x)$, with
\begin{equation}
 \mathcal L_4[(1-x^2)^2H]
 =(1-x^2)^2[-(1-x^2)H''+10xH'+20H].
 \label{core:eq-L4-regularity}
\end{equation}
The boundary terms vanish at both poles. Repeated integration by parts gives, for each positive integer $N$,
\begin{equation}
 |\Gamma^{E,\rm full}_{\ell0}|
 \leq\frac{\sqrt2\,\|\mathcal L_4^NF\|_{L^2(-1,1)}}
 {16[\ell(\ell+1)]^N}.
 \label{core:eq-superalgebraic-collinear}
\end{equation}
The endpoint constants of EE+PP and EP+PE are $\pm2$, respectively, and both first derivatives vanish. Subtracting their respective constants leaves functions in the same smoothness class, so the algebraic tails are $\pm2A_\ell(0)$. The constants in the bounds may grow with the finite distance parameters.

Without distance averaging, $T=2(1-\cos a_cq)$. The Riemann--Lebesgue lemma gives a fixed-order limit of twice the EE coefficient, with squared norm $1/5$. On the other hand, $\cos^2(a_cq)=[1+\cos(2a_cq)]/2$ gives
\begin{equation}
 \mathcal P_{\rm EP+PE}(a_c)\longrightarrow\frac1{10},\qquad
 \mathcal P_{\rm full}(a_c)\longrightarrow\frac3{10}.
 \label{core:eq-oscillatory-norm-limits}
\end{equation}
The cross term between the oscillatory residual and $2K_{\rm EE}$ tends to zero, giving the two noncommuting limits.
\end{proof}

\subsection{Conditions on the distance prior}
\label{core:app-priors}

\begin{lemma}[Endpoint order for a general differential prior]
\label{core:lem-prior-moment}
Suppose $\Delta=\bar\Delta+wZ$, $\chi(t)=\langle e^{\ii tZ}\rangle$, and
$K(q)=q^k[v+o(1)]$, with $K(q)/q^k$ bounded on $[0,2]$. If
$\int_0^\infty t^k|\chi(t)|\dd t<\infty$, then
\begin{equation}
 \int_0^2K(q)e^{\ii\bar\Delta q}\chi(wq)\,\dd q
 =\frac{v}{w^{k+1}}\int_0^\infty t^k\chi(t)\,\dd t+o(w^{-k-1}).
 \label{core:eq-prior-moment-summary}
\end{equation}
\end{lemma}
\begin{proof}
Setting $t=wq$ gives the rescaled integrand
$\mathbf1_{t\leq2w}t^k\chi(t)[K(t/w)/(t/w)^k]e^{\ii\bar\Delta t/w}$.
The bracket is bounded and tends to $v$. Dominated convergence using $t^k|\chi(t)|$ proves the result.
\end{proof}

Finite variance alone cannot replace this weighted integrability condition. For a standard Laplace differential prior, the characteristic function
$\chi(t)=(1+t^2)^{-1}$ fails the $k=2$ condition. Let $P(w)$ denote the left side of Eq.~\eqref{core:eq-prior-moment-summary} and write $K(q)e^{\ii\bar\Delta q}=q^2R(q)$. Then
\begin{equation}
 w^2P(w)=\int_0^2R(q)\dd q-
 \int_0^2\frac{R(q)}{1+w^2q^2}\dd q
 \longrightarrow\int_0^2R(q)\dd q.
 \label{core:prior-laplace}
\end{equation}
Boundedness of $R$ makes the last term vanish. For a nonzero leading coefficient the decay is $w^{-2}$.

\begin{lemma}[Regularity of the averaged full kernel]
\label{core:lem-prior-regularity}
Suppose that for each nonnegative integer $r$ the joint distance distribution satisfies
$\langle|a_ua_v|(1+|a_u|+|a_v|)^r\rangle<\infty$. Then the averaged full-Stokes kernel is smooth, and its angular tail falls faster than every inverse power.
\end{lemma}
\begin{proof}
The apparent denominator in the single-pulsar response can be removed by writing
\begin{equation}
 R_p^A=\frac12p^ip^je^A_{ij}\psi(a_p,A_p),\qquad
 \psi(a,A)=\ii a\int_0^1e^{-\ii atA}\dd t.
 \label{core:prior-entire-transfer}
\end{equation}
$\psi$ is entire in $A$. The $I,V$ kernels are smooth scalars and $Q\pm\ii U$ are smooth spin sections, with
\begin{equation}
 \|\nabla^rK_\sigma(a_u,a_v)\|_\infty
 \leq C_r|a_ua_v|(1+|a_u|+|a_v|)^r.
 \label{core:prior-derivative-bound}
\end{equation}
Averaging therefore commutes with derivatives of every order. Let $k_{\sigma,\ell m}$ be the orthonormal spin-harmonic coefficients of the averaged kernel, and let $\mathcal C_\sigma$ be the self-adjoint Casimir operator with eigenvalue $\ell(\ell+1)$. For each positive integer $N$,
\begin{equation}
 \sum_{\ell>L,m}|k_{\sigma,\ell m}|^2
 \leq\frac{\|\mathcal C_\sigma^N\langle K_\sigma\rangle\|_2^2}
 {[(L+1)(L+2)]^{2N}}.
 \label{core:prior-spectral-bound}
\end{equation}
\end{proof}
Joint priors of compact support or with finite absolute moments of every order satisfy this sufficient condition, including correlated distances. The bound holds at fixed distribution parameters. Its constants need not be uniform in the large-distance limit, and the moment conditions are sufficient rather than necessary.

\section{Uniform four-term approximation and finite transition kernels}
\label{ua:appendix}
\subsection{Construction of the matched response}
In this appendix we use $a=a_u$, $b=a_v$, and $\boldsymbol q=b\bv-a\bu$. The remaining phase scales are those of Eq.~\eqref{ua:q}.
Let $H^X_{uv,\ell m}$ be the Earth-term antenna product including a fixed sky weight, and define
\begin{equation}
 \mathcal T[H](\boldsymbol k)
 =\frac1{8\pi}\int_{S^2}H(\bn)e^{\ii\boldsymbol k\cdot\bn}\dd\Omega.
 \label{ua:T}
\end{equation}
Suppressing the mode indices, the four exact responses are
\begin{align}
 \Gamma_{EE}&=\mathcal T[H_{uv}](0),\nonumber\\
 \Gamma_{EP}&=-e^{\ii b}\mathcal T[H_{uv}](b\bv),\nonumber\\
 \Gamma_{PE}&=-e^{-\ii a}\mathcal T[H_{uv}](-a\bu),\nonumber\\
 \Gamma_{PP}&=e^{\ii\Delta}\mathcal T[H_{uv}](\boldsymbol q).
 \label{ua:four_exact}
\end{align}
We retain the spatial phase and the external phase $\Delta$ separately. In general $\boldsymbol q\cdot\bu=b\cos\gamma-a$ differs from $\Delta$.
For $X=I,E,B$, define in the same fixed sky basis
\begin{equation}
 \Phi^X_{uv}(\boldsymbol k)
 =\tfrac12\mathcal T[H^X_{uu}+H^X_{vv}](\boldsymbol k).
 \label{ua:symmetric}
\end{equation}
Before averaging the coaxial kernels, rotate the entire $m$ multiplet into that basis.
For $0<\gamma<\pi$, choose the midpoint basis
\begin{equation}
 \boldsymbol w=\frac{\bu+\bv}{2\cos(\gamma/2)},\qquad
 \boldsymbol t=\frac{\bv-\bu}{2\sin(\gamma/2)},
 \label{ua:midpoint}
\end{equation}
Let $x=\bn\cdot\boldsymbol w$ and $z=\bn\cdot(\boldsymbol w\times\boldsymbol t)$. With our sign convention for circular polarization, define
\begin{equation}
 G_{\boldsymbol w\boldsymbol t}=-\frac{\ii z(1-x)}{2(1+x)},\qquad
 \mathcal G^{\ell m}_{\boldsymbol w\boldsymbol t}
 =\mathcal T[G_{\boldsymbol w\boldsymbol t}\widetilde Y_{\ell m}].
 \label{ua:G}
\end{equation}
Fix $0\leq\chi\leq1$ equal to one on $[0,1]$ and zero on $[2,\infty)$, and set
\begin{equation}
 \widehat\Phi^X_{uv,A}(\boldsymbol k)=
 \begin{cases}
  \Phi^X_{uv}(\boldsymbol k),&X=I,E,B,\\
  \chi(A^{1/3}\gamma)\sin\gamma\,
  \mathcal G^{\ell m}_{\boldsymbol w\boldsymbol t}(\boldsymbol k),&X=V.
 \end{cases}
 \label{ua:matched_phi}
\end{equation}
The second line is zero at $\gamma=0$ and outside the cutoff support. The approximation in Theorem~\ref{ua:theorem} is
\begin{align}
 \Gamma_U^X={}&\Gamma_{EE}^X
 -e^{\ii b}\widehat\Phi^X_{uv,A}(b\bv)
 -e^{-\ii a}\widehat\Phi^X_{uv,A}(-a\bu)\nonumber\\
 &+e^{\ii\Delta}\widehat\Phi^X_{uv,A}(\boldsymbol q).
 \label{ua:composite}
\end{align}

\subsection{Uniform estimates and proof of the theorem}
\label{ua:oscillatory}
Constants below may depend on the fixed mode and normalization but are independent of $a,b,\bu,\bv$. The sky weight is treated as a smooth covariant tensor on the sphere.

\begin{lemma}[Uniform oscillatory estimate]
\label{ua:oscillation_lemma}
For every pulsar pair and $\boldsymbol k\in\mathbb R^3$,
\begin{equation}
 |\mathcal T[H_{uv}](\boldsymbol k)|\leq C_{\ell m}\omega_0(|\boldsymbol k|),
 \qquad \omega_0(s)=\frac{\log(2+s)}{1+s}.
 \label{ua:h_bound}
\end{equation}
\end{lemma}
\begin{proof}
Let $P_{ab}=\delta_{ab}-n_an_b$ and $u_\perp=P\bu$. The transverse trace-free antenna is
\begin{equation}
 T_u^{ab}=\frac{u_\perp^a u_\perp^b-\tfrac12|u_\perp|^2P^{ab}}
 {2(1+\bn\cdot\bu)}.
 \label{ua:TT}
\end{equation}
The numerator and denominator vanish to the same order at $-\bu$. Direct differentiation gives
\begin{align}
 |T_u|&\leq C,\qquad \|H_{uv}\|_\infty\leq C_{\ell m},\nonumber\\
 |\nabla_nT_u|+|\nabla_uT_u|
 &\leq C[1+d(\bn,-\bu)^{-1}],\label{ua:TT_derivative}\\
 |\nabla_nH_{uv}|
 &\leq C_{\ell m}[1+d(\bn,-\bu)^{-1}+d(\bn,-\bv)^{-1}],
 \label{ua:H_derivative}
\end{align}
where $d$ is geodesic distance on the sphere. The $L^1$ bound for the weak derivative is independent of the separation between the two singular points.
Write $\boldsymbol k=s\widehat{\boldsymbol k}$ and $t=\widehat{\boldsymbol k}\cdot\bn$ and let $X=\nabla t/|\nabla t|^2$.
Away from the two phase poles, $X\cdot\nabla e^{\ii st}=\ii s e^{\ii st}$ and $\operatorname{div}X=0$.
Remove caps of radius $\varepsilon$ and integrate by parts to obtain
\begin{align}
 |\mathcal T[H](\boldsymbol k)|
 &\leq C_{\ell m}(\varepsilon^2+s^{-1})\nonumber\\
 &\quad+\frac{C}{s}\int_{\sin\theta\gtrsim\varepsilon}
 \frac{|\nabla_nH|}{\sin\theta}\dd\Omega,\nonumber\\
 \sup_p\int_{\sin\theta\gtrsim\varepsilon}
 \frac{\dd\Omega}{\sin\theta\,d(\bn,p)}
 &\leq C(1+|\log\varepsilon|).
 \label{ua:convolution}
\end{align}
The last bound follows by splitting the neighborhoods of the phase poles into dyadic annuli. The integral on each annulus is uniformly bounded and there are $O(|\log\varepsilon|)$ annuli.
Set $\varepsilon=s^{-1/2}$ and use the $L^1$ bound for $s\leq1$. Smooth approximation extends the result to weak derivatives.
\end{proof}

\begin{lemma}[Geometric remainder for symmetric coaxial matching]
\label{ua:symmetric_proof}
For $X=I,E,B$,
\begin{equation}
 \|H^X_{uv}-\tfrac12(H^X_{uu}+H^X_{vv})\|_1
 \leq C_{\ell m}\omega_2(\gamma),
 \label{ua:omega}
\end{equation}
where $\omega_2(0)=0$ and the factor is $\omega_2(\gamma)=\gamma^2\log(e/\gamma)$ when $0<\gamma\leq1$ and unity otherwise.
\end{lemma}
\begin{proof}
Denote the corresponding symmetric bilinear contraction by $W_X$. Then
\begin{equation}
 H^X_{uv}-\tfrac12(H^X_{uu}+H^X_{vv})
 =-\tfrac12W_X(T_u-T_v,T_u-T_v).
 \label{ua:quadratic_difference}
\end{equation}
The antenna caps of radius $O(\gamma)$ contribute $O(\gamma^2)$. Outside the caps Eq.~\eqref{ua:TT_derivative} gives $|T_u-T_v|\leq C\gamma/r$, so
$\|H^X_{uv}-(H^X_{uu}+H^X_{vv})/2\|_1
\leq C_{\ell m}[\gamma^2+\gamma^2\int_\gamma^1\dd r/r]$.
Boundedness covers the remaining angular separations.
\end{proof}

\begin{lemma}[Geometric remainder for the first circular variation]
\label{ua:Vproof}
For $0<\gamma\leq1/2$,
\begin{equation}
 \|H^V_{uv}-\sin\gamma\,G\widetilde Y_{\ell m}\|_1
 \leq C_{\ell m}\gamma^2.
 \label{ua:V_L1}
\end{equation}
\end{lemma}
\begin{proof}
Write $c_\gamma=\cos(\gamma/2)$, $s_\gamma=\sin(\gamma/2)$, and $y=\bn\cdot\boldsymbol t$.
The exact circular antenna obeys
\begin{align}
 K_V&=-\frac{\ii[\bu\cdot\bv-(\bn\cdot\bu)(\bn\cdot\bv)]\bn\cdot(\bu\times\bv)}
 {2(1+\bn\cdot\bu)(1+\bn\cdot\bv)}\nonumber\\
 &=-\frac{\ii\sin\gamma\,zN}{2D},\label{ua:V_exact}\\
 N&=1-x^2-s_\gamma^2(1+z^2),\nonumber\\
D&=(1+c_\gamma x)^2-s_\gamma^2y^2.\nonumber
\end{align}
Set $\theta=d(\bn,-\boldsymbol w)$ and take $c=3$.
Inside the cap $\theta\leq c\gamma$, the bounds are $|K_V|\leq C$ and $|\sin\gamma\,G|\leq C\gamma/\theta$. Each integrates to $O(\gamma^2)$.
Outside the cap $D\asymp\theta^4$, and
\begin{align}
 D-(1+x)^2&=2(c_\gamma-1)x(1+x)\nonumber\\
 &\quad+(c_\gamma-1)^2x^2-s_\gamma^2y^2\nonumber\\
 &=O(\gamma^2\theta^2),\nonumber\\
 N-(1-x^2)&=O(\gamma^2),\nonumber\\
 \left|\frac ND-\frac{1-x^2}{(1+x)^2}\right|
 &\leq C\gamma^2\theta^{-4}.\nonumber
\end{align}
Multiplication by $|\sin\gamma\,z|\leq C\gamma\theta$ and integration give
\begin{align}
 \|H^V_{uv}-\sin\gamma\,G\widetilde Y_{\ell m}\|_1
 &\leq C_{\ell m}\!\left[\gamma^2+
 \gamma^3\int_{c\gamma}^{\pi}\theta^{-3}\sin\theta\,\dd\theta\right]\nonumber\\
 &\leq C_{\ell m}\gamma^2.
\end{align}
\end{proof}

\begin{proof}[Proof of Theorem~\ref{ua:theorem}]
For $I,E,B$, Lemmas~\ref{ua:oscillation_lemma} and \ref{ua:symmetric_proof} give
\begin{align}
 |R_{EP}|&\leq C_{\ell m}\min\{\omega_2(\gamma),\omega_0(b)\},\nonumber\\
 |R_{PE}|&\leq C_{\ell m}\min\{\omega_2(\gamma),\omega_0(a)\},\nonumber\\
 |R_{PP}|&\leq C_{\ell m}\min\{\omega_2(\gamma),\omega_0(\kappa)\}.
 \label{ua:sector_bounds}
\end{align}
By $\kappa\geq2A\gamma/\pi$, use $\omega_2$ on $\gamma\leq A^{-1/3}$ and $\omega_0$ on its complement to obtain Eq.~\eqref{ua:global_bound}.
If $\kappa\leq Q$, then $\gamma\leq\pi Q/(2A)$, yielding Eq.~\eqref{ua:coherent_bound}.

For $V$, set $M=\sin\gamma\,G\widetilde Y_{\ell m}$ and $\chi_A=\chi(A^{1/3}\gamma)$. Then
\begin{equation}
 H-\chi_AM=\chi_A(H-M)+(1-\chi_A)H.
 \label{ua:V_split}
\end{equation}
For $A\geq64$, the first term has support on $\gamma\leq2A^{-1/3}\leq1/2$. By Lemma~\ref{ua:Vproof} its $L^1$ norm is $O(A^{-2/3})$.
The second term is supported on $\gamma\geq A^{-1/3}$. By Lemma~\ref{ua:oscillation_lemma} its $PP$ contribution is $O(A^{-2/3}\log A)$ and its $EP,PE$ contribution is $O(A^{-1}\log A)$.
On $\kappa\leq Q$, for sufficiently large $A$, $\chi_A=1$, and all three remainders are $O_{\ell m,Q}(A^{-2})$.
At exact alignment the original and replacement kernels coincide, so the approximation is exact.
At fixed $\gamma\geq\gamma_0>0$, $a,b,\kappa\gtrsim_{\gamma_0}A$. Equation~\eqref{ua:outer} also follows from Lemma~\ref{ua:oscillation_lemma}.
\end{proof}

\subsection{Analytic representation of the finite transition kernels}
\label{ua:finite_kernels}
Let $d=\boldsymbol k\cdot\boldsymbol w$, $r=|\boldsymbol k-d\boldsymbol w|$, and $K=\sqrt{d^2+r^2}$.
For real $d$ and $r>0$, define the following functions using principal branches and paired limits,
\begin{align}
 \mathcal A(d,r)&=\operatorname{Ei}[\ii(d+K)]+\operatorname{Ei}[\ii(d-K)]\nonumber\\
 &\quad-2\gamma_{\!E}-2\log r,\label{ua:Aaux}\\
 \mathcal A(d,0)&=\operatorname{Ei}(2\ii d)-\gamma_{\!E}-\log(2\ii d),\nonumber\\
 M_1&=2\pi e^{-\ii d}\mathcal A,\nonumber\\
 M_2&=2\pi e^{-\ii d}\big\{(\ii d-r^2/2)\mathcal A\nonumber\\
 &\quad+e^{\ii d}[(\ii d-1)\cos K+K\sin K]+1-r^2\big\},
 \label{ua:M}
\end{align}
where $\mathcal A(0,0)=0$. Values on the axis and at the origin are obtained by continuity.

\begin{proposition}[Finite kernel representation at fixed mode]
For $X=E,B$, the coaxial kernel can be written as
\begin{equation}
 H^{X,\ell m}_{ww}(\bn)=\frac{P^{X,\ell m}_w(\bn)}{(1+\bn\cdot\boldsymbol w)^2},
 \label{ua:Ppoly}
\end{equation}
where $P$ is a finite polynomial that includes the sky normalization and has a fourth-order angular zero at $-\boldsymbol w$, with
\begin{equation}
 \mathcal T[H^{X,\ell m}_{ww}](\boldsymbol k)
 =\frac1{8\pi}P^{X,\ell m}_w(-\ii\partial_{\boldsymbol k})M_2(d,r).
 \label{ua:EBfinite}
\end{equation}
Likewise, $G\widetilde Y_{\ell m}=P^V_{wt}/(1+\bn\cdot\boldsymbol w)$ and $P^V_{wt}(-\boldsymbol w)=0$, with
\begin{equation}
 \mathcal G^{\ell m}_{wt}(\boldsymbol k)
 =\frac1{8\pi}P^V_{wt}(-\ii\partial_{\boldsymbol k})M_1(d,r).
 \label{ua:Vfinite}
\end{equation}
These transforms are finite elementary expressions. The coaxial intensity kernel is $(1-x)^2\widetilde Y_{\ell m}/4$. Its transform follows from finitely many derivatives of $4\pi j_0(K)$. In particular,
\begin{equation}
 \mathcal T[H^I_{ww,00}](\boldsymbol k)
 =\frac{j_0(K)}6-\frac{\ii\mu j_1(K)}4-\frac{P_2(\mu)j_2(K)}{12},
 \label{ua:I00finite}
\end{equation}
where $\mu=\widehat{\boldsymbol k}\cdot\boldsymbol w$, with $1/6$ at $K=0$.
\end{proposition}
\begin{proof}
\begin{samepage}
The smooth, tangential, trace-free rank-four tensor harmonic at fixed $E/B$ gives Eq.~\eqref{ua:Ppoly} and its fourth-order angular zero.
The auxiliary function has the convergent integral representation
\begin{align}
 M_1&=2\pi e^{-\ii d}\int_0^2\frac{F(q)-1}{q}\dd q,\nonumber\\
 M_2&=2\pi e^{-\ii d}\int_0^2
 \frac{F(q)-1-(\ii d-r^2/2)q}{q^2}\dd q,\label{ua:M_integrals}\\
 F(q)&=e^{\ii dq}J_0\!\left(r\sqrt{q(2-q)}\right).\nonumber
\end{align}
\end{samepage}
Equation~\eqref{ua:M} can be checked by differentiation with respect to $d$ and termwise integration of the Bessel series in $d=0$.
After subtraction the numerator is $O(q^2)$. Finite-order parameter derivatives are uniform on compact sets and can therefore be taken under the integral.
In $P_w(-\ii\partial_{\boldsymbol k})M_2$, the finite-part subtractions and the undifferentiated $\operatorname{Ei}/\log$ coefficient depend only on the value and second angular derivatives at the south pole. All vanish by the fourth-order angular zero. In particular,
\begin{equation}
 \boldsymbol w\cdot\nabla P+\tfrac12\Delta_\perp P
 =\tfrac12\Delta_{S^2}P=0\qquad(\bn=-\boldsymbol w).
\end{equation}
Hence Eq.~\eqref{ua:EBfinite} equals an ordinary convergent integral and is a finite elementary expression.
The condition $P^V_{wt}(-\boldsymbol w)=0$ likewise removes the subtraction terms and special-function coefficients in $M_1$, giving Eq.~\eqref{ua:Vfinite}.
The intensity expression follows directly from the spherical-harmonic expansion of a plane wave.
\end{proof}

\subsection{Scope of mode estimates and distance averages}
\label{ua:validation}
A fixed-mode error bound does not imply convergence in the norm over the full angular spectrum.
For example, fix generic $0<\gamma<\pi$ and $a=b=A$, and write $D=K_{uv}-(K_{uu}+K_{vv})/2$. The residual intensity kernel satisfies
\begin{align}
 \delta K=D\big[&-e^{-\ii A(1+\bn\cdot\bu)}-e^{\ii A(1+\bn\cdot\bv)}\nonumber\\
 &+e^{\ii A\bn\cdot(\bv-\bu)}\big],\label{ua:L2counter}\\
 \|\delta K\|_{L^2}^2&\longrightarrow3\|D\|_{L^2}^2>0.\nonumber
\end{align}
The three diagonal terms survive and the oscillatory cross terms vanish, even though every fixed harmonic projection can tend to zero.
If the joint distance distribution satisfies $A\geq A_0\geq64$ almost surely, the triangle inequality gives
\begin{equation}
 |\langle\Gamma-\Gamma_U\rangle|
 \leq C_{\ell mX}A_0^{-2/3}\log(2+A_0).
\end{equation}
The distances need not be independent. The low-$A$ tail requires the exact response. A bound on the mean response error is distinct from a distance-mixture likelihood.

\subsection{Numerical comparison of four-term approximations}
\begin{figure*}[!t]
 \centering
 \includegraphics[width=\textwidth]{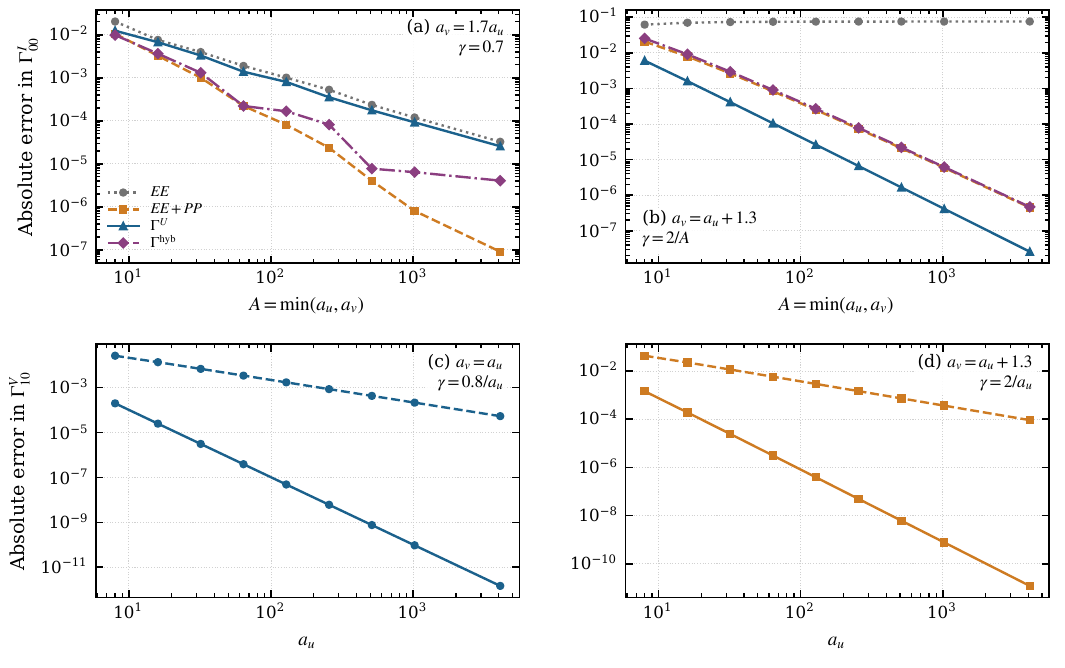}
\caption{Absolute approximation errors relative to the full finite-distance response. (a),(b) Intensity monopole for the four-term composite, hybrid~\eqref{ua:hybrid}, exact $EE+PP$ truncation, and $EE$ approximation. (c),(d) Circular $V_{10}$ response with polar axis normal to the pair plane. Dashed curves show the $EE$ approximation and solid curves include the first-order geometric kernel. All curves use the exact full response as reference.}
 \label{ua:error_figure}
 \label{ua:endpoint_v_figure}
\end{figure*}

When $\Delta$ is finite and $\gamma\propto a_c^{-1}$, $\kappa$ does not grow with the common distance and $PP$ coherence persists. Panels (a) and (b) of Fig.~\ref{ua:error_figure} compare the two paths. Propagation corrections decay along the first, while along the second the finite $PP$ response leaves a persistent error in the $EE$ approximation.

To test the $PP$ approximation separately, consider the hybrid expression, which also satisfies Eq.~\eqref{ua:global_bound},
\begin{equation}
 \Gamma_{\mathrm{hyb}}=\Gamma^{\rm exact}_{EE}
 +\Gamma^{\rm exact}_{EP}+\Gamma^{\rm exact}_{PE}+\Gamma^U_{PP}.
 \label{ua:hybrid}
\end{equation}
The exact $EE+PP$ truncation in the figure shows that retaining $PP$ does not exhaust the finite-distance correction. Moreover, cancellation among terms can make the four-term composite more accurate than the hybrid at individual points. For circular polarization, the exact collinear null can coexist with a nonzero near-collinear slope. In panels (c) and (d) of Fig.~\ref{ua:endpoint_v_figure}, including the leading $O(A^{-1})$ geometric term reduces the replacement remainder to $O(A^{-2})$. A coaxial zero kernel misses the leading response. The asymptotic expansion must use the actual phase combinations such as $\kappa$ and $\kappa\pm\Delta$. No unsuppressed logarithm of distance survives at fixed nonzero angle for the total tensor response. The bounded-kernel assumption need not hold for non-GR responses such as a longitudinal antenna.

\section{Full-spectrum constraints and connection to observations}
\label{core:app-extended}

\subsection{Joint distance averaging and suppression of single-pulsar terms}
The all-sky bound following from Eq.~\eqref{core:eq-PP-variance} is
\begin{equation}
 V_{\rm PP}\leq[2s+2\sin(\gamma/2)\sigma_c]^2.
 \label{core:eq-allsky-protection}
\end{equation}
Common-distance errors cancel in $PP$ but persist in the single-pulsar terms. At exact alignment and equal mean distances, assume independent Gaussian common and differential errors. Increasing $\sigma_c$ monotonically reduces the squared norm $\mathcal P_{\rm EP+PE}$ of the difference between the full kernel and $EE+PP$, while $\mathcal P_{\rm PP}$ remains fixed. When $s\to\infty$,
\begin{equation}
 \frac{\mathcal P_{\rm EP+PE}}{\mathcal P_{\rm PP}}\longrightarrow0
 \quad\Longleftrightarrow\quad\frac{\sigma_c}{s}\longrightarrow\infty.
 \label{core:eq-common-filter-criterion}
\end{equation}
Near alignment, a sufficient window for protecting $PP$ while keeping the distance distribution narrow and positive to good approximation is
\begin{equation}
 s\ll\sigma_c\ll\min(\bar a_c,\gamma^{-1}).
 \label{core:eq-common-window}
\end{equation}
If only the dominant $PP$ norm within the shrinking cap is to be protected, $\gamma\sigma_c\ll1$ can be relaxed to $\gamma\sigma_c/\sqrt{s}\ll1$, provided $s\gamma^2\ll1$ also holds. Appendix~\ref{core:app-distance} gives the proof and decay rates. Selective averaging reflects the correlation structure of the errors. It implies no gain in observational information from less precise distances.

\subsection{Full-Stokes sum rules and spectral partition}
\label{core:sec-joint-angular}
The Stokes kernels share the same two antennas and propagation factor. Their squared norms over the full angular spectrum therefore obey the following identity.
\begin{theorem}[Identity for the full-Stokes squared norms]
\label{core:thm-stokes-budget}
If the endpoint antennas are expressed in a common real polarization basis and all Stokes kernels share a complex transfer factor, then
\begin{equation}
 \sum_{\ell\geq0,m}(|\Gamma^I_{\ell m}|^2+|\Gamma^V_{\ell m}|^2)
 =\sum_{\ell\geq4,m}(|\Gamma^E_{\ell m}|^2+|\Gamma^B_{\ell m}|^2).
 \label{core:eq-Stokes-sum-rule}
\end{equation}
The identity holds for fixed geometry and finite distances and for the squared norm of a distance-averaged kernel. A common finite multipole cutoff need not preserve it.
\end{theorem}
See Appendix~\ref{core:app-spectrum} for the proof. For the $\rho=\bar\Delta=0$ Gaussian average in Theorem~\ref{core:cor-spectral-migration} the normalized continuum electric spectrum is
\begin{equation}
 p_E(k)=\frac{2k}{\sqrt\pi}F_4(k)^2,\qquad
 \int_0^\infty p_E(k)\,\dd k=1,
 \label{core:eq-continuous-E-spectrum}
\end{equation}
where $k=(\ell+1/2)/\sqrt{s}$ and $F_4$ is given in Eq.~\eqref{core:eq-F4}. Equation~\eqref{core:eq-fixed-cutoff-fraction} gives the $s^{-5}$ coefficient at fixed cutoff. A transverse phase redistributes the electric and magnetic fractions on the same angular scale.

\begin{theorem}[Electric and magnetic fractions on the joint scale]
\label{core:thm-joint-angular}
For independent Gaussian common and differential errors, fix the mean radial mismatch $\bar\Delta$, let $s\to\infty$, and take
\begin{equation}
 \lambda=\frac\rho{\sqrt{s}}\quad\hbox{fixed},\qquad
 \gamma\sqrt{s}\longrightarrow0,\qquad
 \frac{\gamma\sigma_c}{\sqrt{s}}\longrightarrow0.
 \label{core:eq-joint-subdomain}
\end{equation}
Then the electric and magnetic continuum spectra on the $k$ scale are nonnegative and obey
$\int_0^\infty[p_E(k;\lambda)+p_B(k;\lambda)]\dd k=1$.
The magnetic fraction of the total squared norm tends to $\mathcal B(\lambda)$ defined in Eq.~\eqref{core:eq-B-fraction}, with
\begin{equation}
 \begin{aligned}
 \mathcal B(\lambda)&=\frac{4-\pi}{\sqrt\pi}\lambda^2+O(\lambda^4)
 &&(\lambda\to0),\\
 \mathcal B(\lambda)&\longrightarrow\frac12&&(\lambda\to\infty).
 \end{aligned}
 \label{core:eq-B-fraction-limits}
\end{equation}
The spherical continuum limit holds at fixed finite $\lambda$. The large-$\lambda$ result is a subsequent limit of the resulting function.
\end{theorem}
\begin{theorem}[Magnetic fraction at fixed transverse mismatch]
\label{core:cor-fixed-rho-budget}
In the resolved limit with radial Gaussian averaging, for fixed bounded $\rho$,
\begin{equation}
 \mathcal P_B(s,\rho)=\rho^2C(s)+O(\rho^4s^{-3}),\qquad
 C(s)\sim\frac{4-\pi}{16s^2}.
 \label{core:eq-fixed-rho-B-total}
\end{equation}
Thus $\mathcal P_B/\mathcal P_{\rm PP}\sim(4-\pi)\rho^2/(\sqrt\pi s)$. At $\rho=0$ the fraction vanishes identically.
\end{theorem}
Full angular integration controls the remainders without interchanging an infinite multipole sum with a termwise expansion (Appendix~\ref{core:app-spectrum}). The transverse phase shifts the local angular wave-vector distribution. The electric and magnetic projections must then be taken relative to the shifted wave vector. The fraction $\mathcal B$ can slightly exceed $1/2$, reaching a maximum of about $0.539$ before approaching equipartition. When $\gamma\sqrt{s}$ is appreciable, Eq.~\eqref{core:eq-finite-geometry-total} gives the finite-angle correction to the total. Finite-distance approximations also require checks on scales such as $\ell^2/\bar a_c$. Scale separation at the spectral peak does not guarantee accuracy throughout the high-order tail.

\subsection{Full spectral tail and scope of the distance prior}
\label{core:sec-regularity-boundaries}
Even single-pulsar terms that are small at fixed mode can determine the high-order tail. At exact alignment, equal mean distances, and finite Gaussian parameters, for every $N$,
\begin{equation}
 \begin{aligned}
 \Gamma^{E,\rm EE+PP}_{\ell0}&=2A_\ell(0)+O_N(\ell^{-2N}),\\
 \Gamma^{E,\rm EP+PE}_{\ell0}&=-2A_\ell(0)+O_N(\ell^{-2N}).
 \end{aligned}
 \label{core:eq-spectral-cancellation}
\end{equation}
The two algebraic tails cancel, leaving a full spectrum that falls faster than every inverse power. At exact alignment and equal distance without distance averaging, let $\mathcal P_{\leq L}(a_c)$ denote the truncated squared norm of the full response. Then
\begin{equation}
 \begin{aligned}
 \lim_{L\to\infty}\lim_{a_c\to\infty}\mathcal P_{\leq L}(a_c)&=\frac15,\\
 \lim_{a_c\to\infty}\lim_{L\to\infty}\mathcal P_{\leq L}(a_c)&=\frac3{10}.
 \end{aligned}
 \label{core:eq-noncommuting-norms}
\end{equation}
The difference comes from the $EP/PE$ residual, which vanishes at each fixed multipole but approaches a total squared norm of $1/10$. Appendix~\ref{core:app-regularization} proves both results. Even a residual with small total norm can affect relative accuracy at very high order.

Gaussian decoherence exponents depend on the prior. The $s^{-3}$ linear polarization law requires weighted integrability of the differential characteristic function. A Laplace prior can instead give an $s^{-2}$ leading term. Variance and endpoint zeros alone do not fix the exponent. A sufficient condition for the averaged tail to retain faster-than-power-law decay is integrability of all angular derivatives of the full kernel over the distance distribution. Gaussian models meet this condition. If their negative-distance tails are appreciable, the actual positive-distance posterior should be used. Appendix~\ref{core:app-priors} gives the general criterion and a counterexample.

\subsection{Response error and parameter estimation}
\label{obs:error_interface}
Evaluating the statistical impact of an ORF correction requires observing windows and noise weights. Let $N$ be the full covariance of the real and imaginary parts of the cross-statistics. In a Gaussian approximation with fixed positive-definite $N$, a locally linearized mean, and no active parameter boundaries,
\begin{equation}
 D^2=\delta\boldsymbol\mu^TN^{-1}\delta\boldsymbol\mu,\qquad
 \delta\boldsymbol\theta^TF\delta\boldsymbol\theta\leq D^2,
 \label{obs:D}
\end{equation}
where $J=\partial\boldsymbol\mu/\partial\boldsymbol\theta$ and $F=J^TN^{-1}J$, and $\delta\boldsymbol\theta$ is the linearized best-fit shift on the identifiable parameter subspace. The condition $D\ll1$ is a conservative local model-error criterion. The quantity $D$ also includes mismatch that cannot be absorbed by parameter shifts and therefore differs from the significance of a parameter bias.

Given a spectrum and polarization templates, first use
\begin{equation}
 \delta C_{uv}(t_i,t_j)=2\operatorname{Re}\int_0^\infty
 S_r(f)\delta\Gamma_{uv}(f)e^{2\pi\ii f(t_i-t_j)}\dd f
 \label{obs:time_cov}
\end{equation}
to construct the covariance difference. Applying $G_u^T\delta C_{uv}G_v$ and the chosen statistic then gives $\delta\boldsymbol\mu$. The factor $S_r$ converts strain to residuals. The columns of $G_i$ span the orthogonal complement of the timing design matrix for that pulsar. For multiple templates, sum over components. An asymptotic order alone therefore cannot set a percentage error for the observable.

\section{Checks of close-pair differencing and a controlled injection}
\label{app:dm_checks}

\subsection{Depth averaging and the cost of a fixed phase}
In Eq.~\eqref{dm:normalized_kernel} write $q=1+\mu$ and define
$\langle h\rangle_W=(3/16)\int_{-1}^{1}(1-\mu)^2h(\mu)\dd\mu$.
Direct integration gives
\begin{equation}
 \langle1\rangle_W=\tfrac12,\quad
 \langle q\rangle_W=\tfrac14,\quad
 \langle q^2\rangle_W=\tfrac15,\quad
 \langle1-\mu^2\rangle_W=\tfrac3{10}.
 \label{dm:weighted_moments}
\end{equation}
For deterministic depth, write $\epsilon_\Delta^2=\rho^2+\Delta^2$. Then $\widehat\Gamma=1+\ii\Delta/4-\Delta^2/10-3\rho^2/40+\cdots$ and
\begin{align}
 1-\operatorname{Re}\widehat\Gamma
 &=\frac{3\rho^2}{40}+\frac{\Delta^2}{10}+O(\epsilon_\Delta^4),\nonumber\\
 1-|\widehat\Gamma|
 &=\frac{3\rho^2}{40}+\frac{11\Delta^2}{160}+O(\epsilon_\Delta^4).
 \label{dm:known_expansion}
\end{align}
Gaussian depth averaging replaces $e^{\ii\Delta q}$ by
$e^{\ii\bar\Delta q-\sigma_\Delta^2q^2/2}$. Expanding the modulus gives Eq.~\eqref{dm:fixed_expansion}. Adjusting the phase at each true depth before averaging gives instead
\begin{equation}
 \langle\lambda_{\rm known}\rangle=
 \frac{3\rho^2}{40}+\frac{11(\bar\Delta^2+\sigma_\Delta^2)}{160}
 +O(\epsilon^4),
 \label{dm:known_average}
\end{equation}
The difference gives Eq.~\eqref{dm:phase_penalty}. The remainder denotes total fourth order in the small phase parameters. It excludes the separate error of the common-sight-line limit.
At finite distance the auto-responses need not equal one. The phase-adjusted equal-weight channel must then use $(g_{uu}+g_{vv})/2-|g_{uv}|$ in place of $1-|g_{uv}|$.

\subsection{Direct check of the full four-term response}
Take pulsar positions $(\pm d/2,0,L)$, with $L=5000$ pc and $f=1/(20\,\mathrm{yr})$. In a fixed normalization for which a distant single pulsar has unit power, we directly evaluate
\begin{equation}
 \lambda_{\rm full}=\frac32\int\frac{\dd\Omega}{4\pi}
 \sum_A\left|\frac{R_u^A-R_v^A}{\sqrt2}\right|^2.
 \label{dm:full_power_check}
\end{equation}
The response uses the true directions, path lengths, and all $EE/EP/PE/PP$ terms. Evaluating the transfer with \texttt{expm1} avoids subtracting two nearly unit correlations.
Table~\ref{dm:full_table} gives the results. The relative change between the two sky grids $6144\times128$ and $8192\times192$ is about $1.2\times10^{-12}$. The correction to the common-sight-line kernel is about $3.4\times10^{-6}$. These are empirical convergence checks at one frequency and three configurations. They do not bound errors across parameter space or after broadband integration.
\begin{table}[t]
\caption{Full difference-channel power and its correction relative to the common-sight-line model. We use the fixed asymptotic single-pulsar normalization without renormalizing by finite-distance auto-responses.}
 \label{dm:full_table}
 \centering
 \begin{tabular}{ccc}
 \toprule
 $d$ (pc)&$\lambda_{\rm full}$&$(\lambda_{\rm full}-\lambda_{\rm coh})/\lambda_{\rm full}$\\
 \midrule
 $0.01$&$7.87430212\times10^{-6}$&$3.42071\times10^{-6}$\\
 $0.10$&$7.87040562\times10^{-4}$&$3.42201\times10^{-6}$\\
 $0.30$&$7.05508877\times10^{-3}$&$3.43252\times10^{-6}$\\
 \bottomrule
 \end{tabular}
\end{table}

\subsection{Local templates after timing projection}
We perform a separate controlled injection with equal, fixed true depths. The main-text noise budget instead averages over a depth prior. Take local pulsar positions $x_u=0,x_v=d$ and a weak-field point mass on a straight flyby trajectory $(b_x,v\tau,h)$, with $b_x=-b/\sqrt2$, $h=b/\sqrt2$, and $\tau=t-0.35T$. The Newtonian pulsar-end Doppler template can be written as~\cite{obs_Cherukupalli2026DM}
\begin{equation}
 s_i(t)=-\frac{GMh}{cv^2B_i^2}\sqrt{B_i^2+v^2\tau^2},\qquad
 B_i^2=(b_x-x_i)^2+h^2,
 \label{dm:doppler_template}
\end{equation}
up to an affine function, and satisfies $\dd^2s_i/\dd t^2=-a_{z,i}/c$. This injection isolates the pulsar-end Doppler term. A complete gauge-invariant timing signal would also require treatment of other contributions. We omit the Earth, Shapiro, and Einstein terms and do not average over a flyby population.

Take $T=20$ yr, $129$ uniformly spaced samples, $b=0.003$ pc, and $v=300$ km/s. For each pulsar the white-noise variance is unity, the one-sided spectrum is $2\Delta t$, and
$\mathcal G(f)=10^3(fT)^{-13/3}$. We use the equal-depth transverse kernel in Eq.~\eqref{dm:normalized_kernel} and integrate from $0.002/T$ to the Nyquist frequency.
Let the columns of $Z$ span the orthogonal complement of the timing design matrix $M$. Projecting both the signal and the covariance gives
\begin{equation}
 {\rm SNR}^2=(Z^T\boldsymbol s)^T
                   (Z^T\mathsf C Z)^{-1}(Z^T\boldsymbol s).
 \label{dm:projected_snr}
\end{equation}
For identical sampling and timing fits, projection commutes with the sum and difference transformation. The joint ${\rm SNR}^2$ is the sum of the two channel contributions.

For a quadratic timing fit at $d=0.003,0.01,0.1$ pc, the difference channel retains fractions $0.42759,0.49826,0.50000$ of template power, respectively. The joint signal-to-noise ratios are $10.863,11.705,10.208$ times those of \emph{one reference pulsar in the unpolarized background} with the same noise and fit. At $d\to0$ differencing cancels the template as well. A local signal affecting just one pulsar retains half its template power in the normalized difference. For this example a cubic fit reduces the single-pulsar projected template power to $12.1\%$ of its quadratic-fit value. With the corresponding colored covariance, the single-pulsar signal-to-noise ratio is $0.7023$ times its value for the quadratic fit. These values describe the controlled injection. They do not give the distant-pair or population benchmark of the original proposal or a dark matter abundance sensitivity for an observed cluster.

Doubling the number of frequency integration points from $160+800$ to $320+1600$ gives a relative change below $4.7\times10^{-12}$ in the signal-to-noise gain at the seven tested separations. Halving the low-frequency cutoff changes the gain by less than $2.5\times10^{-8}$. An independent $\widehat\Gamma(\rho,0)=1/2+j_0(\rho)/2+j_2(\rho)/8$ calculation checks the angular integral, and inversion of the full two-pulsar covariance checks the sum and difference calculation. If the true depth changes, the signal geometry and emission times must change with it.

\bibliographystyle{apsrev4-2}
\bibliography{references}

@article{Belgacem2020,
  author = {Belgacem, Enis and Kamionkowski, Marc},
  title = {Chirality of the gravitational-wave background and pulsar-timing arrays},
  journal = {Phys. Rev. D},
  volume = {102},
  pages = {023004},
  year = {2020},
  doi = {10.1103/PhysRevD.102.023004},
  eprint = {2004.05480},
  archivePrefix = {arXiv},
  primaryClass = {astro-ph.CO}
}

@article{Boitier2021,
  author = {Bo{\^i}tier, Adrian and Tiwari, Shubhanshu and Jetzer, Philippe},
  title = {Analytic series expansion of the overlap reduction function for gravitational wave search with pulsar timing arrays},
  journal = {Phys. Rev. D},
  volume = {103},
  pages = {064044},
  year = {2021},
  doi = {10.1103/PhysRevD.103.064044},
  eprint = {2011.13405},
  archivePrefix = {arXiv},
  primaryClass = {gr-qc}
}

@article{Curylo2026,
  author = {Cury{\l}o, Ma{\l}gorzata and Thrane, Eric and Lasky, Paul D. and Gaynor, Dawson S.},
  title = {A comprehensive framework for phase-coherent mapping of the gravitational-wave sky with pulsar timing arrays},
  journal = {Astrophys. J.},
  volume = {1009},
  pages = {166},
  year = {2026},
  eprint = {2604.19073},
  archivePrefix = {arXiv},
  primaryClass = {astro-ph.HE}
}

@article{Detweiler1979,
  author = {Detweiler, S.},
  title = {Pulsar timing measurements and the search for gravitational waves},
  journal = {Astrophys. J.},
  volume = {234},
  pages = {1100--1104},
  year = {1979},
  doi = {10.1086/157593}
}

@article{FreireRidolfi2018,
  author = {Freire, P. C. C. and Ridolfi, A.},
  title = {An algorithm for determining the rotation count of pulsars},
  journal = {Mon. Not. R. Astron. Soc.},
  volume = {476},
  pages = {4794--4805},
  year = {2018},
  doi = {10.1093/mnras/sty524},
  eprint = {1802.07211},
  archivePrefix = {arXiv}
}

@article{NG15CW,
  author = {Agazie, Gabriella and others},
  collaboration = {NANOGrav},
  title = {The {NANOGrav} 15 yr Data Set: Bayesian Limits on Gravitational Waves from Individual Supermassive Black Hole Binaries},
  journal = {Astrophys. J. Lett.},
  volume = {951},
  pages = {L50},
  year = {2023},
  doi = {10.3847/2041-8213/ace18a},
  eprint = {2306.16222},
  archivePrefix = {arXiv},
  primaryClass = {astro-ph.HE}
}

@misc{NG15CWData,
  author = {{The NANOGrav Collaboration}},
  title = {Continuous Wave Analysis of the {NANOGrav} 15-Year Dataset},
  howpublished = {Zenodo, version 1.0.0},
  year = {2023},
  doi = {10.5281/zenodo.8067506},
  note = {{DOI}: \href{https://doi.org/10.5281/zenodo.8067506}{\mbox{10.5281/zenodo.8067506}}}
}

@misc{NG15NoiseData,
  author = {{The NANOGrav Collaboration}},
  title = {Noise Spectra and Stochastic Background Sensitivity Curve for the {NG15}-year Dataset},
  howpublished = {Zenodo, version 1.0.0},
  year = {2023},
  doi = {10.5281/zenodo.8092346},
  note = {{DOI}: \href{https://doi.org/10.5281/zenodo.8092346}{\mbox{10.5281/zenodo.8092346}}}
}

@article{NG15Timing,
  author = {Agazie, Gabriella and others},
  collaboration = {NANOGrav},
  title = {The {NANOGrav} 15 yr Data Set: Observations and Timing of 68 Millisecond Pulsars},
  journal = {Astrophys. J. Lett.},
  volume = {951},
  pages = {L9},
  year = {2023},
  doi = {10.3847/2041-8213/acda9a},
  eprint = {2306.16217},
  archivePrefix = {arXiv},
  primaryClass = {astro-ph.HE}
}

@misc{NG15TimingData,
  author = {{The NANOGrav Collaboration}},
  title = {The {NANOGrav} 15-Year Data Set},
  howpublished = {Zenodo, version 2.1.0},
  year = {2025},
  doi = {10.5281/zenodo.16051178},
  note = {{DOI}: \href{https://doi.org/10.5281/zenodo.16051178}{\mbox{10.5281/zenodo.16051178}}}
}

@article{Ridolfi2016,
  author = {Ridolfi, A. and others},
  title = {Long-term observations of the pulsars in {47 Tucanae} -- {I}. A study of four elusive binary systems},
  journal = {Mon. Not. R. Astron. Soc.},
  volume = {462},
  pages = {2918--2933},
  year = {2016},
  doi = {10.1093/mnras/stw1850},
  eprint = {1607.07248},
  archivePrefix = {arXiv},
  primaryClass = {astro-ph.HE}
}

@article{Sazhin1978,
  author = {Sazhin, M. V.},
  title = {Opportunities for detecting ultralong gravitational waves},
  journal = {Sov. Astron.},
  volume = {22},
  pages = {36--38},
  year = {1978}
}

@misc{joint_angular_limit_JAL_DLMF18,
 author={{NIST Digital Library of Mathematical Functions}},
 title={Section 18.11: Relations to Other Functions, Mehler--Heine Formulas},
 url={https://dlmf.nist.gov/18.11},
 year={2026},
 note={Equations 18.11.1 and 18.11.5; accessed 2026-10-20}
}

@article{next_JimenezCruz2024,
  author = {Jim{\'e}nez Cruz, N. M. and Malhotra, Ameek and Tasinato, Gianmassimo and Zavala, Ivonne},
  title = {Measuring the circular polarization of gravitational waves with pulsar timing arrays},
  journal = {Phys. Rev. D}, volume = {110}, pages = {103505}, year = {2024},
  doi = {10.1103/PhysRevD.110.103505}, eprint = {2406.04957}, archivePrefix = {arXiv}, primaryClass = {astro-ph.CO}
}

@article{obs_Chen2026GC,
  author = {Chen, W. and others},
  title = {Fifteen New Millisecond Pulsars in {47 Tucanae}},
  journal = {Astron. Astrophys.},
  volume = {710},
  pages = {A130},
  year = {2026},
  doi = {10.1051/0004-6361/202659650},
  eprint = {2605.06492},
  archivePrefix = {arXiv},
  primaryClass = {astro-ph.HE}
}

@misc{obs_Cherukupalli2026DM,
  author = {Cherukupalli, Abhiram and Lee, Vincent S. H. and Berghaus, Kim V. and Zurek, Kathryn M.},
  title = {Close Pulsar Pairs See Dark Matter Substructure through the Gravitational-Wave Background},
  year = {2026},
  eprint = {2609.38305},
  archivePrefix = {arXiv},
  primaryClass = {astro-ph.CO},
  doi = {10.48550/arXiv.2609.38305},
  url = {https://arxiv.org/abs/2609.38305}
}

@article{obs_Maiorano2021GC,
  author = {Maiorano, Michele and de Paolis, Francesco and Nucita, Achille},
  title = {Including millisecond pulsars inside the core of globular clusters in pulsar timing arrays},
  journal = {Eur. Phys. J. Plus},
  volume = {136},
  pages = {1087},
  year = {2021},
  doi = {10.1140/epjp/s13360-021-02098-0},
  eprint = {2110.11046},
  archivePrefix = {arXiv},
  primaryClass = {astro-ph.GA}
}

@article{obs_Prager2017GC,
  author = {Prager, Brian J. and Ransom, Scott M. and Freire, Paulo C. C. and Hessels, Jason W. T. and Stairs, Ingrid H. and Arras, Phil and Cadelano, Mario},
  title = {Using Long-Term Millisecond Pulsar Timing to Obtain Physical Characteristics of the Bulge Globular Cluster {Terzan 5}},
  journal = {Astrophys. J.},
  volume = {845},
  pages = {148},
  year = {2017},
  doi = {10.3847/1538-4357/aa7ed7},
  eprint = {1612.04395},
  archivePrefix = {arXiv},
  primaryClass = {astro-ph.SR}
}

@article{obs_Reardon2024Near,
  author = {Reardon, Daniel J. and others},
  title = {The Neutron Star Mass, Distance, and Inclination from Precision Timing of the Brilliant Millisecond Pulsar {J0437--4715}},
  journal = {Astrophys. J. Lett.},
  volume = {971},
  pages = {L18},
  year = {2024},
  doi = {10.3847/2041-8213/ad614a},
  eprint = {2407.07132},
  archivePrefix = {arXiv},
  primaryClass = {astro-ph.HE}
}

@article{obs_Smith2024GC,
  author = {Smith, Peter J. and H{\'e}nault-Brunet, Vincent and Dickson, Nolan and Gieles, Mark and Baumgardt, Holger},
  title = {Probing Populations of Dark Stellar Remnants in the Globular Clusters {47 Tuc} and {Terzan 5} Using Pulsar Timing},
  journal = {Astrophys. J.},
  volume = {975},
  pages = {268},
  year = {2024},
  doi = {10.3847/1538-4357/ad77bc},
  eprint = {2407.06274},
  archivePrefix = {arXiv},
  primaryClass = {astro-ph.GA}
}

@article{obs_Urquhart2026GC,
  author = {Urquhart, Ryan and Strader, Jay and Chomiuk, Laura and Ransom, Scott M. and Heinke, Craig O. and Bahramian, Arash and Maccarone, Thomas J.},
  title = {New Deep Radio Continuum Imaging Still Indicates a Large Reservoir of Undiscovered Millisecond Pulsars in {Terzan 5}},
  journal = {Mon. Not. R. Astron. Soc.},
  volume = {547},
  pages = {stag276},
  year = {2026},
  doi = {10.1093/mnras/stag276},
  eprint = {2602.10199},
  archivePrefix = {arXiv},
  primaryClass = {astro-ph.HE}
}

@misc{paper_AndrianovKopeikin2026,
  author = {Andrianov, S. A. and Kopeikin, S. M.},
  title = {Gravitational-Wave Sky Mapping with Pulsar Timing Arrays: The Full Earth-Pulsar Response and Resolution Operational Regimes},
  year = {2026}, eprint = {2607.19329}, archivePrefix = {arXiv}, primaryClass = {astro-ph.HE},
  url = {https://arxiv.org/abs/2607.19329}
}

@article{paper_Anholm2009,
  author = {Anholm, Melissa and Ballmer, Stefan and Creighton, Jolien D. E. and Price, Larry R. and Siemens, Xavier},
  title = {Optimal strategies for gravitational wave stochastic background searches in pulsar timing data},
  journal = {Phys. Rev. D}, volume = {79}, pages = {084030}, year = {2009},
  doi = {10.1103/PhysRevD.79.084030}, eprint = {0809.0701}, archivePrefix = {arXiv}, primaryClass = {gr-qc}
}

@article{paper_BernardoLiuNg2024,
  author = {Bernardo, Reginald Christian and Liu, Guo-Chin and Ng, Kin-Wang},
  title = {Correlations for an anisotropic polarized stochastic gravitational wave background in pulsar timing arrays},
  journal = {J. Cosmol. Astropart. Phys.}, volume = {2024}, number = {04}, pages = {034}, year = {2024},
  doi = {10.1088/1475-7516/2024/04/034}, eprint = {2312.03383}, archivePrefix = {arXiv}, primaryClass = {gr-qc}
}

@article{paper_CPTA2023Evidence,
  author = {Xu, Heng and others},
  title = {Searching for the Nano-Hertz Stochastic Gravitational Wave Background with the {Chinese Pulsar Timing Array} Data Release {I}},
  journal = {Res. Astron. Astrophys.}, volume = {23}, pages = {075024}, year = {2023},
  doi = {10.1088/1674-4527/acdfa5}, eprint = {2306.16216}, archivePrefix = {arXiv}, primaryClass = {astro-ph.HE}
}

@article{paper_ChuLiuNg2021,
  author = {Chu, Yu-Kuang and Liu, Guo-Chin and Ng, Kin-Wang},
  title = {Observation of polarized stochastic gravitational-wave background in pulsar-timing-array experiments},
  journal = {Phys. Rev. D}, volume = {104}, pages = {124018}, year = {2021},
  doi = {10.1103/PhysRevD.104.124018}, eprint = {2107.00536}, archivePrefix = {arXiv}, primaryClass = {gr-qc}
}

@article{paper_EPTA2023Evidence,
  author = {Antoniadis, J. and others},
  title = {The second data release from the {European Pulsar Timing Array}. {III}. Search for gravitational wave signals},
  journal = {Astron. Astrophys.}, volume = {678}, pages = {A50}, year = {2023},
  doi = {10.1051/0004-6361/202346844}, eprint = {2306.16214}, archivePrefix = {arXiv}, primaryClass = {astro-ph.HE}
}

@article{paper_EstabrookWahlquist1975,
  author = {Estabrook, F. B. and Wahlquist, H. D.},
  title = {Response of {Doppler} spacecraft tracking to gravitational radiation},
  journal = {Gen. Relativ. Gravit.}, volume = {6}, pages = {439--447}, year = {1975},
  doi = {10.1007/BF00762449}
}

@article{paper_Freire2017Cluster,
  author = {Freire, P. C. C. and others},
  title = {Long-term Observations of the Pulsars in {47 Tucanae} -- {II}. Proper Motions, Accelerations and Jerks},
  journal = {Mon. Not. R. Astron. Soc.}, volume = {471}, pages = {857--876}, year = {2017},
  doi = {10.1093/mnras/stx1533},
  eprint = {1706.04908}, archivePrefix = {arXiv}, primaryClass = {astro-ph.HE}
}

@article{paper_Gair2014,
  author = {Gair, Jonathan R. and Romano, Joseph D. and Taylor, Stephen and Mingarelli, Chiara M. F.},
  title = {Mapping gravitational-wave backgrounds using methods from {CMB} analysis: Application to pulsar timing arrays},
  journal = {Phys. Rev. D}, volume = {90}, pages = {082001}, year = {2014},
  doi = {10.1103/PhysRevD.90.082001}, eprint = {1406.4664}, archivePrefix = {arXiv}, primaryClass = {gr-qc}
}

@article{paper_HellingsDowns1983,
  author = {Hellings, R. W. and Downs, G. S.},
  title = {Upper limits on the isotropic gravitational radiation background from pulsar timing analysis},
  journal = {Astrophys. J. Lett.}, volume = {265}, pages = {L39--L42}, year = {1983},
  doi = {10.1086/183954}
}

@article{paper_KatoSoda2016,
  author = {Kato, Ryo and Soda, Jiro},
  title = {Probing circular polarization in stochastic gravitational wave background with pulsar timing arrays},
  journal = {Phys. Rev. D}, volume = {93}, pages = {062003}, year = {2016},
  doi = {10.1103/PhysRevD.93.062003}, eprint = {1512.09139}, archivePrefix = {arXiv}, primaryClass = {gr-qc}
}

@article{paper_KumarEtAl2024,
  author = {{Anil Kumar}, Neha and {\c{C}al{\i}{\c{s}}kan}, Mesut and Sato-Polito, Gabriela and Kamionkowski, Marc and Ji, Lingyuan},
  title = {Linear polarization of the stochastic gravitational-wave background with pulsar timing arrays},
  journal = {Phys. Rev. D}, volume = {110}, pages = {043501}, year = {2024},
  doi = {10.1103/PhysRevD.110.043501}, eprint = {2312.03056}, archivePrefix = {arXiv}, primaryClass = {astro-ph.CO}
}

@article{paper_KumarKamionkowski2024,
  author = {{Anil Kumar}, Neha and Kamionkowski, Marc},
  title = {Efficient Computation of Overlap Reduction Functions for Pulsar Timing Arrays},
  journal = {Phys. Rev. Lett.}, volume = {133}, pages = {151401}, year = {2024},
  doi = {10.1103/PhysRevLett.133.151401}, eprint = {2311.14159}, archivePrefix = {arXiv}, primaryClass = {astro-ph.CO}
}

@article{paper_LiuNg2022,
  author = {Liu, Guo-Chin and Ng, Kin-Wang},
  title = {Timing-residual power spectrum of a polarized stochastic gravitational-wave background in pulsar-timing-array observation},
  journal = {Phys. Rev. D}, volume = {106}, pages = {064004}, year = {2022},
  doi = {10.1103/PhysRevD.106.064004}, eprint = {2201.06767}, archivePrefix = {arXiv}, primaryClass = {gr-qc}
}

@article{paper_Mingarelli2013,
  author = {Mingarelli, C. M. F. and Sidery, T. and Mandel, I. and Vecchio, A.},
  title = {Characterizing gravitational wave stochastic background anisotropy with pulsar timing arrays},
  journal = {Phys. Rev. D}, volume = {88}, pages = {062005}, year = {2013},
  doi = {10.1103/PhysRevD.88.062005}, eprint = {1306.5394}, archivePrefix = {arXiv}, primaryClass = {astro-ph.HE}
}

@article{paper_MingarelliMingarelli2018,
  author = {Mingarelli, Chiara M. F. and Mingarelli, Angelo B.},
  title = {Proving the short-wavelength approximation in pulsar timing array gravitational-wave background searches},
  journal = {J. Phys. Commun.}, volume = {2}, pages = {105002}, year = {2018},
  doi = {10.1088/2399-6528/aae06d}, eprint = {1806.06979}, archivePrefix = {arXiv}, primaryClass = {astro-ph.IM}
}

@article{paper_MingarelliSidery2014,
  author = {Mingarelli, Chiara M. F. and Sidery, Trevor},
  title = {Effect of small interpulsar distances in stochastic gravitational wave background searches with pulsar timing arrays},
  journal = {Phys. Rev. D}, volume = {90}, pages = {062011}, year = {2014},
  doi = {10.1103/PhysRevD.90.062011}, eprint = {1408.6840}, archivePrefix = {arXiv}, primaryClass = {astro-ph.HE}
}

@article{paper_NANOGrav2023Anisotropy,
  author = {Agazie, Gabriella and others}, collaboration = {NANOGrav},
  title = {The {NANOGrav} 15 yr Data Set: Search for Anisotropy in the Gravitational-wave Background},
  journal = {Astrophys. J. Lett.}, volume = {956}, pages = {L3}, year = {2023},
  doi = {10.3847/2041-8213/acf4fd}, eprint = {2306.16221}, archivePrefix = {arXiv}, primaryClass = {astro-ph.HE}
}

@article{paper_NANOGrav2023Evidence,
  author = {Agazie, Gabriella and others}, collaboration = {NANOGrav},
  title = {The {NANOGrav} 15 yr Data Set: Evidence for a Gravitational-wave Background},
  journal = {Astrophys. J. Lett.}, volume = {951}, pages = {L8}, year = {2023},
  doi = {10.3847/2041-8213/acdac6}, eprint = {2306.16213}, archivePrefix = {arXiv}, primaryClass = {astro-ph.HE}
}

@article{paper_PPTA2023Evidence,
  author = {Reardon, Daniel J. and others},
  title = {Search for an Isotropic Gravitational-wave Background with the {Parkes Pulsar Timing Array}},
  journal = {Astrophys. J. Lett.}, volume = {951}, pages = {L6}, year = {2023},
  doi = {10.3847/2041-8213/acdd02}, eprint = {2306.16215}, archivePrefix = {arXiv}, primaryClass = {astro-ph.HE}
}

@article{paper_SatoPolitoKamionkowski2022,
  author = {Sato-Polito, Gabriela and Kamionkowski, Marc},
  title = {Pulsar-timing measurement of the circular polarization of the stochastic gravitational-wave background},
  journal = {Phys. Rev. D}, volume = {106}, pages = {023004}, year = {2022},
  doi = {10.1103/PhysRevD.106.023004}, eprint = {2111.05867}, archivePrefix = {arXiv}, primaryClass = {astro-ph.CO}
}

@article{paper_TaylorGair2013,
  author = {Taylor, Stephen R. and Gair, Jonathan R.},
  title = {Searching for anisotropic gravitational-wave backgrounds using pulsar timing arrays},
  journal = {Phys. Rev. D}, volume = {88}, pages = {084001}, year = {2013},
  doi = {10.1103/PhysRevD.88.084001}, eprint = {1306.5395}, archivePrefix = {arXiv}, primaryClass = {gr-qc}
}

@article{paper_ZhouLi2026,
  author = {Zhou, Kun and Li, Jin},
  title = {Full analytic expressions of overlap reduction functions for anisotropies of the stochastic gravitational-wave background with pulsar timing arrays},
  journal = {Phys. Rev. D}, volume = {114}, pages = {044003}, year = {2026},
  doi = {10.1103/34y2-pn7d}, eprint = {2608.07095}, archivePrefix = {arXiv}, primaryClass = {gr-qc}
}

@article{struct_TPHazboun2019,
 author={Hazboun, Jeffrey S. and Romano, Joseph D. and Smith, Tristan L.},
 title={Realistic sensitivity curves for pulsar timing arrays},
 journal={Phys. Rev. D}, volume={100}, pages={104028}, year={2019},
 doi={10.1103/PhysRevD.100.104028}, eprint={1907.04341}, archivePrefix={arXiv}, primaryClass={gr-qc}
}

@article{struct_TPvanHaasteren2013,
 author={van Haasteren, Rutger and Levin, Yuri},
 title={Understanding and analysing time-correlated stochastic signals in pulsar timing},
 journal={Mon. Not. Roy. Astron. Soc.}, volume={428}, pages={1147--1159}, year={2013},
 doi={10.1093/mnras/sts097}, eprint={1202.5932}, archivePrefix={arXiv}, primaryClass={astro-ph.IM}
}

@article{uniform_asymptotics_Hu2022,
 author={Hu, Yu and Wang, Pan-Pan and Tan, Yu-Jie and Shao, Cheng-Gang},
 title={Full analytic expression of overlap reduction function for gravitational wave background with pulsar timing arrays},
 journal={Phys. Rev. D},volume={106},pages={024005},year={2022},
 doi={10.1103/PhysRevD.106.024005},eprint={2205.09272},archivePrefix={arXiv}, primaryClass={gr-qc}
}

@article{uniform_asymptotics_uniform_NG15Noise2023,
  author = {Agazie, Gabriella and others},
  collaboration = {NANOGrav},
  title = {The {NANOGrav} 15-Year Data Set: Detector Characterization and Noise Budget},
  journal = {Astrophys. J. Lett.},
  volume = {951},
  pages = {L10},
  year = {2023},
  doi = {10.3847/2041-8213/acda88},
  eprint = {2306.16218},
  archivePrefix = {arXiv},
  primaryClass = {astro-ph.HE}
}

@CONTROL{apsrev42Control,author="08",editor="1",pages="1",title="0",year="1"}
\end{document}